\documentclass[a4paper,11pt]{article}
\pdfoutput=1 

\usepackage{jheppub} 

\usepackage[T1]{fontenc} 

\usepackage{ifpdf}
\usepackage{subfigure}
\newcommand{\nn}{\nonumber}
\newcommand{\lsim}{\mathrel{\mathop{\kern 0pt \rlap
  {\raise.2ex\hbox{$<$}}}
  \lower.9ex\hbox{\kern-.190em $\sim$}}}
\newcommand{\gsim}{\mathrel{\mathop{\kern 0pt \rlap
  {\raise.2ex\hbox{$>$}}}
  \lower.9ex\hbox{\kern-.190em $\sim$}}}

\newcommand{\be}{\begin{equation}}
\newcommand{\ee}{\end{equation}}
\newcommand{\bea}{\begin{eqnarray}}
\newcommand{\eea}{\end{eqnarray}}

\def\ptmiss{\not\!\!{p_T}}

\title{\boldmath  Perspectives on a Supersymmetric Extension of the Standard Model  
with a $Y=0$ Higgs Triplet and a Singlet  at the LHC }

\author[a]{Priyotosh Bandyopadhyay}

\author[a,b]{Claudio Corian\`o}
\author[a]{Antonio Costantini}

\affiliation[a]{Dipartimento di Matematica e Fisica "Ennio De Giorgi", \\ Universit\`a del Salento and INFN-Lecce, \\ Via Arnesano, 73100 Lecce, Italy}

\affiliation[b]{STAG Research Centre and Mathematical Sciences,\\ University of Southampton, Southampton SO17 1BJ, UK}

\emailAdd{priyotosh.bandyopadhyay@le.infn.it}
\emailAdd{claudio.coriano@le.infn.it}
\emailAdd{antonio.costantini@le.infn.it}

\abstract{We investigate a supersymmetric extension of the Minimal Supersymmetric Standard Model (MSSM), called the 
TNMSSM, containing a $SU(2)$ Higgs triplet $(\hat{T})$ of $Y=0$ hypercharge and a singlet superfields $(\hat{S})$ in the corresponding superpotential. The model can be viewed, equivalently, as an extension of the NMSSM with the addition of a $\hat{T}-\hat{S}$ interaction and of an extra coupling of the triplet to the two Higgs doublets of the NMSSM. In this scenario the Higgs particle spectrum at tree-level gets additional mass contributions from the triplet and singlet scalar components respect to the MSSM, which are  particularly enhanced at low $\tan{\beta}$. We calculate the one-loop Higgs masses for the neutral physical Higgs bosons by a Coleman-Weinberg effective potential approach. In particular, we investigate separately the impact of the radiative corrections due to the electroweak, gauge-gaugino-higgsino, fermion-sfermion and Higgs self-interactions to the Higgs masses. Due to the larger number of scalars and of triplet and singlet couplings, the Higgs corrections can be larger than the strong corrections. This reduces the amount of fine-tuning required to fit the recent Higgs data. Using the expressions of the beta-functions of the model, we show that the large triplet singlet coupling remains perturbative up to
$\sim10^{8-10}$ GeV. The model is also characterized by a light pseudoscalar in the spectrum, which is a linear combination of the triplet, doublet and singlet CP-odd components. We discuss the production and decay signatures of the Higgs bosons in this model, including scenarios with hidden Higgses, which could be investigated at the LHC in the current run. }

\begin{document}
\maketitle
\flushbottom

\section{Introduction}
With the recent discovery of the Higgs boson at the Large Hadron Collider, the mechanism responsible for the breaking of the electroweak symmetry has finally been uncovered and it has been shown to involve at least one scalar field along the lines of the Standard Model (SM) description. This discovery has removed, at least in part, previous doubts about the real existence of a scalar with Higgs-like properties in our Universe. 
Both the CMS \cite{CMS, CMS2} and the ATLAS \cite{ATLAS} experimental collaborations have confirmed the discovery of a Higgs boson, by an analysis of the $\gamma \gamma, ZZ^*,$  and $WW^*$ decay channels of the Higgs particle - as predicted by the Standard Model (SM) - at a confidence level of more than $5\sigma$, except for the $WW^*$ decay rate, which has been recorded with a $4.7\sigma$ accuracy by CMS \cite{CMS2}. The fermionic decay modes, instead, have still to reach the $5\sigma$ accuracy, and show some disagreement in the results elaborated by the two experimental collaborations. Clearly, the disagreement of the experimental results with the predictions from the SM opens the possibility of further investigation of the Higgs sector. \\
  For such reasons, it is widely believed that the SM is not a complete theory, being not able, for instance, to account for the 
 neutrino masses, but also for being affected, in the scalar sector, by the gauge hierarchy problem \cite{hierarchy}.
 The widespread interest in the study of a possible supersymmetric extension of this model has always being motivated with the goal of finding a natural and elegant solution to this problem. In fact, supersymmetry protects the Higgs mass from the undesired quadratic divergences introduced by the radiative corrections in the scalar sector of the SM, and it does so  by the inclusion of superpartners. \\
In the minimal supersymmetric extension of the SM (MSSM) the conditions of analyticity of the superpotential and of absence of the gauge anomalies require a minimal extensions of the scalar sector with two Higgs superfields, in the forms of $SU(2)$ doublets carrying opposite hypercharges $(Y)$. Supersymmetric extensions are, in general, characterized by a large set of additional parameters which render their phenomenological study quite involved. For this reason, in the near past, the interest has turned towards models, such as the constrained minimal supersymmetric extension (cMSSM/mSUGRA), with only 5 new parameters, generated at a large supergravity scale, quite close to the Planck scale \cite{cMSSM}. 

Unlike the SM case, in the MSSM the tree-level  mass of the lightest Higgs ($h_1$)  $m_{h_1}$ is not a free parameter but it is constrained to lay below the mass of the Z gauge boson, $m_Z$  ($m_{h_1}\leq m_Z$).   This constraint has been in tension with the results of the experimental searches at LEP-2  which have failed to detect any CP-even Higgs below $m_Z$ and which had established a lower bound of $114.5$ GeV for the SM Higgs boson \cite{LEPb}. With the recent discovery of a CP-even Higgs boson around 125 GeV \cite{CMS, CMS2, ATLAS} the resolution of this conflict is, therefore, mandatory. 

Even in this unfavourable situation, supersymmetric scenarios remain popular due their naturalness and for having a dark matter candidate in theories with a conserved $R$-parity symmetry \cite{rparity}. To avoid the conflict between the MSSM prediction for the Higgs and 
the LHC results, one needs to consider the effect of the radiative corrections which could lift the bound on the Higgs mass in this model. It has been shown - and it is now well known - that in the case of the MSSM the significant radiative corrections come from the stop-top corrections, specially at low $\tan{\beta}$, due to large Yukawa couplings and to the presence of colour charges. This has triggered analysis envisioning scenarios with a heavy stop, which require a very high supersymmetric (SUSY) scale for the most constrained supersymmetric models like mSUGRA/cMSSM, AMSB, etc \cite{cMSSMb}. In the case of the phenomenological MSSM (pMSSM) there are two possibilities: a very large third generation SUSY mass scale and/or a large splitting between the stop mass eigenstates \cite{pMSSMb}. The second case leads to large soft trilinear  couplings $\gsim 2$ TeV \cite{pMSSMb}, which brings back the fine-tuning problem in a different way. \\
A possible way to address the fine tuning problem is to consider an extended Higgs sector.  In this respect, there are some choices which could resolve it, based on the inclusion of one singlet \cite{nmssm} and of one or more triplet superfields of appropriate hypercharges \cite{tssmHiggsb}. In particular, the addition of a $Y=0$ hypercharge superfield gives large tree-level as well as 
one-loop corrections to the Higgs masses, and relaxes the fine tuning problem of the MSSM by requiring a lower SUSY mass scale \cite{tssmyzero, DiChiara:2008rg}. There are some special features of these extensions which are particularly interesting and carry specific signatures.  For instance, the addition of a ($Y=0,\pm 2$ hypercharge) - ($SU(2)$ triplet) Higgs sector induces $H^\pm-W^\mp-Z$ couplings mediated by the non-zero vacuum expectation value (vev) of the Higgs triplet, due to the breaking of the custodial symmetry \cite{tssmch1prime,tssmch1}. Other original features of the $Y=\pm 2$ hypercharge triplets are the presence of doubly charged Higgs in the spectrum \cite{tssmch2}. There are also other significant constraints which are typical of these scenarios, and which may help in the experimental analysis. In the supersymmetric Higgs triplet extension, the vev of the triplet $v_T$ is highly constraints by the $\rho$ parameter \cite{rho}, which leads to $v_T\lesssim 5$ GeV in the case of $Y=0$ triplets. In the same case, this value of 
$v_T$ can account for the value of the mixing parameter $\mu_D$ of the 2-Higgs doublets (or $\mu$-term), which remains small in the various possible scenarios. Another dynamical way to generate a $\mu_D$ term is by adding a SM gauge singlet superfield to the spectrum \cite{nmssm}, as in the NMSSM. Thus a triplet-singlet extended supersymmetric SM  built on the superpotential of the MSSM, can address both the fine tuning issue and resolve, at the same time, the problem of the $\mu$-term of the two Higgs doublets \cite{tnssm, FileviezPerez:2012gg, Agashe:2011ia}. \\
 In this work we are going to investigate these extensions, presenting results on the spectrum of these models.  
    We will also see that the addition of a discrete symmetry in this model removes the mass terms from the superpotential and its continuum limit generates a Nambu-Goldstone pseudoscalar particle in the spectrum, characterising some of its most significant features. A more thorough analysis of this specific aspect will be presented elsewhere. The goal of our study is to investigate the allowed region of their parameter space in view of the experimental constraints emerging from the recent experimental results at the LHC.\\ 
 Our work is organized as follows. We start by introducing a scale invariant superpotential which is MSSM-like, but with the inclusion of a $Y=0$ Higgs triplet and of an extra SM gauge singlet superfield. In section \ref{model} we detail the model. In section \ref{treel}  we investigate both the possibility of generating a tree-level Higgs mass around 125 GeV due to the extra contributions from the triplet and the singlet, and the possibility of hidden Higgs bosons. Section \ref{storngw} describes the strong and weak sectors of the model, followed by a discussion of the strong and weak   contributions to the radiative corrections of the Higgs masses in section~\ref{onel}. We address in section \ref{beta} the perturbative running of the couplings as we vary the scale of the theory, and in section \ref{finet} we examine the issue of fine-tuning in this model. A study of the light pseudoscalar state  is investigated in section \ref{axion}. Finally, in section \ref{Hdata} we take into account the constraints from the LHC Higgs data  and LEP data to find out the phenomenological parameter space. In section \ref{pheno} 
the new possible production and decay channels are given along with the signatures that could be tested at the LHC and at the future colliders.  In section \ref{dis}  we present our conclusions, summarizing our results and elaborate on possible extensions of this work. We have left to an appendix both the expressions of the renormalization group equations of the dimensionless couplings and the tree-level vertices of the Higgs sector.

\section{The Model}\label{model}
We consider a scale invariant superpotential $W_{TNMSSM}$ with an extended Higgs
sector containing a $Y=0$ SU(2) triplet $\hat{T}$ and a SM gauge singlet ${\hat S}$ (see \cite{tnssm, FileviezPerez:2012gg}) on top of the superpotential of the MSSM. We recall that the inclusion of the singlet superfield on the superpotential of the MSSM realizes the NMSSM superpotential. We prefer to separate the complete superpotential of the model into a MSSM part, 
\begin{equation}
W_{MSSM}= y_t \hat U \hat H_u\!\cdot\! \hat Q - y_b \hat D \hat H_d\!\cdot\! \hat Q - y_\tau \hat E \hat H_d\!\cdot\! \hat L\ ,
\label{spm}
 \end{equation}
 where ''$\cdot$'' denotes a contraction with the Levi-Civita symbol $\epsilon^{ij}$, with $\epsilon^{12}=+1$, and combine the singlet superfield $(\hat{S})$ and the triplet contributions into a second superpotential 
\begin{equation}
W_{TS}=\lambda_T  \hat H_d \cdot \hat T  \hat H_u\, + \, \lambda_S S  \hat H_d \cdot  \hat H_u\,+ \frac{\kappa}{3}S^3\,+\,\lambda_{TS} S  \textrm{Tr}[T^2]
\label{spt}
 \end{equation}
 with 
 \begin{equation}
 W_{TNMSSM}=W_{MSSM} + W_{TS}.
 \end{equation}
  The triplet and doublet superfields are given by 

\begin{equation}\label{spf}
 \hat T = \begin{pmatrix}
       \sqrt{\frac{1}{2}}\hat T^0 & \hat T_2^+ \cr
      \hat T_1^- & -\sqrt{\frac{1}{2}}\hat T^0
       \end{pmatrix},\qquad \hat{H}_u= \begin{pmatrix}
      \hat H_u^+  \cr
       \hat H^0_u
       \end{pmatrix},\qquad \hat{H}_d= \begin{pmatrix}
      \hat H_d^0  \cr
       \hat H^-_d
       \end{pmatrix}.
 \end{equation}
 Here $\hat T^0$ is a complex neutral superfield, while  $\hat T_1^-$ and $\hat T_2^+$ are the charged Higgs superfields. Note that $(\hat{T}_1^-)^*\neq \hat{T}_2^+$. 
 Only the MSSM Higgs doublets couple to the fermion multiplet via Yukawa coupling as in Eq.~(\ref{spm}), while the singlet and the triplet superfields generate the supersymmetric $\mu_D$ term after their neutral parts  acquire  vevs, as shown in Eq.~(\ref{spt}).
 
 In any scale invariant supersymmetric theory with a cubic superpotential, the complete Lagrangian with the soft SUSY breaking terms has an accidental  $Z_3$ symmetry, the invariance after the multiplication of all the components of the chiral superfield by the phase $e^{2\pi i/3}$.
Such terms are given by
 \bea\nn
V_{soft}& =&m^2_{H_u}|H_u|^2\, +\, m^2_{H_d}|H_d|^2\, +\, m^2_{S}|S|^2\, +\, m^2_{T}|T|^2\,+\, m^2_{Q}|Q|^2 + m^2_{U}|U|^2\,+\,m^2_{D}|D|^2 \\ \nn
&&+(A_S S H_d.H_u\, +\, A_{\kappa} S^3\, +\, A_T H_d.T.H_u \, +\, A_{TS} S Tr(T^2)\\ 
 &&\,+\, A_U U H_U . Q\, +\, \, A_D D H_D . Q + h.c),
\label{softp}
 \eea
while the D-terms are given by 
 \begin{equation}
 V_D=\frac{1}{2}\sum_k g^2_k ({ \phi^\dagger_i t^a_{ij} \phi_j} )^2 .
 \label{dterm}
 \end{equation}
 
 In this article we assume that all the coefficients involved in the Higgs sector are real in order to preserve CP invariance. The breaking of the $SU(2)_L\times U(1)_Y$ electroweak symmetry is obtained by giving real vevs to the neutral components
 of the Higgs fields
 \be
 <H^0_u>=\frac{v_u}{\sqrt{2}}, \, \quad \, <H^0_d>=\frac{v_d}{\sqrt{2}}, \quad ,<S>=\frac{v_S}{\sqrt{2}} \, \quad\, <T^0>=\frac{v_T}{\sqrt{2}},
 \ee
 which give mass to the $W^\pm$ and $Z$ bosons
 \be
 m^2_W=\frac{1}{4}g^2_L(v^2 + 4v^2_T), \, \quad\ m^2_Z=\frac{1}{4}(g^2_L \, +\, g^2_Y)v^2, \, \quad v^2=(v^2_u\, +\, v^2_d) .
 \ee
 and also generate the $ \mu_D=\frac{\lambda_S}{\sqrt 2} v_S+ \frac{\lambda_T}{2} v_T$ term.
 
 The non-zero triplet contribution to the $W^\pm$ mass leads to a deviation of the tree-level expression of the $\rho$ parameter
 \be
 \rho= 1+ 4\frac{v^2_T}{v^2} .
 \ee
 Thus the triplet vev is strongly  constrained by the global fit on the measurement of the $\rho$ parameter \cite{rho}
 \be
 \rho =1.0004^{+0.0003}_{-0.0004} ,
 \ee 
 which restricts its value to $v_T \leq 5	$ GeV.  In our numerical analysis we have chosen  $v_T =3	$ GeV.

\section{Tree-level Higgs masses}\label{treel}

To determine the tree-level mass spectrum, we first consider the tree-level minimisation conditions, 
\be\label{mnc}
\partial_{\Phi_i}V|_{vev}=0; \quad V=V_D\, +\,V_F\,+\,V_{soft}, \quad <\Phi_{i,r}>=\frac{v_{i}}{\sqrt 2},\quad \Phi_i=H^0_{u},H^0_{d}, S, T^0,
\ee
where we have defined the vacuum parameterizations of the fields in the Higgs sector as 
\be
H^0_u=\frac{1}{\sqrt{2}}(H^0_{u,r} + i H^0_{u,i}), \quad H^0_d=\frac{1}{\sqrt{2}}(H^0_{d,r} + i H^0_{d,i}), \quad S=\frac{1}{\sqrt{2}}(S_r + i S_i), \quad T^0=\frac{1}{\sqrt{2}}(T^0_{r} + i T^0_{i}). 
\ee
from which the soft-breaking masses are derived in the form

\begin{align}\label{mnc2}
m^2_{H_u}=& \frac{v_d}{2\,v_u} \left(\sqrt{2} A_S v_S-v_T \left(A_T+\sqrt{2} v_S \lambda _T \lambda_{TS}\right)+\lambda _S \left(\kappa  v_S^2+v_T^2 \lambda_{TS}\right)\right)\nn\\
&-\frac{1}{2}\left(\lambda _S^2 \left(v_d^2-v_S^2\right)+ \frac{1}{2}\lambda _T^2\left(v_d^2+v_T^2\right)+\sqrt{2} \lambda _S v_S \lambda _T v_T\right)\nn\\
&+\frac{1}{8}(v_d^2 - v_u^2)
   \left(g_L^2+g_Y^2\right),
\end{align}
\begin{align}
m^2_{H_d}=& \frac{v_u}{2\,v_d} \left(\sqrt{2} A_S v_S-v_T \left(A_T+\sqrt{2} v_S\lambda _T \lambda _{TS}\right)+\lambda _S \left(\kappa  v_S^2+v_T^2 \lambda _{TS}\right)\right)\nn\\
   &-\frac{1}{2} \left(\lambda _S^2 \left(v_u^2+v_S^2\right)+\frac{1}{2}\lambda _T^2 \left(v_u^2+v_T^2\right)-
   \sqrt{2} \lambda _S v_S \lambda _T v_T\right)\nn\\
   &+\frac{1}{8}(v_u^2 - v_d^2)
   \left(g_L^2+g_Y^2\right),
\end{align}
\begin{align}
m^2_S=& \frac{1}{2 \sqrt{2} v_S}\left(v_T \left(\lambda _T \left(\lambda _S
   \left(v_d^2+v_u^2\right)-2 v_d v_u \lambda _{TS}\right)-2 A_{TS}
   v_T\right)+2 A_S v_d v_u\right)\nn\\
   &-\frac{A_{\kappa} v_S}{\sqrt{2}}+\kappa 
   v_d v_u \lambda _S-\frac{1}{2} \lambda _S^2 \left(v_d^2+v_u^2\right)-\kappa ^2
   v_S^2-\kappa  v_T^2 \lambda _{TS}-2 v_T^2 \lambda _{TS}^2,
\end{align}
\begin{align}
m^2_T=& \frac{1}{4 v_T}\left(\sqrt{2} v_S \lambda _T
   \left(\lambda _S \left(v_d^2+v_u^2\right)-2 v_d v_u \lambda _{TS}\right)-2
   A_T v_d v_u\right)-\sqrt{2} A_{TS} v_S\nn\\
   &+\lambda _{TS} \left(v_d
   v_u \lambda _S-v_S^2 \left(\kappa +2 \lambda _{TS}\right)\right)-\frac{1}{4} \lambda _T^2
   \left(v_d^2+v_u^2\right)-v_T^2 \lambda _{TS}^2.
\end{align}

\begin{figure}[t]
\begin{center}
\includegraphics[width=0.75\linewidth]{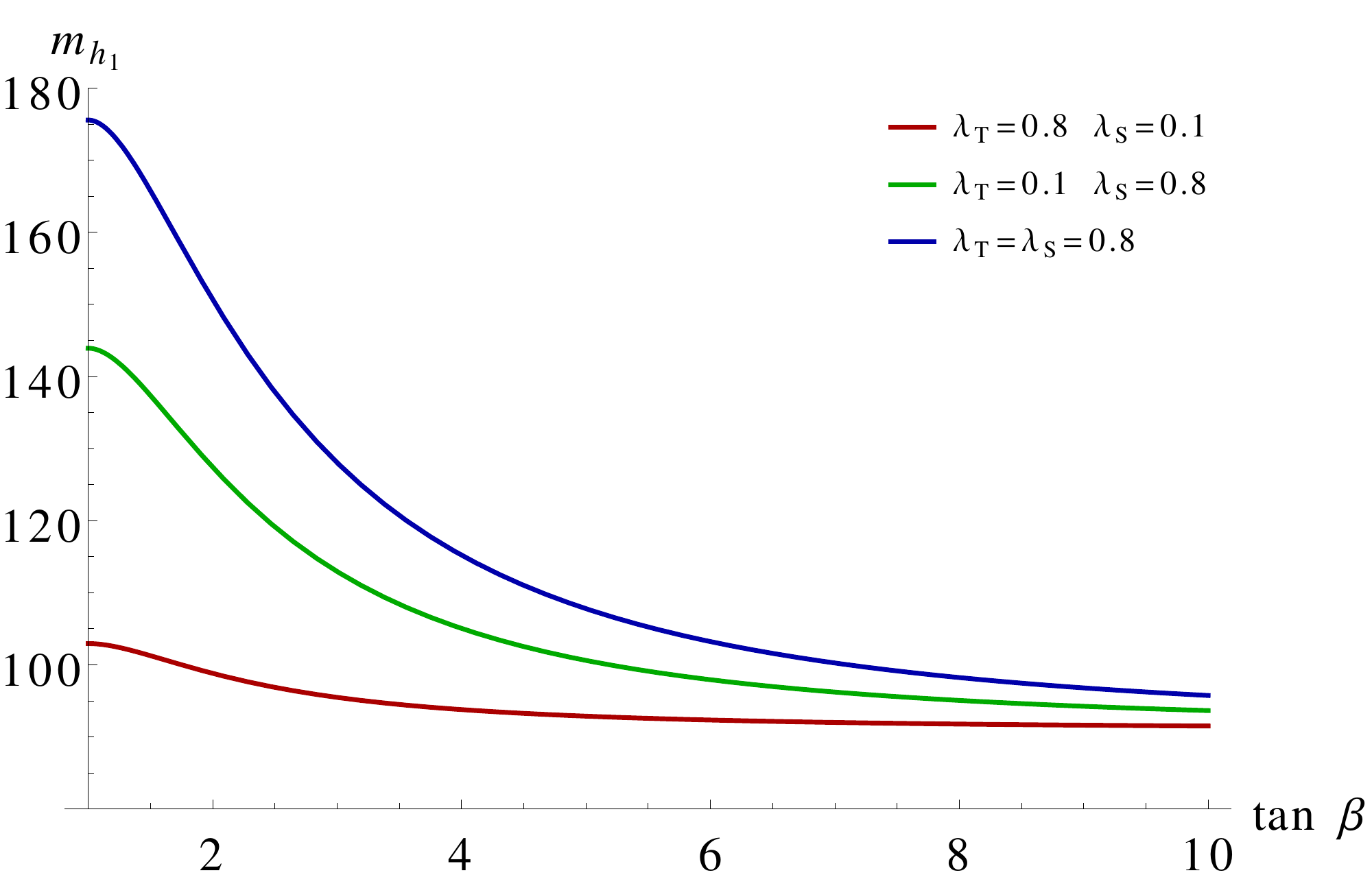}
\caption{Tree-level lightest CP-even Higgs mass maximum values with respect to $\tan{\beta}$ for 
 (i) $\lambda_T=0.8,\, \lambda_S=0.1$ (in red), (ii)$\lambda_T=0.1,\, \lambda_S=0.8$ (in green ) and (iii) $\lambda_T=0.8,\, \lambda_S=0.8$ (in blue).}\label{mht}
\end{center}
\end{figure}

It can be shown that the second derivative of the potential with respect to the fields satisfy the tree-level stability constraints. The neutral CP-even mass matrix in this case is $4$-by-$4$, since the mixing terms involve the two $SU(2)$ Higgs doublets, the scalar singlet $S$ and the neutral component of the Higgs triplet.  After electroweak symmetry breaking, the neutral Goldstone gives mass to the $Z$ boson and the charged Goldstone bosons give mass to the $W^\pm$ boson. Being the Lagrangean CP-symmetric, we are left with four CP-even, three CP-odd  and three charged Higgs bosons as shown below
 \bea\label{hspc}
  \rm{CP-even} &&\quad \quad  \rm{CP-odd} \quad\quad   \rm{charged}\nn \\
 h_1, h_2, h_3, h_4 &&\quad \quad a_1, a_2, a_3\quad \quad h^\pm_1, h^\pm_2, h^\pm_3. 
 \eea
The neutral Higgs bosons are combination of doublets, triplet and singlet, whereas the charged Higgses are a combination of doublets and triplet only. We will denote with $m_{h_i}$ the corresponding mass eigenvalues, assuming that one of them will coincide with the 125 GeV Higgs $(h_{125})$ boson detected at the LHC. The scenarios that we consider do not assume that this is the lightest eigenvalue which is allowed in the spectrum of the theory. Both scenarios with lighter and heavier undetected Higgs states will be considered. In particular, we will refer to those in which one or more Higgses with a mass lower than 125 GeV is present, to {\em hidden Higgs} scenarios.  \\
At tree-level the maximum value of the lightest neutral Higgs has additional contributions from the triplet and the singlet sectors respectively. The numerical value of the upper bound on the lightest CP-even Higgs can be extracted from the relation
\be\label{hbnd}
m^2_{h_1}\leq m^2_Z(\cos^2{2\beta} \, +\, \frac{\lambda^2_T}{g^2_L\,+\,g^2_Y }\sin^2{2\beta}\, +\, \frac{2\lambda^2_S}{g^2_L\,+\,g^2_Y }\sin^2{2\beta}), \qquad \tan\beta=\frac{v_u}{v_d},
\ee
which is affected on its right-hand-side by two additional contributions from the triplet and singlet. These can raise the allowed tree-level Higgs mass. Both 
contributions are proportional to $\sin{2\beta}$, and thus they can be large for a low value of $\tan{\beta}$, as shown
in Figure~\ref{mht}. The plots indicate that for higher values of $\lambda_{T,S}$ a lightest 
tree-level Higgs boson mass of $\sim 125$ GeV can be  easily achieved. For general parameters,
the required quantum corrections needed in order to raise the mass bound are thus much smaller compared to the MSSM. In the case of the MSSM, as we have already mentioned, at tree-level  $m_h\leq m_Z$, and we need a correction $\gsim  35$ GeV to match the experimental value of the discovered Higgs boson mass, which leads to a fine-tuning of the SUSY parameters. In fact, this requires that the allowed parameter space of the MSSM is characterized either by large SUSY masses or by large splittings among the mass eigenvalues.
In fact, this requires that the allowed parameter space of the MSSM is characterized either by large SUSY masses or by large splittings among the mass eigenvalues.
We have first investigated  the tree-level mass spectrum for the Higgs bosons and analysed the prospect of a $\sim 125$ GeV Higgs boson along with the hidden Higgs scenarios.  We
have looked for tree-level mass eigenvalues where at least one of them corresponds to the Higgs
discovered at the LHC. For this purpose we have performed an initial scan of the parameter space  
\bea\label{parat}
|\lambda_{T, S, TS}| \leq 1, \quad |\kappa|\leq 3, \quad |v_s|\leq 1 \, \rm{TeV}, \quad 1\leq \tan{\beta}\leq 10, 
\eea
and searched for a CP-even Higgs boson around $100-150$ GeV, assuming that at least one of the 4 eigenvalues $m_{h_i}$ will fall within  the interval $123$ GeV $\leq m_{h_i}\leq 127$ GeV at one-loop.

\begin{figure}[t]
\begin{center}
\mbox{\hskip -15 pt\subfigure[]{\includegraphics[width=0.55\linewidth]{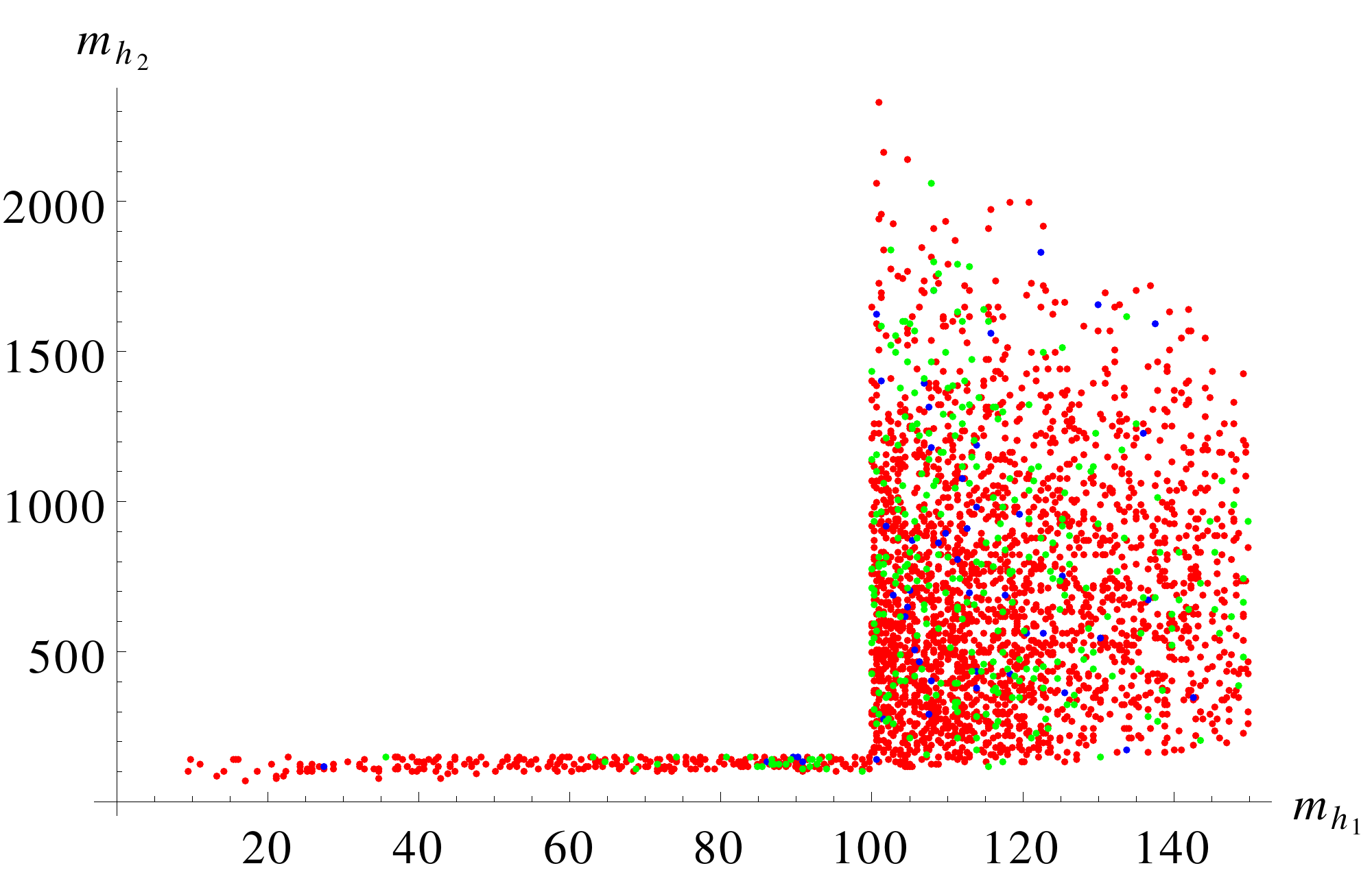}}
\subfigure[]{\includegraphics[width=0.55\linewidth]{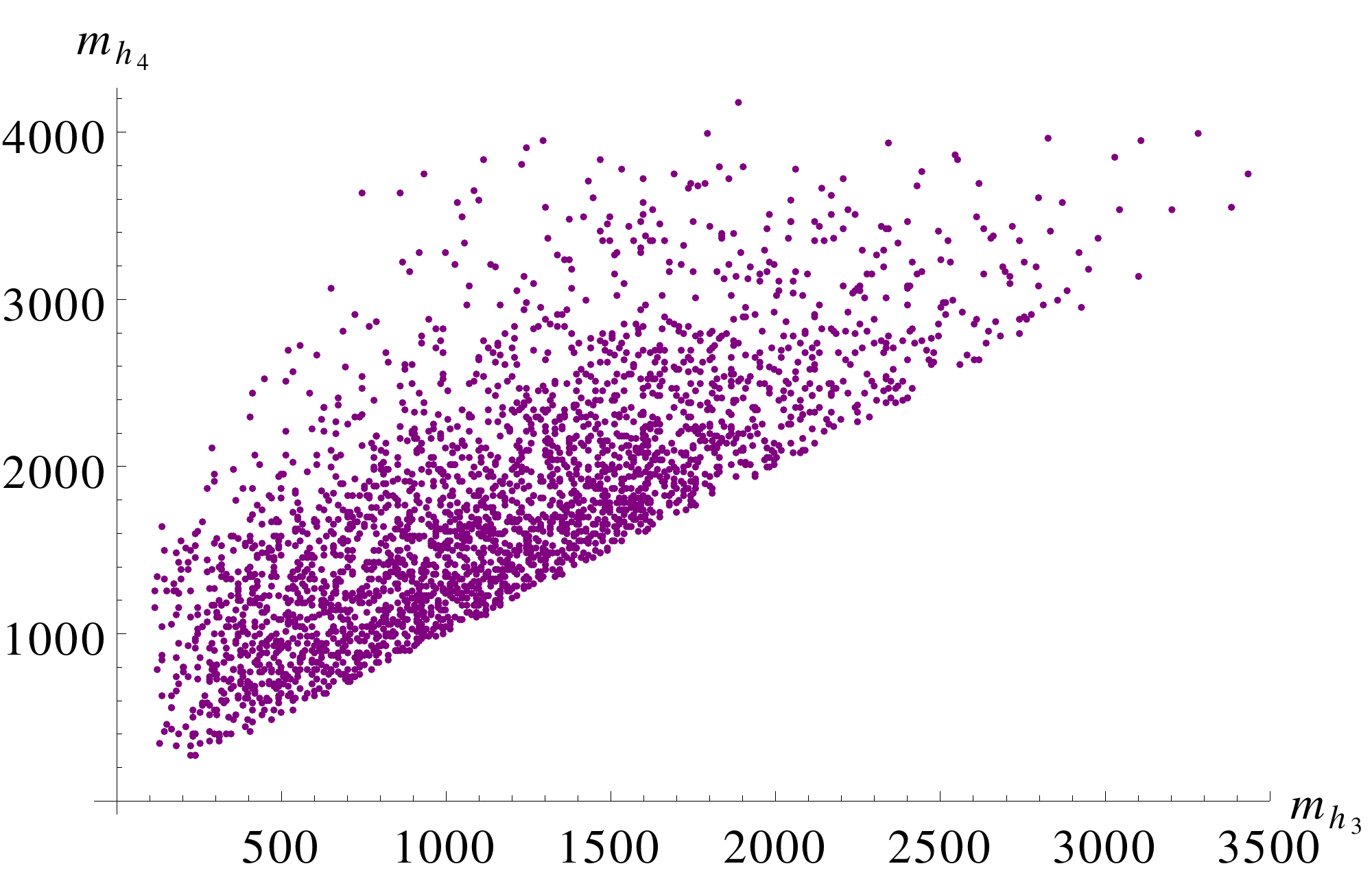}}}
\caption{Tree-level CP-even Higgs mass correlations (a) $m_{h_1}$ vs $m_{h_2}$ and (b) $m_{h_3}$ vs $m_{h_4}$, where we have a candidate $\sim 125$ GeV Higgs boson. The colors refers to the character of the $h_1$ mass eigenstate, describing the weights of the doublet, singlet and triplet  contributions in their linear combinations. Red points are $>90\%$ doublets-like, the green points are either $\geq 90\%$ triplet-like or singlet-like and blue points are mixtures of
doublet and triplet/singlet components. The linear combinations corresponding to green points are chosen to satisfy the constraints from LEP onto Z and lepton final states. }\label{hihj}
\end{center}
\end{figure}

 Figure~\ref{hihj}(a) presents the mass correlations between $m_{h_1}$ and $m_{h_2}$, where we have a CP-even neutral Higgs boson in the $100\leq m_{h_i}\leq 150$ GeV range. The candidate Higgs boson around $125$  GeV will be determined at one loop level by including positive and negative radiative corrections in the next section. The mass correlation plot
at tree-level shows that there are solutions with very light $h_1$, $m_{h_1}\leq 100$ GeV, which should be confronted with LEP data \cite{LEPb}.  At LEP were conducted searches for the Higgs boson via the $e^+ e^- \to Z h$ and 
$e^+ e^- \to  h_1 h_2$ channels (in models with multiple Higgs bosons) and their fermionic decay modes ($h\to b\bar{b},\tau \bar{\tau}$ and $Z\to \ell \ell$). The higher centre of mass energy at LEP II  (210 GeV) allowed to set a lower bound of 114.5 on the SM-like Higgs boson and of 93 GeV for the MSSM-like Higgs boson in the maximal mixing scenario \cite{LEPb}. Interestingly, neither the triplet (in our case) nor the singlet type Higgs boson couple to $Z$ or to leptons (see  Eq.~(\ref{spt})), and as such they are not excluded by LEP data.

We mark such points with $\geq 90\%$ triplet/singlet components, which can evade the LEP bounds, in green. In Figure~\ref{hihj}(a) one can immediately realize that the model allows for some very light Higgs bosons ($m_{h_1}\leq 100$ GeV). We expect that the possibility of such a hidden Higgs would be explored at the LHC with 14 TeV centre of mass energy, whereas the points where $h_1$ is mostly a doublet ($\geq 90\%$) could be ruled out by the LEP data. The points with the mixed scenario for $h_1$ (with doublet, triplet and singlet) are marked in blue. We remark that a triplet of non-zero hypercharge will not easily satisfy the constraints from LEP, due to its coupling to the $Z$ boson. 

For the points with $m_{h_1/a_1}\leq 100$ GeV which are mostly doublet (red ones) it is very hard to satisfy the LEP bounds \cite{LEPb}. This is because, being doublet like, such $h_1$ would have been produced at LEP and decayed to the fermionic pairs, which have been searched extensively at LEP. On the other hand the singlet and triplet like points (green points) are very difficult to produce at LEP due to the non-coupling to $Z$ boson, which was one of the dominant production channel. This is true for both $e^+e^- \to Zh_1$ and  $e^+e^- \to h_1a_1$. Such triplet and singlet like points will reduce the decay widths in charged lepton pair modes due to non-coupling with fermions. These make the green points more suitable candidate for the hidden Higgs bosons, both for the CP-even and CP-odd. However such parameter space would be highly constrained from the data of the discovered Higgs boson around 125 GeV at the LHC. So far the discovery of the Higgs boson at the LHC has reached  $5\sigma$ or more in the channels  $h_{125}\to \gamma\gamma, WW^*, ZZ^*$. Effectively this could be satisfied by the candidate Higgs around 125 GeV which is mostly doublet like and its decay branching fractions should be within the uncertainties give by  CMS and ATLAS experiments at the LHC. Such requirements rule out vast number of parameter points, including some the triplet and/or signet like hidden Higgs boson(s). In section~\ref{Hdata} we consider such constrains coming from the Higgs data at LHC and the existing data from LEP. 

\begin{figure}[t]
\begin{center}
\mbox{\hskip -15 pt\subfigure[]{\includegraphics[width=0.55\linewidth]{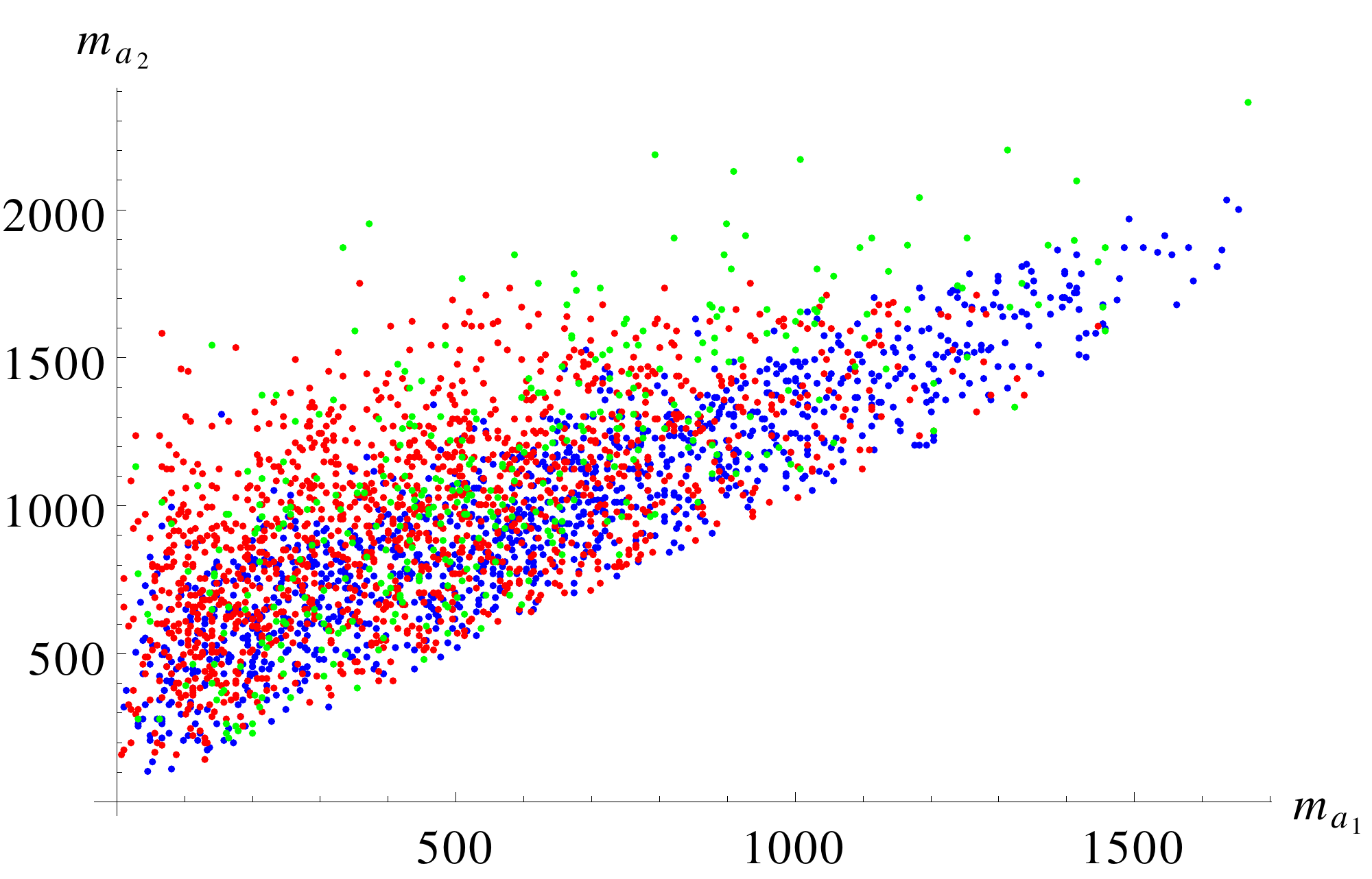}}
\subfigure[]{\includegraphics[width=0.55\linewidth]{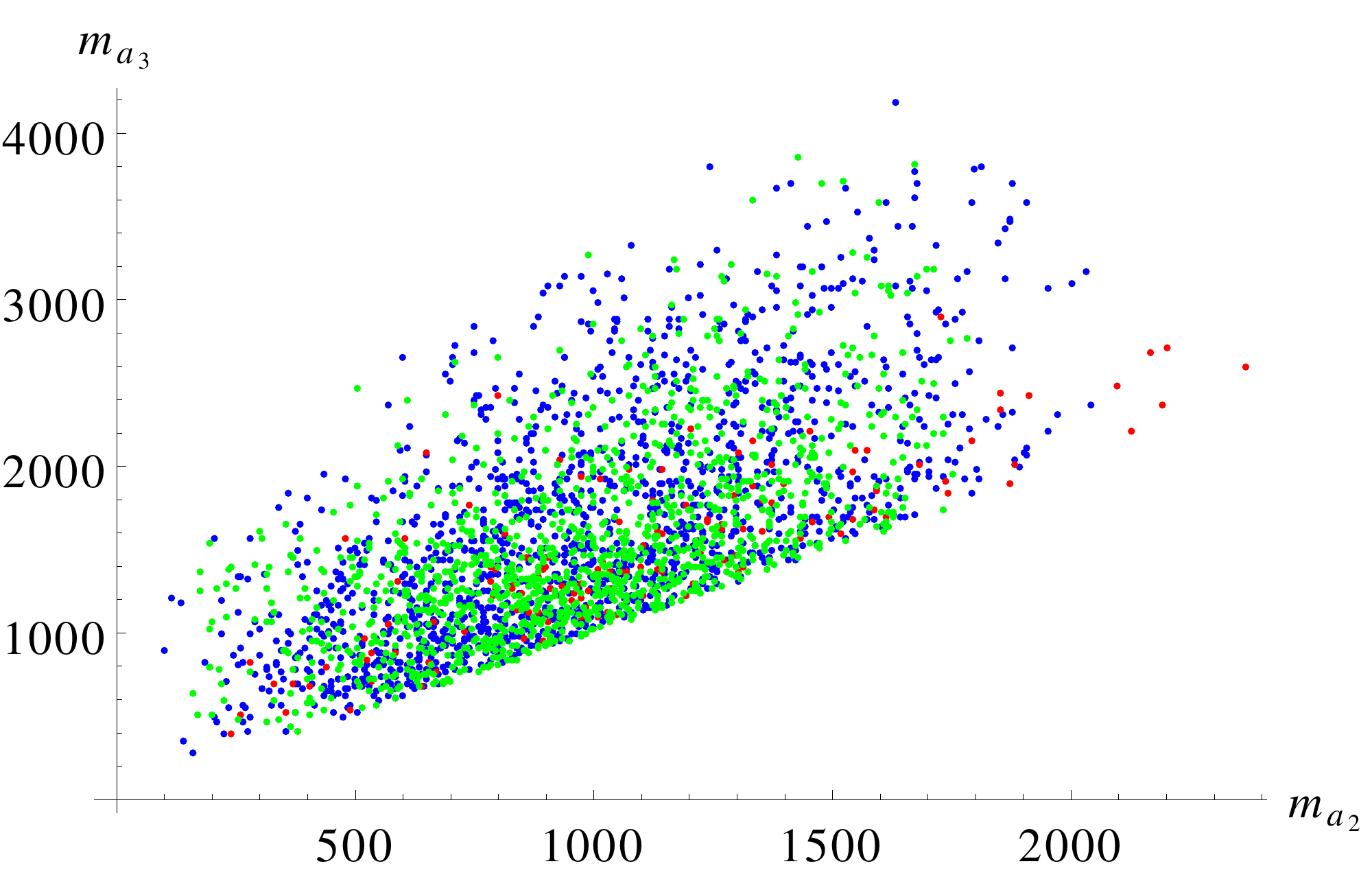}}}

\caption{Tree-level CP-odd Higgs mass correlations (a) $m_{a_1}$ vs $m_{a_2}$ and (b) $m_{a_2}$ vs $m_{a_3}$, where we have a candidate $\sim 125$ GeV Higgs boson. The red points are $>90\%$ doublets-like and the green points are $\geq 90\%$ triplet-like. The blue points are mixtures of
doublet and triplet components for $a_1$ in (a) and for  $a_2$ in (b) respectively.}\label{aiaj}
\end{center}
\end{figure}

\begin{figure}[t]
\begin{center}
\mbox{\hskip -15 pt\subfigure[]{\includegraphics[width=0.55\linewidth]{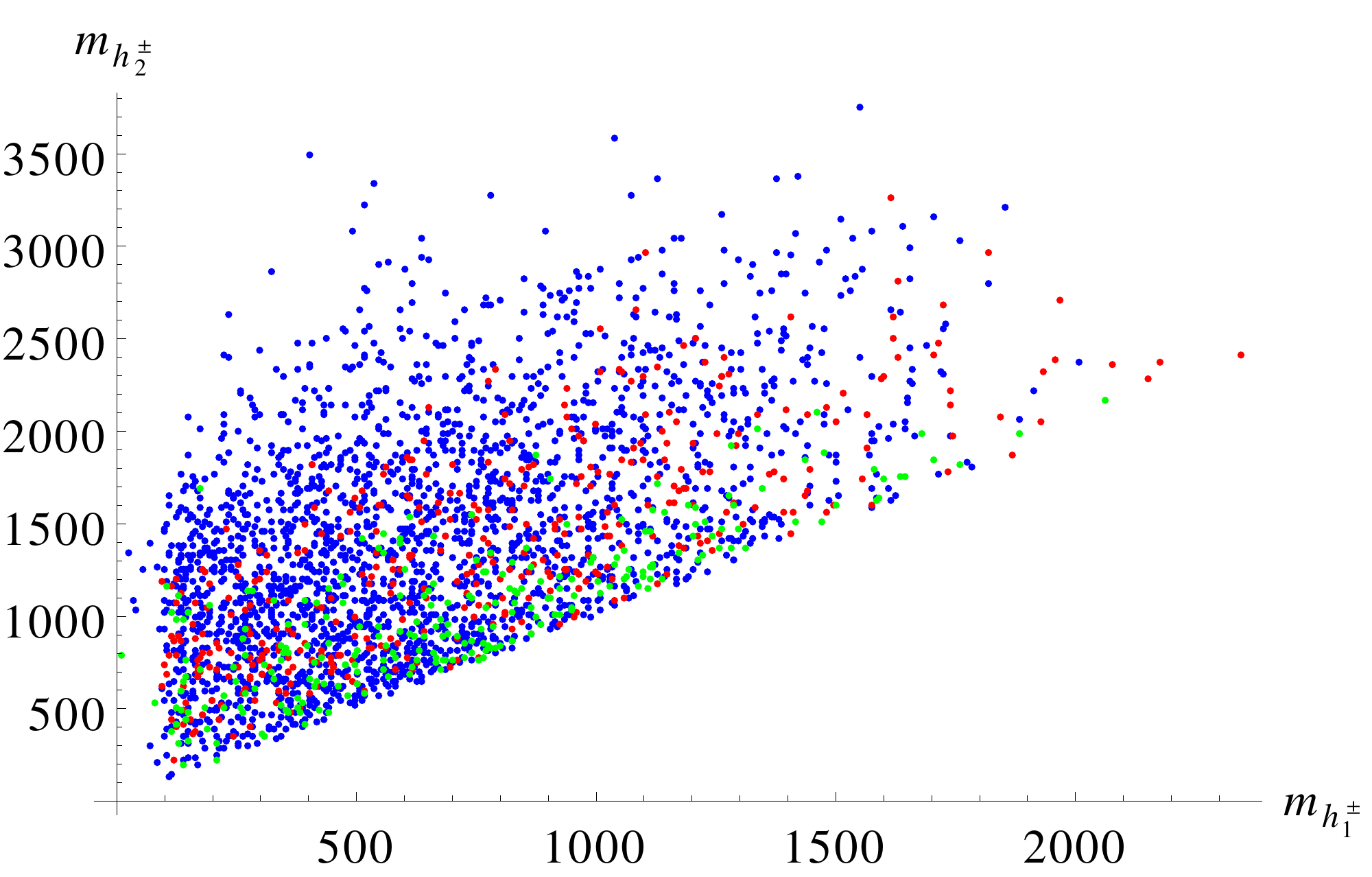}}
\subfigure[]{\includegraphics[width=0.55\linewidth]{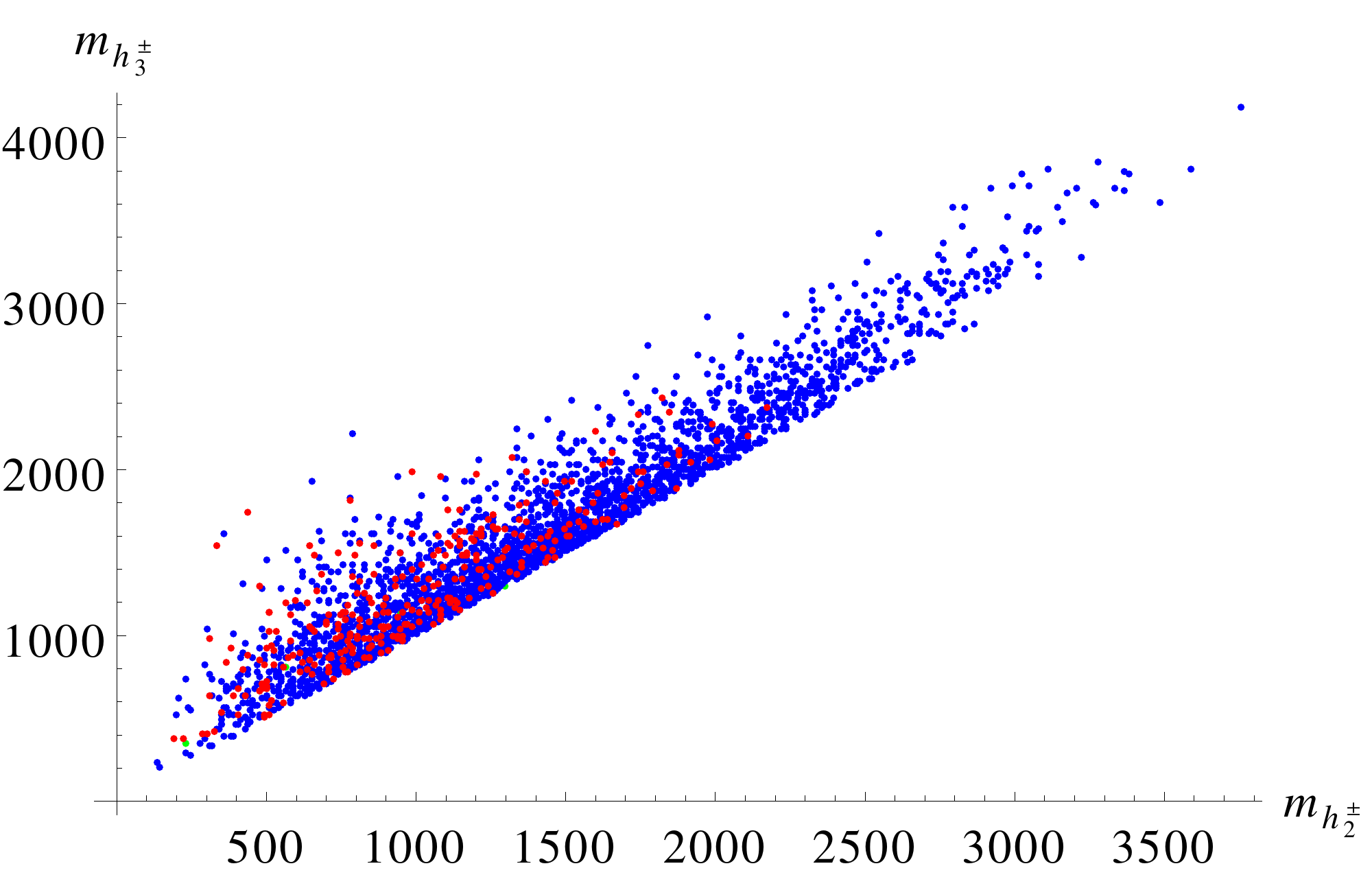}}}

\caption{Tree-level charged Higgs mass correlations (a) $m_{h^\pm_1}$ vs $m_{h^\pm_2}$ and (b) $m_{h^\pm_2}$ vs $m_{h^\pm_3}$, where we have a candidate $\sim 125$ GeV Higgs boson. The red points are $>90\%$ doublets-like and the green points are $\geq 90\%$ triplet- or singlet-like. The blue points are mixture of doublet and triplets/singlets, for $h^\pm_1$ in (a) and for  $h^\pm_2$ in (b) respectively.}\label{chichj}
\end{center}
\end{figure}

Figure~\ref{hihj}(b) shows the mass correlation between ${h_3}$ and $h_4$ for the the same region (\ref{parat}) of the parameter  space.  We see that although there are points characterized by a mass $m_{h_3}$ lighter than 500 GeV, states with $m_{h_4}\leq 500$ GeV are less probable. 
Figure~\ref{aiaj} shows the mass correlations of the CP-odd neutral Higgs bosons. Specifically, Figure~\ref{aiaj}(a) presents the analysis of the mass correlation between $a_1$ and $a_2$. The plot shows that there exists the possibility of having a pseudo-scalar $a_1$ lighter than 100 GeV, accompanied by a CP-even $\sim 125$ GeV Higgs boson.  Note that a very light pseudoscalar Higgs in the MSSM gets strong bounds from LEP \cite{LEPb}. In this case, for a high $\tan{\beta}$, the pair production process $e^+e^- \to h A$, where $A$ is the pseudoscalar of the MSSM, is the most useful one, providing limits in the vicinity of 93 GeV for $m_A$ \cite{LEPb}.  In the TNMSSM instead, if the light 
pseudoscalar Higgs bosons are either of triplet or singlet type then they do not couple to the $Z$, which makes it easier for these states to satisfy the LEP bounds. For this purpose, points which are mostly-triplet or -singlet ($90\%$) have been marked in green; points which are mostly-doublet ($90\%$) in red, whereas the mixed points have been marked in blue as before. Certainly, mass eigenvalues labelled in green would be much more easily allowed by the LEP data, but they would also be able to evade the recent bounds from the LHC $H\tau \tau $ decay mode for a pseudoscalar Higgs \cite{Htautau}. This occurs because neither the triplet nor the singlet Higgs boson couple to fermions (See Eq.~(\ref{spt})).
Figure~\ref{aiaj}(b) presents the correlation between $a_2$ and $a_3$ where the same colour code applies for the structure of $a_2$. As one can easily realize from the figure, there are plenty of green coloured points which represent  triplet/singlet type $a_2$ states, which can easily evade the recent bounds on pseudoscalar states derived at the LHC  \cite{Htautau}. 
Figure~\ref{chichj} shows the correlation of the three charged Higgs bosons for the region in parameter space where we can have a $\sim 125$ GeV Higgs candidate.  Figure~\ref{chichj}(a) shows that there are allowed points for a charged Higgs of light mass  ($m_{h^\pm_1} \lsim 200$ GeV) correlated with a heavier charged Higgs $h^\pm_2$.  Only Higgses of doublet and triplet type  can contribute to the charged Higgs sector. We have checked the structure of the lightest charged Higgs $h^\pm_1$ in Figure~\ref{chichj}(a), where the red points correspond to $\geq 90\%$ doublet, while the green points correspond to  $\geq 90\%$ triplet and the blue points to doublet-triplet mixed states. Charged Higgs bosons which are mostly triplet-like in their content (the green points) do not couple to the fermions (see Eq.~(\ref{spt})), and thus can easily evade the bounds on the light charged Higgs derived at the LHC from the $H^\pm  \to \tau \nu$ decay channel \cite{chHb}. This kind of triplet charged Higgs boson would also be hard to produce from the conventional decay of the top quark and the new production modes as well as the decay modes will open up due to the new vertex $h^\pm_i - Z-W^\mp$ \cite{tssmch1}. Thus vector boson fusion (VBF) with the production of a single charged Higgs is a possibility due to a non-zero  $h^\pm_i - Z-W^\mp$ vertex \cite{tssmch1}. Apart from the $h^\pm_i \to ZW^\pm$ channels, the $h^\pm_i \to  a_1(h_1)W^\pm$ channels are also allowed, for very light neutral Higgs bosons ($a_1/h_1$). Figure~\ref{chichj}(b) presents the correlation between $m_{h^\pm_2}$ and $m_{h^\pm_3}$. We have used for $h^\pm_2$ the same colour conventions as in the previous plots. We see that there are only few triplet type $h^\pm_2$ (green points), most of the allowed mass points being doublet-triplet mixed states (blue points).

\section{Strong and weak sectors}\label{storngw}

The TNMSSM scenario has an additional triplet which is colour singlet and electroweak charged and a singlet superfields (see Eq.~(\ref{spt})) not charged under $SU(3)_c\times SU(2)_L\times U(1)_Y$. Therefore, the strong sector of the model is the same of the MSSM, but supersymmetric F-terms affect the fermion mass matrices, and contribute to the off-diagonal terms.  It generates additional terms in the stop mass matrix from the triplet and singlet vevs, which will be shown below. These terms are proportional to $\lambda_T v_T$ and $\lambda_S v_S$ respectively, and allow to generate an effective $\mu_D$-term in the model.  The triplet contribution is of course restricted, due to the bounds coming from the $\rho$ parameter \cite{rho}. Thus, a large effective $\mu_D$ term can be spontaneously generated by the vev of the singlet, $v_S$. 

\begin{figure}[hbt]
\begin{center}
{\includegraphics[width=0.75\linewidth]{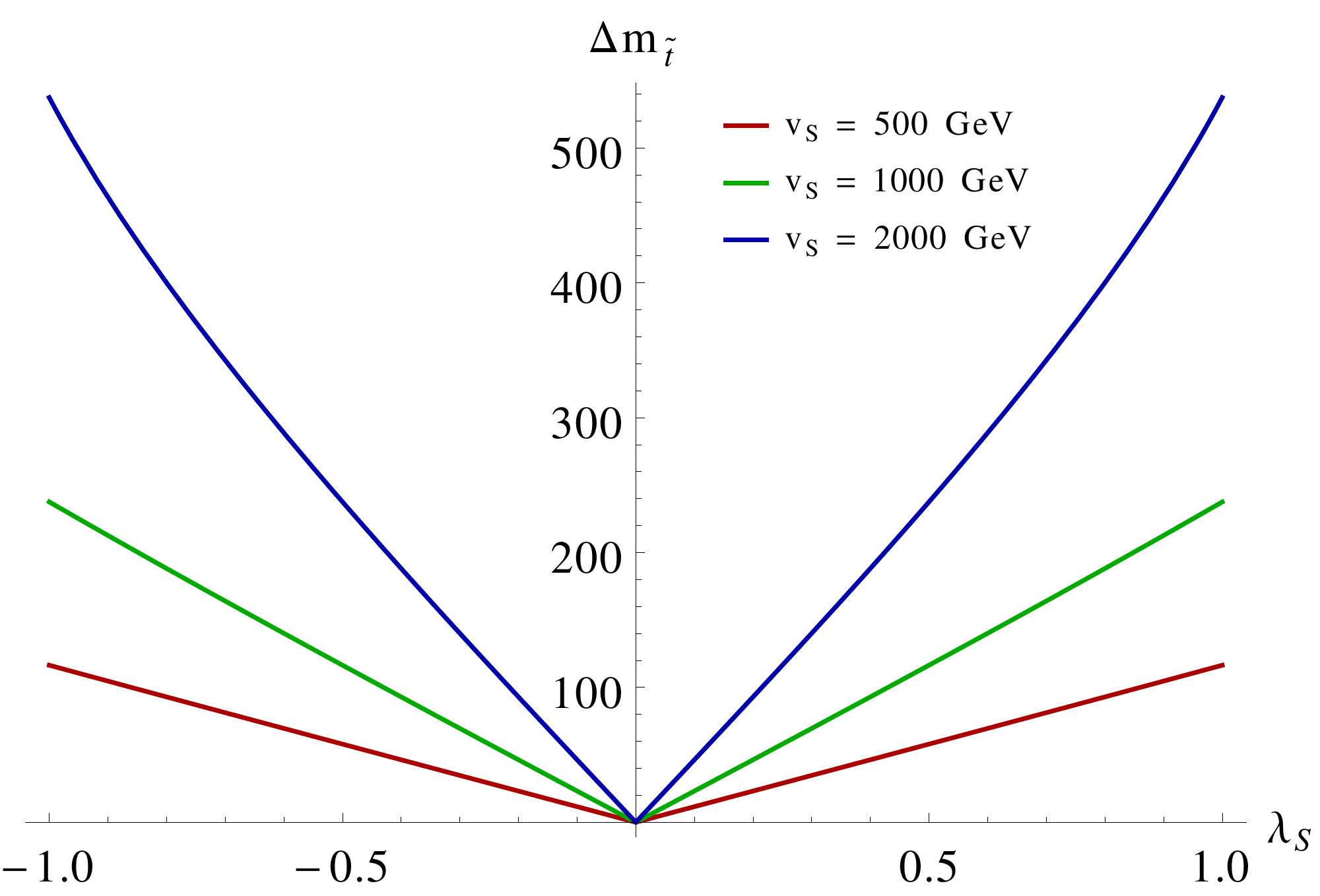}}
\caption{The mass splitting between the stop mass eigen states ($\tilde{t}_{2,1}$) vs $\lambda_S$
for $A_t=0$ with $v_S=500, 1000, 2000$ GeV respectively.}\label{stm}
\end{center}
\end{figure}

Figure~\ref{stm} shows the mass splitting between the $\tilde{t}_2$ and $\tilde{t}_1$ stops versus $\lambda_S$, for several $v_S$ choices and with $A_t=0$. Large mass splittings can be generated without a large parameter $A_t$, by a suitably large $v_S$, which is a common choice if the singlet is gauged respect to an extra $U(1)'$ \cite{uprime}, due the mass bounds for the additional gauge boson $Z'$ \cite{zprime}. The mass matrices for the stop and the sbottom are given by 

\bea\label{stop}
M_{\tilde{t}}=\left(
\begin{array}{cc}
m^2_t+m^2_{Q_3}+\frac{1}{24} \left(g_Y^2-3g_L^2\right) \left(v_u^2-v_d^2\right)\qquad &  \frac{1}{\sqrt{2}}
   A_t v_u+\frac{Y_t v_d}{2} \left( \frac{v_T \lambda _T}{\sqrt 2}- v_S \lambda _S\right) \\
   \\
\frac{1}{\sqrt{2}}
   A_t v_u+\frac{Y_t v_d}{2} \left( \frac{v_T \lambda _T}{\sqrt 2}- v_S \lambda _S\right) & m_t^2+m^2_{\bar{u}_3}+\frac{1}{6} \left(v_d^2-v_u^2\right) g_Y^2
\end{array}
\right)
\eea

\bea\label{sbt}
M_{\tilde{b}}=\left(
\begin{array}{cc}
m_b^2+m^2_{Q_3}+\frac{1}{24} \left(g_Y^2+3g_L^2\right) \left(v_u^2-v_d^2\right)\qquad & \frac{1}{\sqrt{2}}
   A_b v_d+\frac{Y_b v_u}{2} \left( \frac{v_T \lambda _T}{\sqrt 2}- v_S \lambda _S\right) \\
   \\
 \frac{1}{\sqrt{2}}
   A_b v_d+\frac{Y_b v_u}{2} \left( \frac{v_T \lambda _T}{\sqrt 2}- v_S \lambda _S\right) & m_b^2+m^2_{\bar{d}_3}+\frac{1}{12} \left(v_u^2-v_d^2\right) g_Y^2 \\
\end{array}
\right)
\eea

In the electroweak sector the neutralino ($\tilde{\chi}^0_{i=1,..6}$ ) and chargino ($\tilde{\chi}^\pm_{i=1,2,3}$ ) sector are enhanced due to
the extra Higgs fields in the superpotential given in (\ref{spt}). The neutralino sector is now composed of
$\tilde{B},\, \tilde{W}_3, \,\tilde{H}_u, \, \tilde{H}_d, \, \tilde{T}_0, \, \tilde{S}$. The corresponding mass matrix is thus now 6-by-6 and given by  

\bea\label{ntln}
M_{\tilde{\chi}^0}&=\left(
\begin{array}{cccccc}
 M_1 & 0 & -\frac{1}{2} g_Y v_d & \frac{1}{2}g_Y v_u & 0 & 0 \\
 0 & M_2 & \frac{1}{2}g_L v_d & -\frac{1}{2} g_L v_u & 0 & 0 \\
 -\frac{1}{2} g_Y v_d & \frac{1}{2}g_L v_d & 0 & \frac{1}{2}v_T \lambda _T-\frac{1}{\sqrt{2}}v_S \lambda_S & \frac{1}{2}v_u \lambda _T & -\frac{1}{\sqrt{2}}v_u \lambda _S \\
 \frac{1}{2}g_Y v_u & -\frac{1}{2} g_L v_u & \frac{1}{2}v_T \lambda _T-\frac{1}{\sqrt{2}}v_S \lambda_S & 0 & \frac{1}{2}v_d \lambda _T & -\frac{1}{\sqrt{2}}v_d \lambda _S \\
 0 & 0 & \frac{1}{2}v_u \lambda _T & \frac{1}{2}v_d \lambda _T & \sqrt{2} v_S \lambda _{TS} & \sqrt{2}
   v_T \lambda _{TS} \\
 0 & 0 & -\frac{1}{\sqrt{2}}v_u \lambda _S & -\frac{1}{\sqrt{2}}v_d \lambda _S & \sqrt{2} v_T
   \lambda _{TS} & \sqrt{2} \kappa  v_S. \\
\end{array}
\right)\nn\\
\eea
The triplino ($\tilde{T}_0$) and the 
singlino ($\tilde{S}$) masses and mixings are spontaneously generated by the corresponding vevs.
The triplino and singlino are potential dark matter candidates and have an interesting phenomenology as they do not couple directly to the fermion superfields.  The doublet-triplet(singlet) mixing is very crucial
in determining the rare decay rates as well as the dark matter relic densities.

Unlike the neutralino sector, the singlet superfield does not contribute to the chargino mass matrix, and hence the MSSM chargino mass matrix is extended by the triplets only. The chargino mass matrix in the basis of $\tilde{W}^+, \tilde{H}^+_{u}, \tilde{T}^+_{2} (\tilde{W}^-, \tilde{H}^-_{d}, \tilde{T}^-_{1} )$ takes the form

 \bea\label{chn}
M_{\tilde{\chi}^\pm}=\left(
\begin{array}{ccc}
 M_2 & \frac{1}{\sqrt{2}}g_L v_u & -g_L v_T \\
 \frac{1}{\sqrt{2}}g_L v_d & \frac{1}{\sqrt{2}}v_S \lambda _S+\frac{1}{2}v_T \lambda _T & \frac{1}{\sqrt{2}}v_u
   \lambda _T \\
 g_L v_T & -\frac{1}{\sqrt{2}}v_d \lambda _T & \sqrt{2} v_S \lambda _{TS} \\
\end{array}
\right).
\eea
The chargino decays also have an interesting phenomenology due to the presence of a doublet-triplet mixing. 

\section{Higgs masses at one-loop}\label{onel}

 To study the effect of the radiative correction to the Higgs masses, we calculate the one-loop
Higgs mass for the neutral Higgs bosons via the Coleman-Weinberg effective 
potential given in Eq.~(\ref{cwe}) 
\begin{align}\label{cwe}
V_{\rm CW}=\frac{1}{64\pi^2}{\rm STr}\left[ \mathcal{M}^4
\left(\ln\frac{\mathcal{M}^2}{\mu_r^2}-\frac{3}{2}\right)\right],
\end{align}
where $\mathcal{M}^2$ are the field-dependent mass matrices, $\mu_r$ is the renormalization scale, and the supertrace includes a factor of $(-1)^{2J}(2J+1)$ for each particle of spin J in the loop. We have omitted additional charge and colour factors which should be appropriately included. The corresponding one-loop contribution to the neutral Higgs mass matrix  is given by Eq.~(\ref{1Lmh})
\begin{align}
(\Delta\mathcal{M}^2_h)_{ij}
&=\left.\frac{\partial^2 V_{\rm{CW}}(\Phi)}{\partial \Phi_i\partial \Phi_j}\right|_{\rm{vev}}
-\frac{\delta_{ij}}{\langle \Phi_i\rangle}\left.\frac{\partial V_{\rm{CW}}(\Phi)}{\partial \Phi_i}\right|_{\rm{vev}}
\nn\\
&=\sum\limits_{k}\frac{1}{32\pi^2}
\frac{\partial m^2_k}{\partial \Phi_i}
\frac{\partial m^2_k}{\partial \Phi_j}
\left.\ln\frac{m_k^2}{\mu_r^2}\right|_{\rm{vev}}
+\sum\limits_{k}\frac{1}{32\pi^2}
m^2_k\frac{\partial^2 m^2_k}{\partial \Phi_i\partial \Phi_j}
\left.\left(\ln\frac{m_k^2}{\mu_r^2}-1\right)\right|_{\rm{vev}}
\nonumber\\
&\quad-\sum\limits_{k}\frac{1}{32\pi^2}m^2_k
\frac{\delta_{ij}}{\langle \Phi_i\rangle}
\frac{\partial m^2_k}{\partial \Phi_i}
\left.\left(\ln\frac{m_k^2}{\mu_r^2}-1\right)\right|_{\rm{vev}}\ ,\quad \Phi_{i,j}=H^0_{u,r},H^0_{d,r},S_r,T^0_r\ .
\label{1Lmh}
\end{align}

Here, $m^2_k$ is the set of eigenvalues of the field-dependent mass matrices given in the equation above, and we remind that the real components of the neutral Higgs fields are defined as 
\bea
&H^0_u=\frac{1}{\sqrt{2}}(H^0_{u,r} + i H^0_{u,i}), \quad H^0_d=\frac{1}{\sqrt{2}}(H^0_{d,r} + i H^0_{d,i}),\nn\\
&S=\frac{1}{\sqrt{2}}(S_r + i S_i), \quad T^0=\frac{1}{\sqrt{2}}(T^0_{r} + i T^0_{i}). 
\eea.

 For simplicity we drop the supertrace expressions in Eq. (\ref{1Lmh}), but for each particle the supertrace coefficient should be taken into account.

Having characterized the entire sector of the TNMSSM, we gear up for the numerical evaluation of the one-loop neutral Higgs masses in the model. We have already seen in Eq.~\ref{hbnd} that for low $\tan{\beta}$ the contribution of the radiative corrections required in order to reach the $\sim 125$ GeV Higgs mass, overcoming the tree-level bound in (\ref{hspc}), is reduced. This is due to the additional Higgs and higgsinos running in the loops. In our analysis we have chosen the following subregion of the parameter space 
\bea\label{scan}
&|\lambda_{T, S, TS}| \leq 1, \quad |\kappa|\leq 3, \quad |v_s|\leq 1 \, \rm{TeV}, \quad 1\leq \tan{\beta}\leq 10,\nn\\
&|A_{T, S, TS, U, D}|\leq 500,\qquad|A_\kappa|\leq1500, \qquad  m^2_{Q_3, \bar{u}_3, \bar{d}_3}\leq1000,\\
&65\leq|M_{1, 2}|\leq1000,\nn
\eea
that we have used in the computation of the Higgs boson mass. In this scan, we have included the radiative corrections to the mass eigenvalues at one-loop order of the neutral sector and retained only those sets of eigenvalues which contain 
one 125 GeV CP-even Higgs. We have selected the range  $65\leq|M_{1, 2}|\leq1000$ in order to avoid the constraints on the Higgs invisible decay and use $\mu_r=500$ GeV for the numerical calculation.

\begin{figure}[]
\begin{center}

\mbox{\hskip -15 pt\subfigure[]{\includegraphics[width=0.55\linewidth]{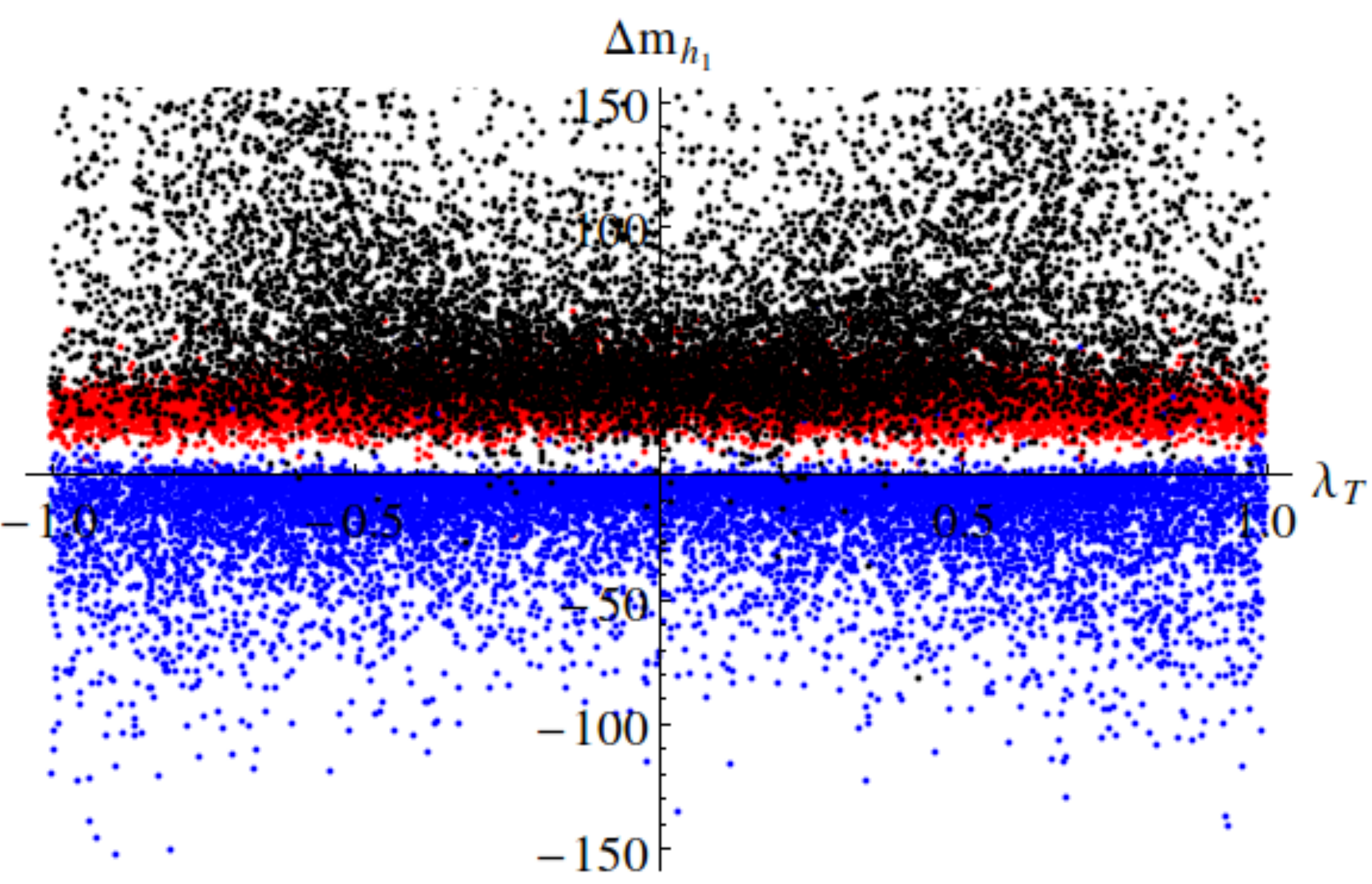}}
\subfigure[]{\includegraphics[width=0.55\linewidth]{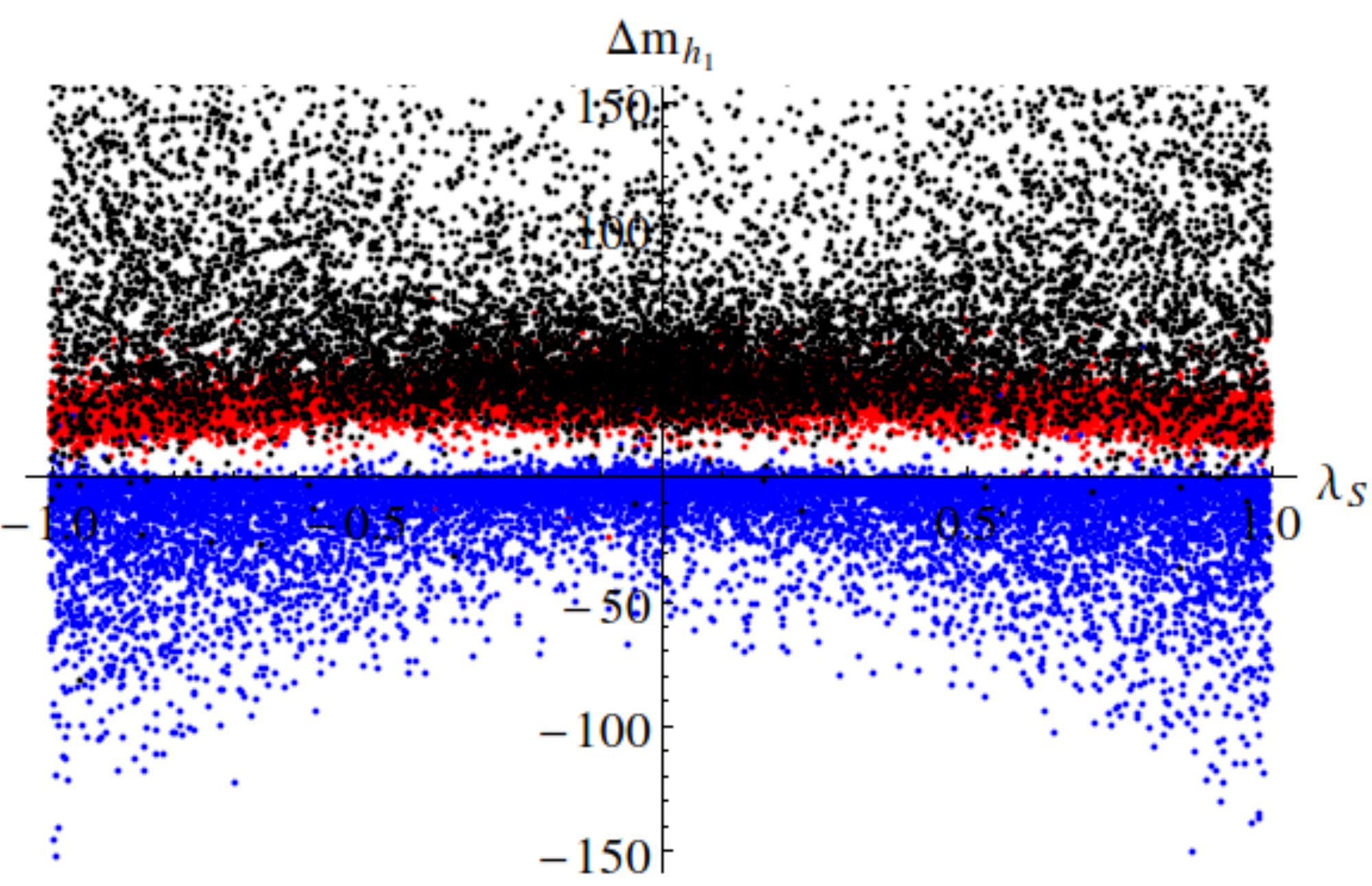}}}

\mbox{\subfigure[]{\includegraphics[width=0.7\linewidth]{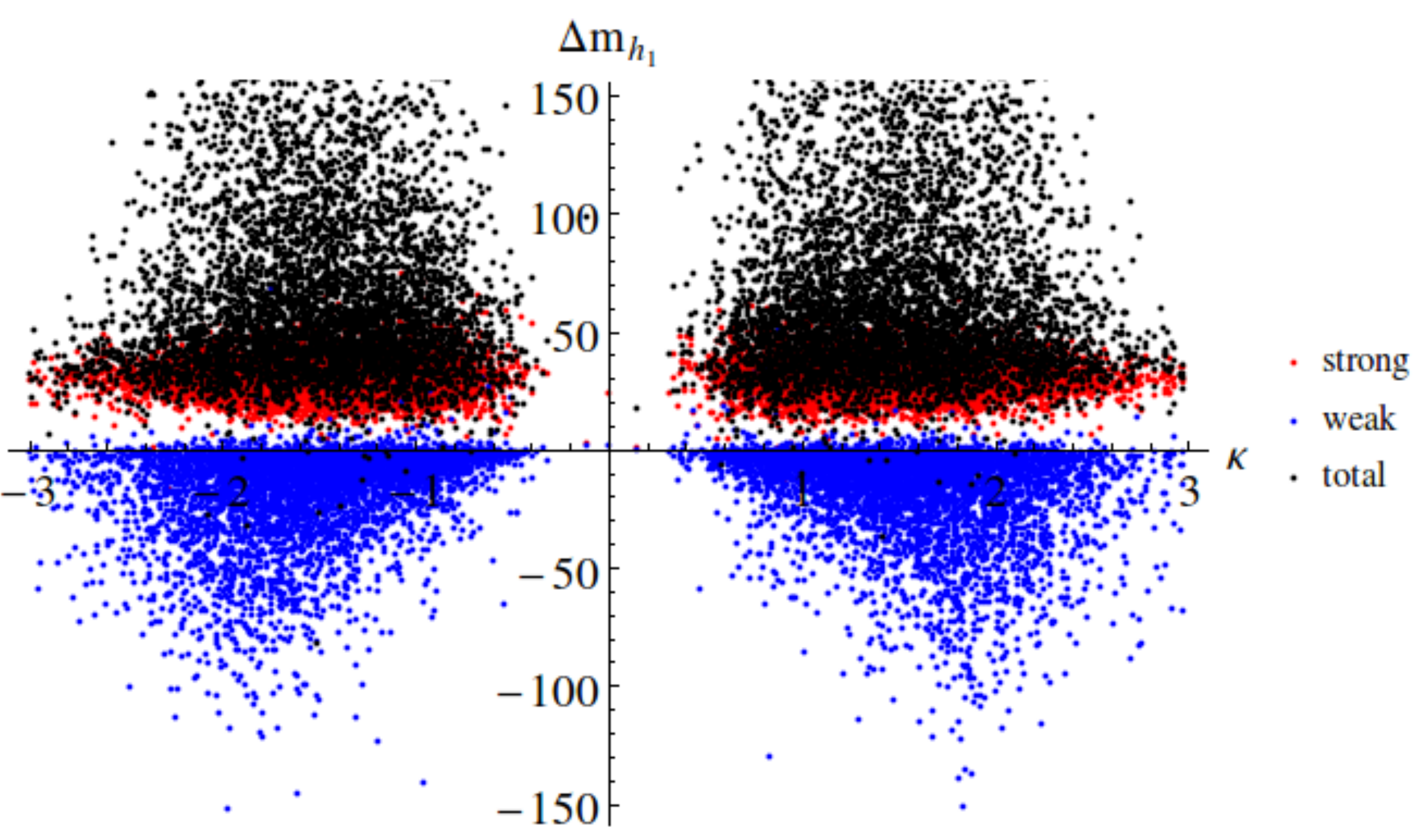}}
}

\caption{The radiative corrections at one-loop 
for $m_{h_1}$ vs (a) $\lambda_T$, (b) $\lambda_S$ and (c) $\kappa$. The red points are
only with strong (top-stop, bottom-sbottom) corrections, blue points are with weak corrections without the Higgs bosons (higgsinos, gauge boson and gauginos), (c) black points are the total (strong +weak + Higgs bosons) corrections. }\label{dmh1vsp}
\end{center}
\end{figure}

Figure~\ref{dmh1vsp} shows the radiative corrections to $m_{h_1}$ as $\Delta m_{h_{1}}=m^{1-\rm{loop}}_{h_1} - m^{\rm{tree}}_{h_1}$, plotted against (a) $\lambda_T$, (b) $\lambda_S$ and (c) $\kappa$ respectively. The red points show the corrections to $m_{h_1}$ from the strong sector, due to the contributions generated by top-stop and bottom-sbottom running in the loops.
The blue points include the corrections from the weak sector with gauge bosons, gaugino and higgsino, and the black points take into account the total corrections which include strong, weak and the contributions from the Higgs sector. As one can deduce from the plots, the corrections (top-stop, bottom-sbottom) coming from the strong interactions are independent of the triplet and singlet Higgs couplings, as expected, with a maximum split of 50 GeV respect to the tree-level mass eigenvalue.\\
In the triplet-singlet extension we have four CP-even, three CP-odd neutral Higgs bosons and three charged Higgs bosons as shown in Eq.~(\ref{hspc}). These enhance both the Higgs and higgsino contributions to the radiative correction. The weak corrections (blue points) are dominated by the large number of higgsinos which contribute negatively to the mass and tend to increase for large values of the Higgs couplings ($\lambda_{T,S}$ and $\kappa$).\\
Finally, the black points show the sum of all the sectors, which are positive in sign, due to the large number of scalars contributing in the loop, with an extra factor of two for the charged Higgs bosons. This factor of two originates from the CW expression of the potential, and accounts for their multiplicity $(\pm)$. Such scalar contributions increase with the values of the corresponding couplings $\lambda_T, \lambda_S, \kappa$. 
From Figure~\ref{dmh1vsp} one can immeditaley notice that the electroweak radiative corrections could be sufficient in order to fulfill the requirement of a $\sim 125$ GeV Higgs mass, without any contribution from the strong sector.

\begin{figure}[t]
\begin{center}
\mbox{\hskip -15 pt
\subfigure[]{\includegraphics[width=0.55\linewidth]{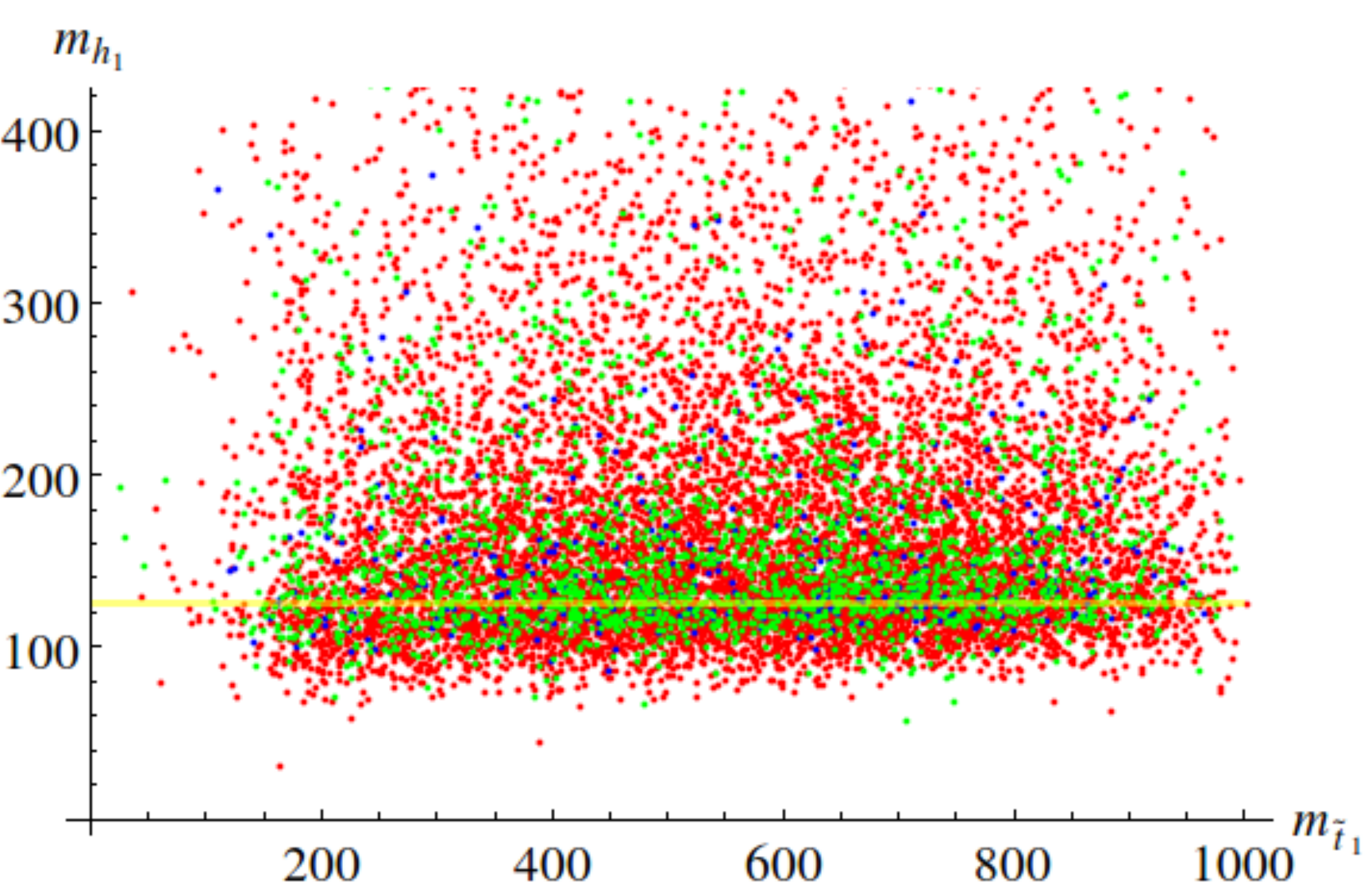}}
\subfigure[]{\includegraphics[width=0.55\linewidth]{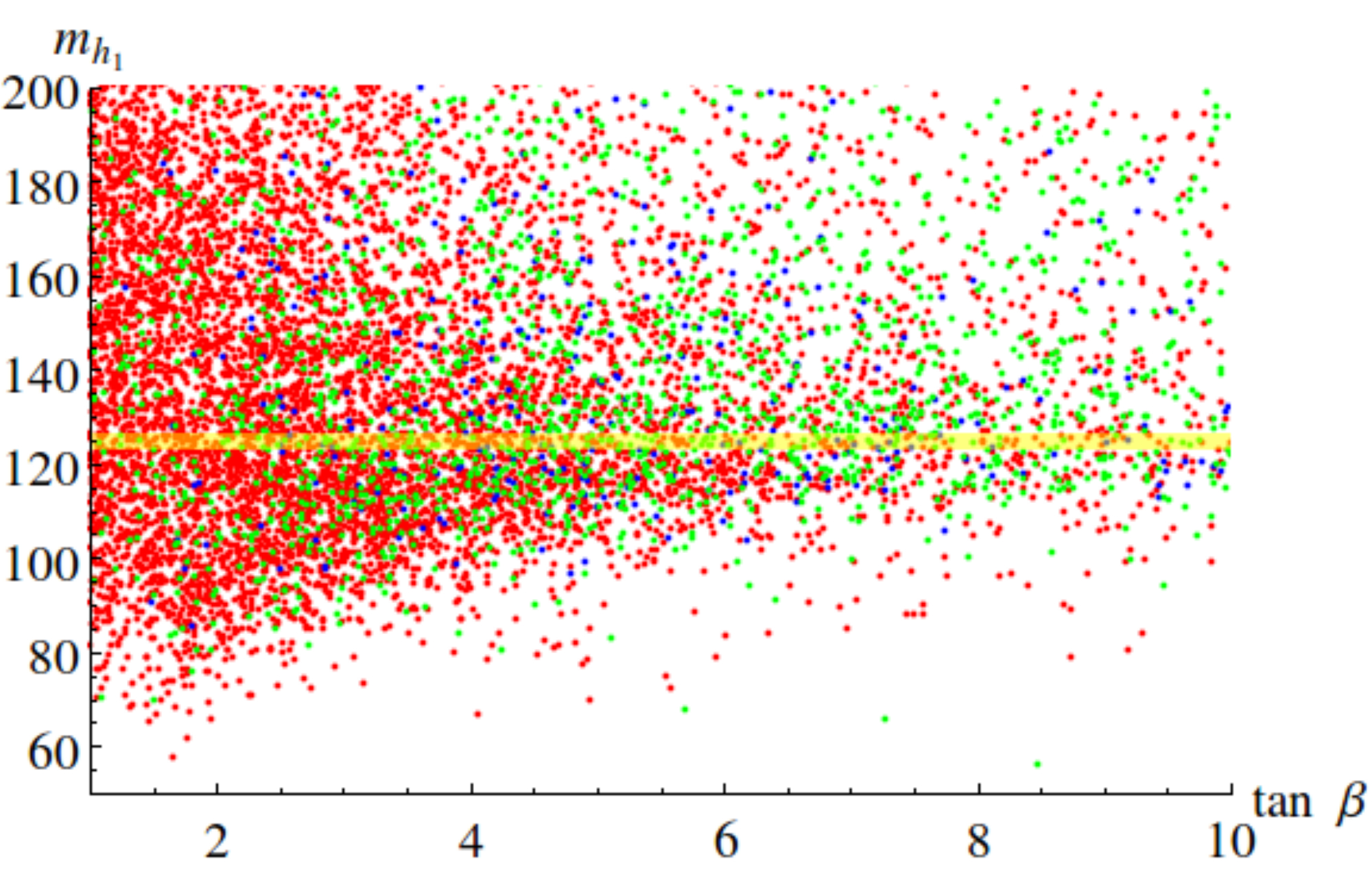}}}

\caption{The variation of  the one-loop lightest CP-even Higgs mass $m_{h_1}$ with (a) the lightest stop mass $m_{\tilde{t}_1}$, and (b) with $\tan{\beta}$, respectively. The yellow band shows the candidate Higgs mass
$123\leq m_{h_1} \leq 127$ GeV. }\label{mh1lvsp}
\end{center}
\end{figure}

To illustrate this point, in Figure~\ref{mh1lvsp}(a) we have plotted the lightest CP-even neutral Higgs mass at one-loop versus the lighter stop mass ($m_{\tilde{t}_1}$). We have used the same color coding conventions of the tree-level analysis. The red points are mostly doublets ($\geq 90\%$), the green points are mostly triplet/singlet($\geq 90\%$) and blue points are mixed ones, as explained in section~\ref{treel}.  The yellow band shows the Higgs mass range $123\leq m_{h_1} \leq 127$ GeV. 
We notice that a $\sim 125$ GeV CP-even neutral Higgs could be achieved by requiring a stop of very 
low mass, as low as 100 GeV. This is due to the presence of additional tree-level and radiative corrections from the Higgs sectors.  Thus, in the case of extended SUSY scenarios like the TNMSSM, 
the discovery of a $\sim 125$ GeV Higgs boson does not put a stringent lower bound on the required SUSY mass scale, and one needs to rely on direct SUSY searches for that.

In Figure~\ref{mh1lvsp}(b) we present the dependency of the one-loop corrected Higgs mass of the lightest CP-even neutral Higgs on $\tan{\beta}$. The distribution of points is clearly concentrated at low values of $\tan{\beta} \lsim 4$. This is due to the additional contributions on the tree-level Higgs masses, which are maximal in the same region of $\tan{\beta}$ (see Eq.~(\ref{hbnd})). It is then clear that an extended Higgs sector reduces the amount of fine-tuning  \cite{tssmyzero} needed in order to reproduce the mass of the discovered Higgs boson, compared to constrained supersymmetric scenarios. The latter, in general, require much larger supersymmetric mass scales beyond the few TeV \cite{cMSSMb} region. Compared to the pMSSM, this also represents an improvement, as it does not require large mixings in the stop masses in order to have the lighter stop mass below a TeV \cite{pMSSMb}.
\begin{figure}[thb]
\begin{center}
\mbox{\hskip -15 pt
\subfigure[]{\includegraphics[width=0.55\linewidth]{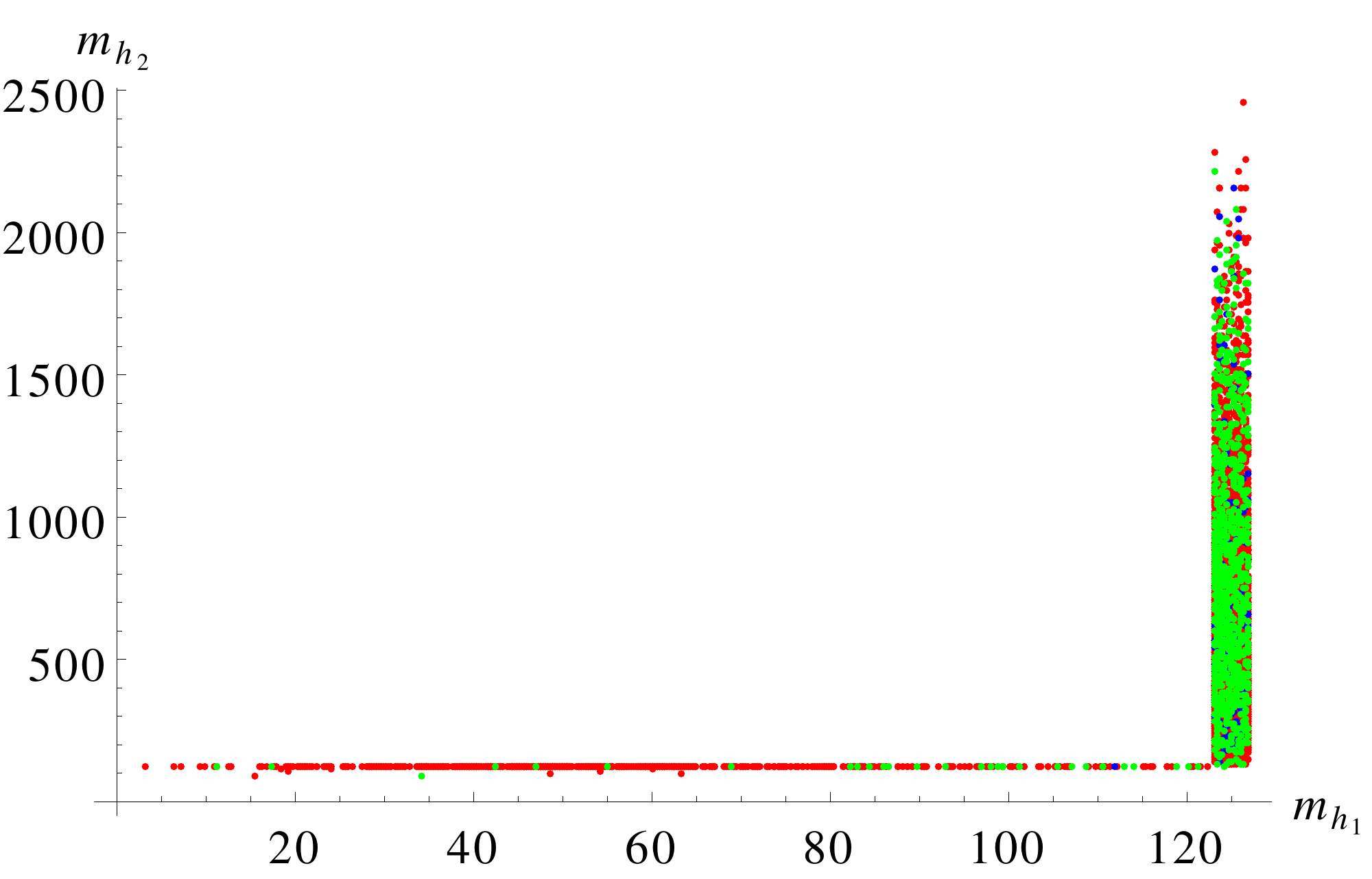}}
\subfigure[]{\includegraphics[width=0.55\linewidth]{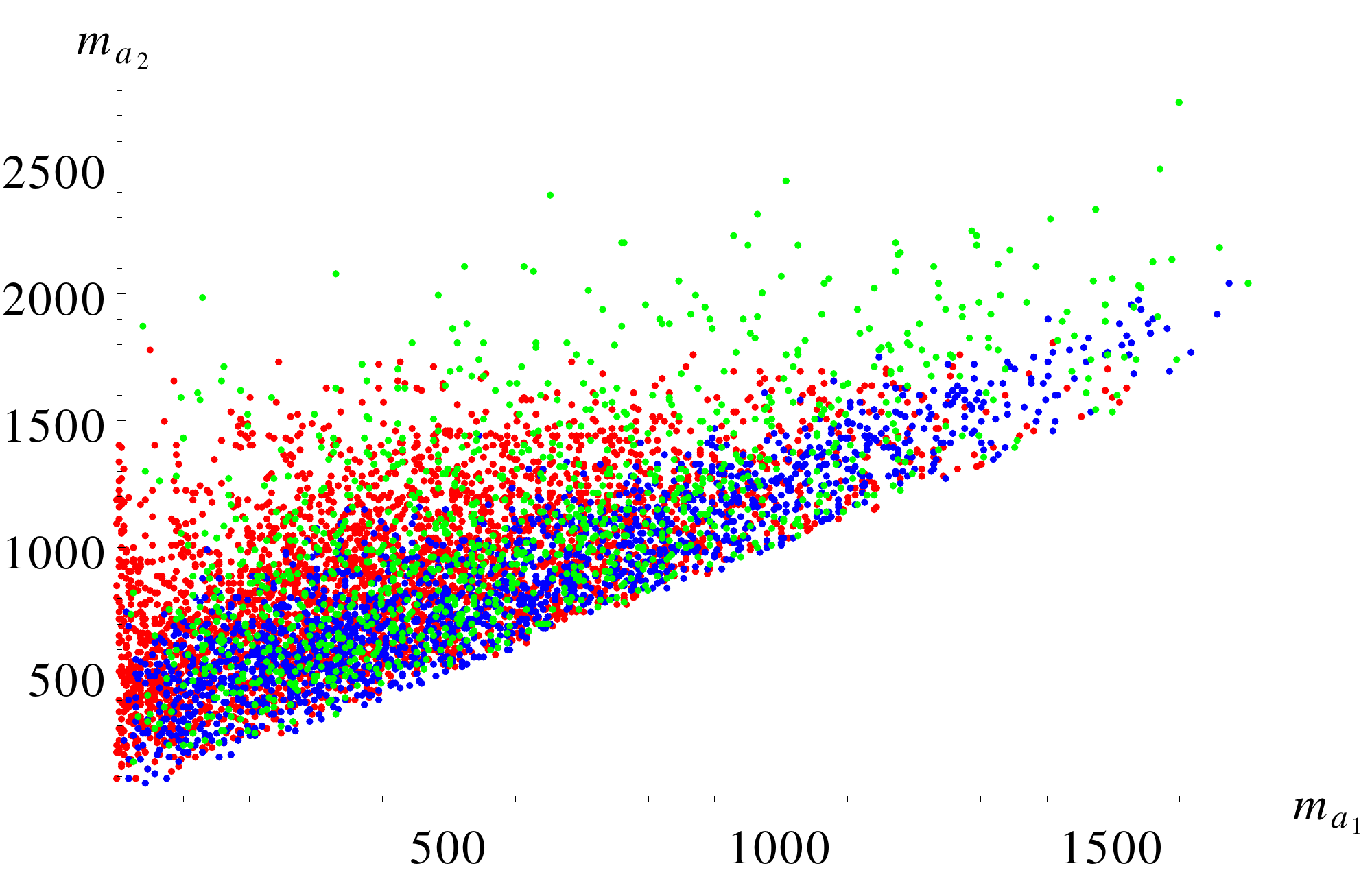}}}
\caption{The mass correlations at one-loop (a) $m_{h_1}-m_{h_2}$ and (b) $m_{a_1}-m_{a_2}$
where we have a CP-even candidate Higgs in the mass range $123\leq m_{h_i} \leq 127$ GeV.  Red, blue and green points are defined as in Figure 2. }\label{mhimhj}
\end{center}
\end{figure}
\subsection{Hidden Higgs bosons}
Next we investigate the case in which we have one or more hidden Higgs bosons, lighter in mass than 125 GeV,  scalars and/or pseudoscalars.  In Figure~\ref{mhimhj} we present the mass correlations at one-loop for (a) $m_{h_1}-m_{h_2}$ and (b) $m_{a_1}-m_{a_2}$, where we have a CP-even candidate Higgs boson in the mass range $123\leq m_{h_i} \leq 127$ GeV. The red points are mostly doublets ($\geq 90\%$), green points are mostly triplets/singlets ($\geq 90\%$) and blue points are mixed ones, as already explained. The green points have a high chance of evading the LEP bounds \cite{LEPb}, showing that the possibility 
of having a hidden scalar sector is realistic, even after taking into account the radiative corrections to the mass spectrum. A closer inspection of  Figure~\ref{mhimhj}(a) 
reveals that there are points where both $m_{h_1}$ and $m_{h_2}$ are less than $100$ GeV, showing that there is the possibility of having two CP-even hidden Higgs bosons. In that case $h_3$ is the candidate Higgs of $\sim 125$ GeV. Similarly, Figure~\ref{mhimhj}(b) shows the possibility of having two hidden pseduoscalars. The arguments mentioned in section~\ref{treel} will apply to the Higgs masses at one-loop as well. These imply that for $m_{h_1/a_1}\leq 123$ GeV, the green points  could evade the bounds from LEP and LHC, the red points would be ruled out and the blue points need to be carefully confronted with the data. In section~\ref{Hdata} we analyse such scenarios in detail. The lightest pseudoscalar present in the spectrum, as we are going to discuss below, can play a significant role in cosmology. In fact, it is crucial 
in enhancing the dark matter annihilation cross-section, which is needed in order to get the correct dark matter relic in the universe \cite{Arina:2014yna}.

\section{$\beta$-fuctions and the running of the couplings}\label{beta}
We have implemented the model in SARAH (version 4.5.5) \cite{sarah} in order to generate the vertices and the model files for CalcHep \cite{calchep}, and generated the $\beta$ functions for the dimensionless couplings and the other soft parameters. The $\beta$ functions for $\lambda_{T, S, TS}$, $\kappa, g_Y, g_L, g_c, y_{t, b}$ are given in the appendix \ref{RGs}.

\begin{figure}[t]
\begin{center}
\mbox{\subfigure[]{\hskip -15 pt
\includegraphics[width=0.55\linewidth]{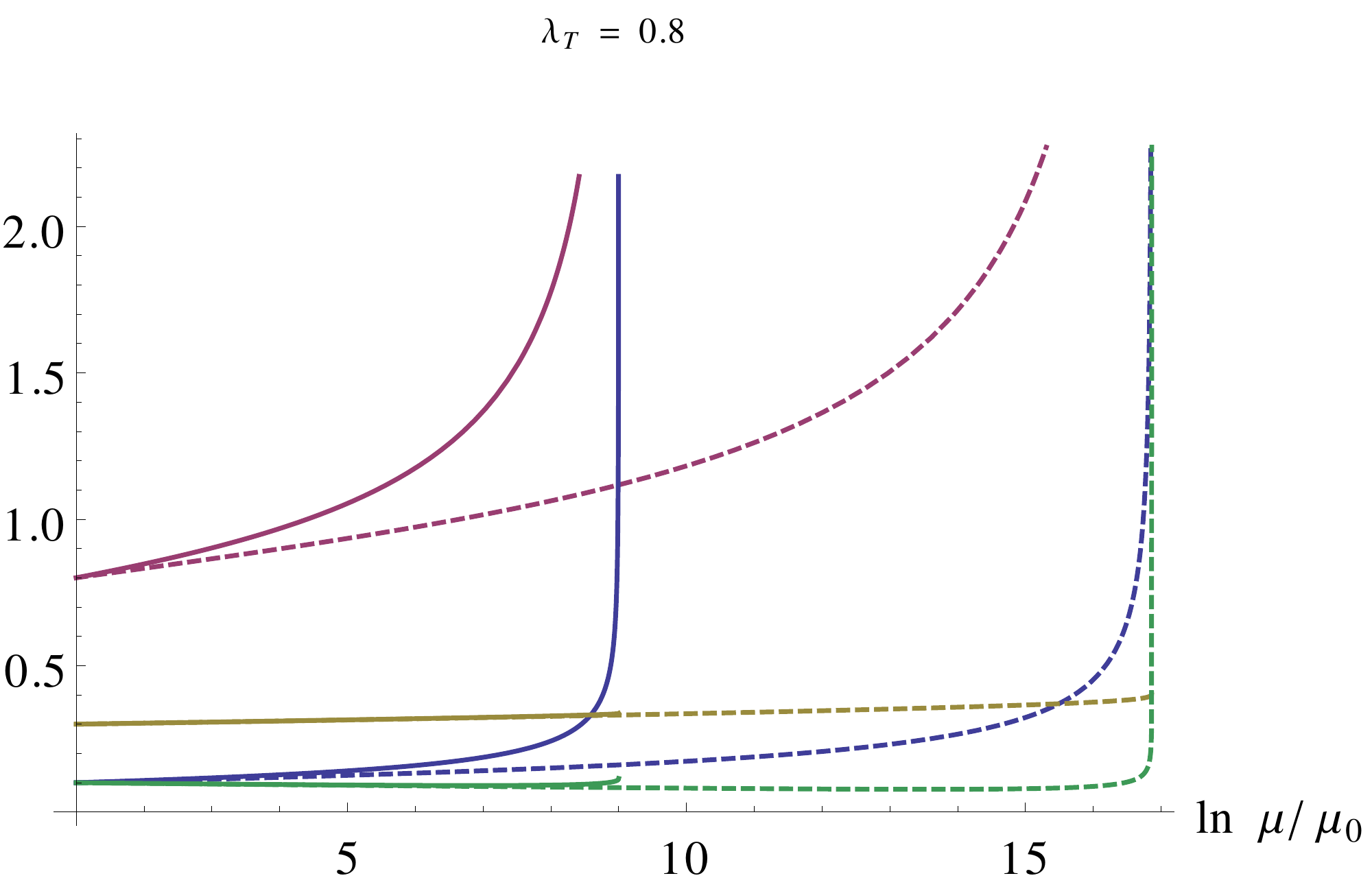}}
\subfigure[]{\includegraphics[width=0.55\linewidth]{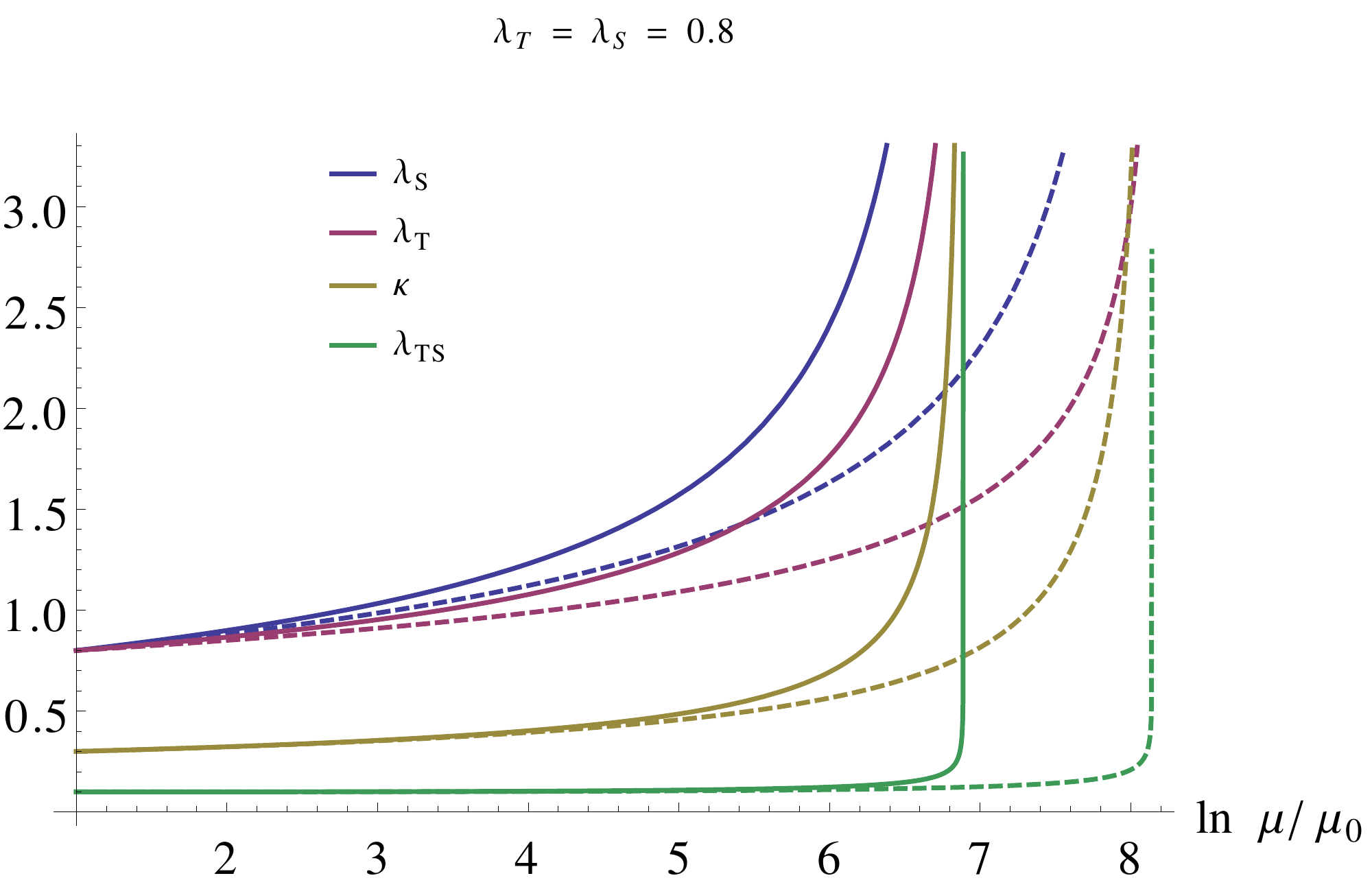}}}
\mbox{\subfigure[]{\hskip -15 pt
\includegraphics[width=0.55\linewidth]{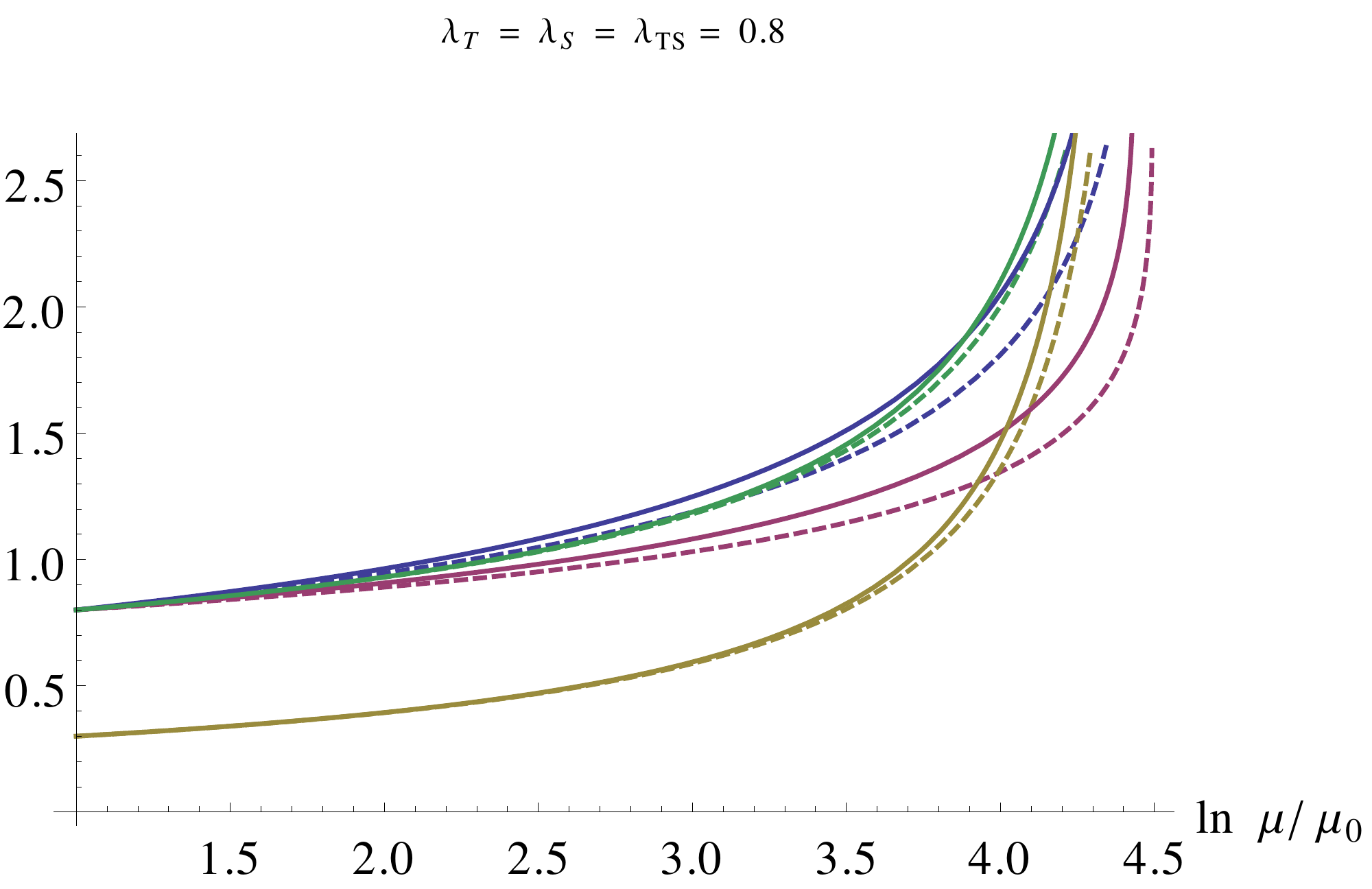}}
\subfigure[]{\includegraphics[width=0.55\linewidth]{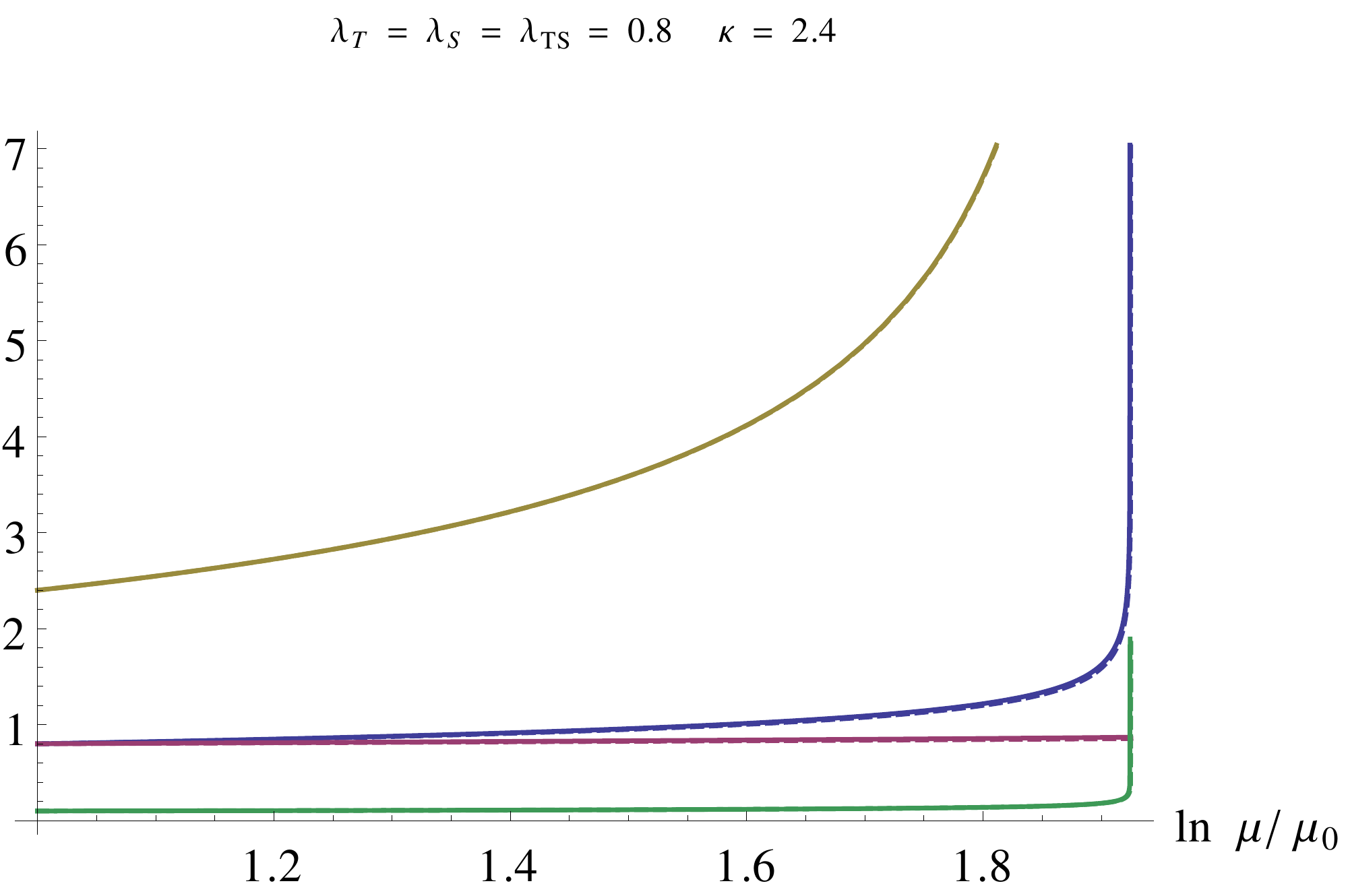}}}
\caption{The running of the dimensionless Higgs couplings $\lambda_{T, S, TS}$
and $\kappa $ with the log of the ratio the scales ($\ln{\mu/\mu_0}$) for $\tan{\beta}=1.5$ (solid lines) and $\tan{\beta}=10$ (dashed lines), where $\mu_0=M_Z$.  We have checked the corresponding variations for (a)$\lambda_{T}=0.8$,  (b)$\lambda_{T, S}=0.8$,  (c)$\lambda_{T, S, TS}=0.8$ and  (d)$\lambda_{T, S,}=0.8, \kappa=2.4$  chosen at scale $\mu_0$  respectively.}\label{rgv}
\end{center}
\end{figure}

To analyse the perturbativity of the couplings we have selected four different scenarios and identified the cut-off scale ($\Lambda$) in  the renormalization group evolution, where one of the coupling hits the Landau pole and becomes non-perturbative ($\lambda_i(\Lambda)=4\pi$). Figure~\ref{rgv}(a) presents a mostly-triplet scenario at the electroweak scale as we choose $\lambda_T=0.8$, $\lambda_{S,TS}=0.1, \kappa=0.3$ at the scale $\mu_0=M_Z$, for $\tan{\beta}=1.5$ (solid lines) and $\tan{\beta}=10$ (dashed lines). For lower values of $\tan{\beta}$ ($\tan{\beta}=1.5$) the triplet coupling $\lambda_T$ becomes non-perturbative already at scale of $\Lambda \sim 10^{9-10}$ GeV, similarly to the behaviour shown in the triplet-extended MSSM \cite{FileviezPerez:2012gg, nardini}. For larger values of $\tan{\beta}$ ($\tan{\beta}=10$) all the couplings remain perturbative up to the (Grand Unification) GUT scale ($\Lambda \sim 10^{16}$ GeV).

Figure~\ref{rgv}(b) presents the case where $\lambda_{T, S}=0.8$ at $\mu_0=M_Z$.  We see that although the  $\tan{\beta}$ dependency becomes less pronounced, the theory becomes non-perturbative at a relatively lower scale $\Lambda \sim 10^{8}$ GeV.

From Figure~\ref{rgv}(c) it is evident that on top of $\lambda_T$ and $\lambda_S$ if we also choose $\lambda_{TS}=0.8$ at $\mu_0=M_Z$, the $\tan{\beta}$ dependency almost disappears. In this case the 
theory becomes more constrained with a cut-off scale $\Lambda \sim 10^{6}$ GeV. 

Finally, Figure~\ref{rgv}(d) illustrates the effect of a larger $\kappa$ value, the singlet self-coupling, with $\kappa=2.4$ at $\mu_0=M_Z$. The perturbative behaviour of the theory comes under question at a scale as low as $10^4$ GeV.  Such a large value of $\kappa$ at the electroweak scale thus restricts the upper scale of the theory to lay below 10 TeV, unless one extends the theory with an extra sector\footnote{For the scan in Eq.~\ref{scan} we select $|\kappa|\leq 3 $. The theoretical perturbativity of the parameter points have to be checked explicitly.}.  Choosing relatively lower values of $\lambda_{TS}$ and $\kappa$ would allow the theory to stay perturbative until $10^{8-10}$ GeV even with
$\lambda_{T, S}$ as large as $0.8$. The choice of larger values of $\lambda_{T, S}$ increases the tree-level contributions to the Higgs mass (see Eq.~(\ref{hbnd})) as well as the radiative corrections, via the additional Higgs bosons exchanged in the loops. Both of these contributions reduce the amount of supersymmetric fine-tuning, assuming a Higgs boson of $\sim 125$ GeV in the spectrum,  by a large amount, respect both to a normal and to a constrained MSSM scenario. Obviously, the addition of the triplet spoils the gauge coupling unification under the renormalization group evolution. This features is already evident in the triplet-extended MSSM \cite{FileviezPerez:2012gg, nardini}. 

\section{Fine-tuning}\label{finet}
The minimisation conditions given in Eq.~(\ref{mnc2}) relate the $Z$ boson mass to
the soft breaking parameters in the form  
\bea
M_Z^2&=&\mu^2_{\rm{soft}}-\mu_{\text{eff}}^2\\
\mu_{\text{eff}}&=&v_S \lambda_S-\frac{1}{\sqrt 2}v_T\lambda_T, \quad \mu^2_{\rm{soft}}=2\frac{m_{H_d}^2-\tan^2\beta\, m_{H_u}^2}{\tan^2\beta -1}.
\eea
It is also convenient to introduce the additional parameter
\bea\label{ft}
\mathcal{F}&=&\left|\ln\frac{\mu^2_{\rm{soft}}-\mu_{\text{eff}}^2}{\mu^2_{\text{soft}}}\right|,
\eea
characterizing the ratio between $M^2_Z$ and $\mu^2_{\text{soft}}$, which can be considered a measure of the fine-tuning. Unlike the MSSM, here the $\mu_{\text{eff}}$ parameter is generated spontaneously by the singlet and triplet vevs. Notice that while the triplet contribution is bounded by the $\rho$ parameter \cite{rho}, the singlet vev is unbounded and it may drive $\mu_{\text{eff}}$ to a large value. Similarly, the 
soft parameters $m_{H_u, H_d}$, which are determined by the minimisation condition (\ref{mnc2}), can be very large, and thus can make $\mu^2_{\text{soft}}$ also large. Finally, to reproduce the $Z$ boson mass we need large cancellations between these terms, which leads to the well know fine-tuning problem of the MSSM and of other supersymmetric scenarios. 

\begin{figure}[bht]
\begin{center}
\mbox{\subfigure[]{\hskip -15 pt
\includegraphics[width=0.55\linewidth]{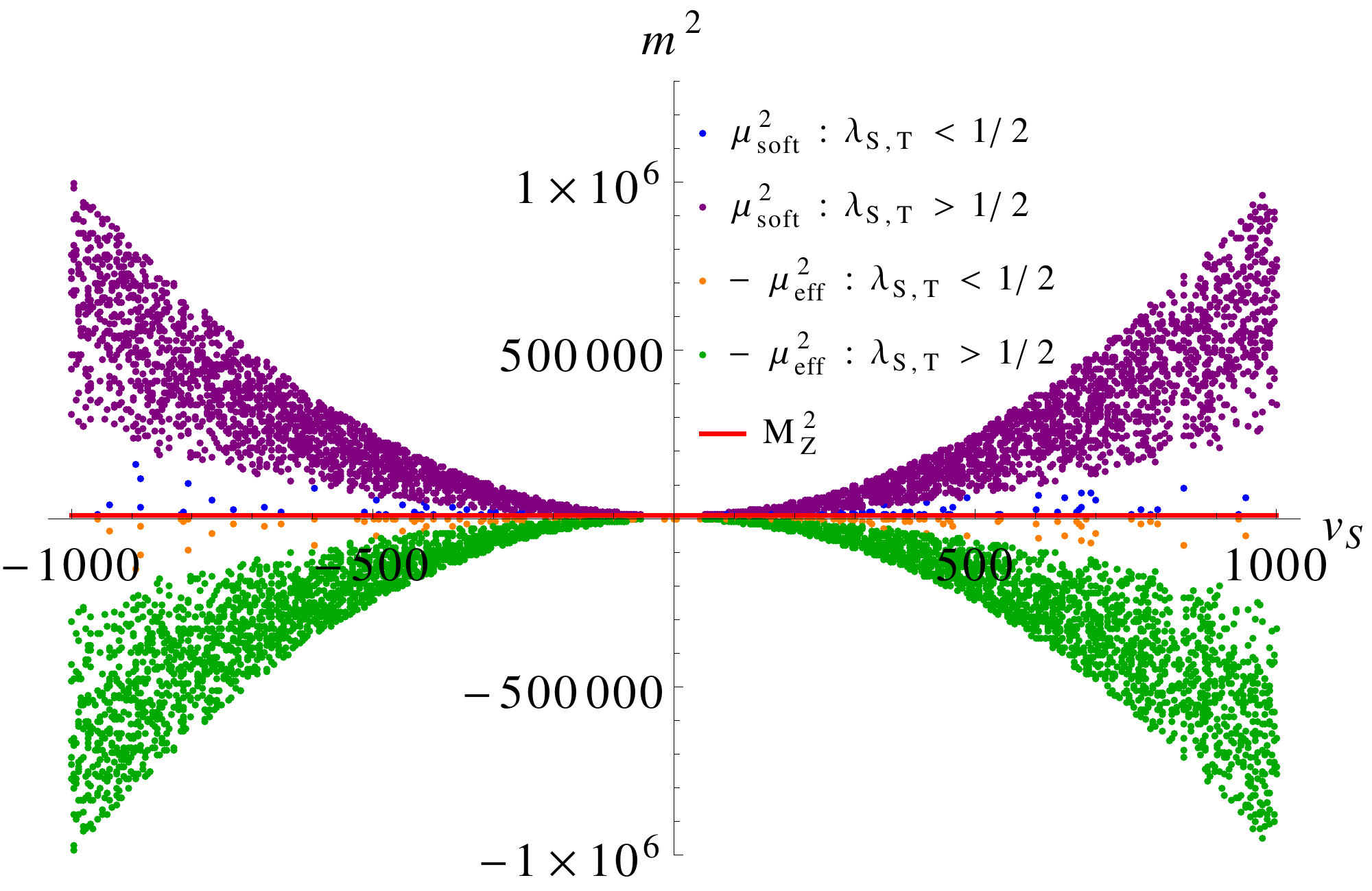}}
\subfigure[]{\includegraphics[width=0.55\linewidth]{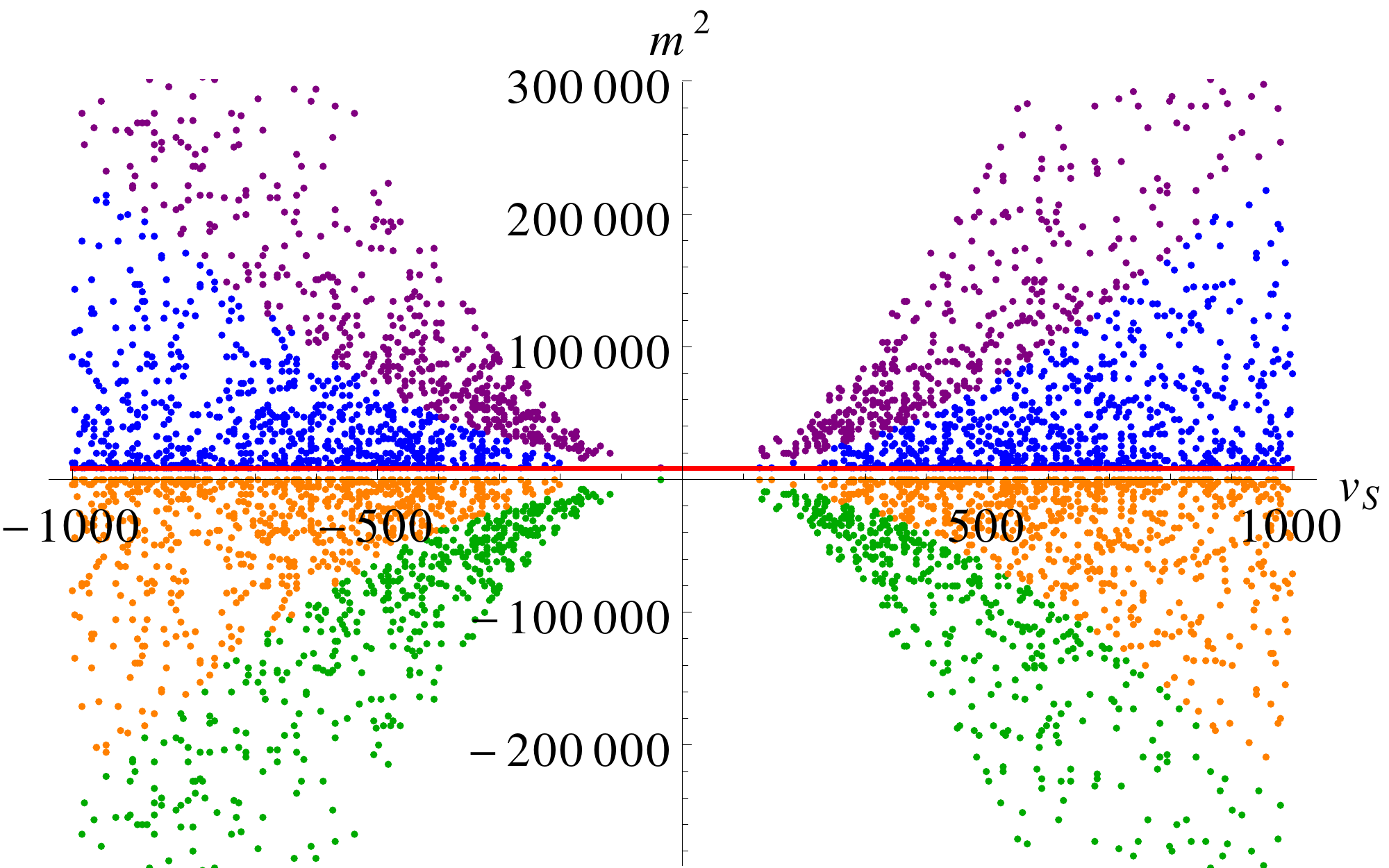}}}
\caption{The (a)tree-level and (b) one-loop level fine-tuning measures $\mu_{\text{soft}}$ and  $-\mu^2_{\text{eff}}$ versus the singlet vev $v_S$ for a candidate Higgs of mass between $120\leq m_{h_1} \leq 130$ GeV respectively. The violet points represent $\mu^2_{\rm{soft}}$ values $\lambda_{S,T}\geq 0.5$ and blue points represent $\mu^2_{\rm{soft}}$ values $\lambda_{S,T}<0.5$. The green points represent $\mu^2_{\rm{eff}}$ values $\lambda_{S,T}\geq 0.5$ and the orange points represent $\mu^2_{\rm{eff}}$ values $\lambda_{S,T}< 0.5$. The red line shows the $Z$ boson mass $M_Z$.}\label{fnt}
\end{center}
\end{figure}
We show in Figure~\ref{fnt}(a) plots of $\mu^2_{\text{soft}}$ and  $-\mu^2_{\text{eff}}$ versus the singlet vev $v_S$ for tree-level 
candidate Higgs masses in the interval $120\leq m_{h_1} \leq 130$ GeV. Figure~\ref{fnt}(b) presents the same plots, but with $m_{h_1}$, the candidate Higgs mass, calculated at one-loop. The violet points represent $\mu^2_{\rm{soft}}$ values for which $\lambda_{S,T}\geq 0.5$, and the points in blue refer to values of $\mu^2_{\rm{soft}}$ with $\lambda_{S,T}<0.5$. The green points mark values of $\mu^2_{\rm{eff}}$ with $\lambda_{S,T}\geq 0.5$, and the orange points refer to $\mu^2_{\rm{eff}}$ values with $\lambda_{S,T}< 0.5$. We see that for low $\lambda_{T,S}$ both $\mu^2_{\text{soft}}$ and  $-\mu^2_{\text{eff}}$  (blue and orange points) are small, so that the required cancellation needed in order to reproduce the $Z$ boson mass is also small. This leads to less fine-tuning, measured by $\mathcal{F}< 1$. Unfortunately, such points are small in numbers in the tree-level case,
since they require the extra contributions from the triplet and the singlet in order to reproduce the $\sim 125$ GeV Higgs mass. For  $\lambda_{T,S}\geq 0.5$ both $\mu^2_{\text{soft}}$ and  $-\mu^2_{\text{eff}}$  (the violet and green points) are both very large, leading to large cancellations and thus to a fine-tuning parameter $\mathcal{F}\sim 5$ for $\mu^2_{\text{soft}},-\mu^2_{\text{eff}}  \sim 10^6$. \\
Comparing Figure~\ref{fnt}(a) and Figure~\ref{fnt}(b) we see that the tree-level Higgs mass needs more fine-tuning as $\mu^2_{\text{soft}, \text{eff}} \sim 10^6$  for large $\lambda_{T, S}$. The situation improves significantly at one-loop due to the contributions from the radiative corrections. This is due to the fact that there are more solutions with low values of $\lambda_T$, $\lambda_{T, S}<0.5$, compared to tree-level and, on top of this,  (for high and low $\lambda_{T,S}$) the required fine-tuning is reduced ($\mathcal{F}\lesssim 2$). This fine-tuning measure is a theoretical estimation but it is constrained from the lightest chargino mass bound from LEP ($m_{\tilde{\chi}^\pm_1}> 104$ GeV), which results in $\mu_{\text{eff}}>104$ GeV and $\mathcal{F}\gsim 0.2$.

We have performed a run of $m^2_{H_u, H_d}$ using the corresponding $\beta$-functions for large $\lambda_{T,S}$, from electroweak scale ($M_Z$) up to a high-energy scale $\sim 10^{9,10}$ GeV, where the couplings become non-perturbative. It can be shown that $m^2_{H_u, H_d}, \mu^2_{\text{soft}}$  do not blow up unless the couplings $\lambda_{T,S}$ hit the Landau pole. The requirement of perturbativity of the evolution gives stronger bounds on the range of validity of the theory and the fine-tuning parameter is a good indicator at the electroweak scale.

In the case of MSSM, the large effective quartic coupling 
comes from the storng SUSY sectors which also increase $m^2_{H_u}$ and other parameters. However the situation changes in the case of extended Higgs sectors,
which gives additional tree-level as well as quantum corrections to the Higgs masses.
These reduce the demand for larger $m^2_{H_u}$. In our case there is a singlet and a triplet which contribute largely at the tree-level for low $\tan\beta$ and also contribute at the quantum level. In the case of tree-level Higgs mass, the extra tree-level contributions demand very large $\lambda_{T,S} \sim 0.8$, which in turn make
$\mu_{\text{eff}}$ very large and so the fine-tuning. However in the case of Higgs mass at one-loop, the extra contributions from the extended Higgs sectors are shared by both tree-level and quantum corrections, which reduces the requirement of large $\lambda_{T,S}$. This reduces  $\mu_{\text{eff}}$  and so the fine-tuning $\mathcal{F}$.

\section{A light pseudoscalar in the spectrum}\label{axion}
In the limit when the $A_i$ parameters in Eq.~(\ref{softp}) go to zero, the  discrete $Z_3$ symmetry of the Lagrangian is promoted to  a continuos $U(1)$ symmetry given by Eq.~(\ref{csmy}). 
\begin{figure}[thb]
\begin{center}
\mbox{\subfigure[]{ \hskip -15 pt
\includegraphics[width=0.55\linewidth]{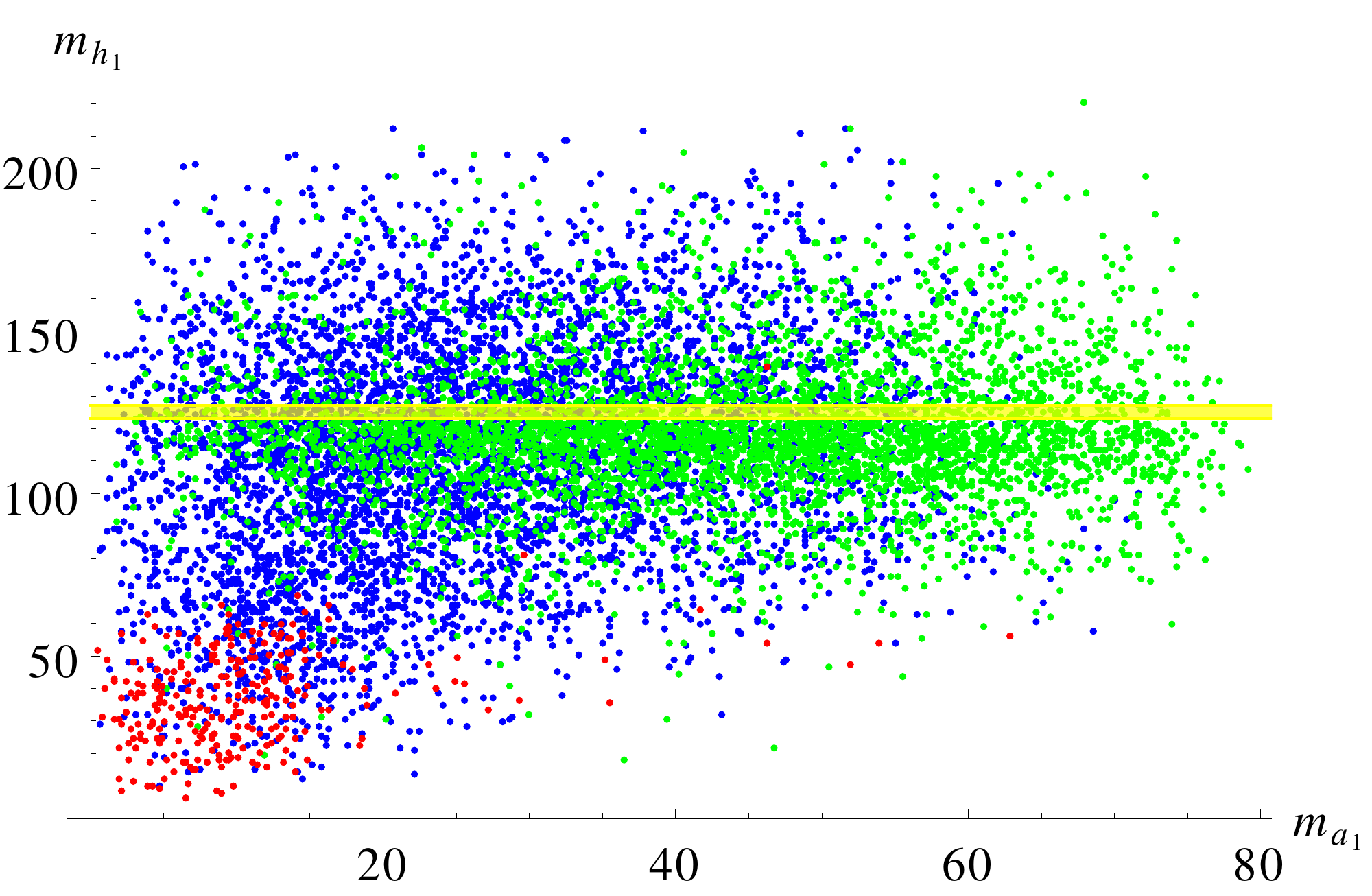}}
\subfigure[]{\includegraphics[width=0.55\linewidth]{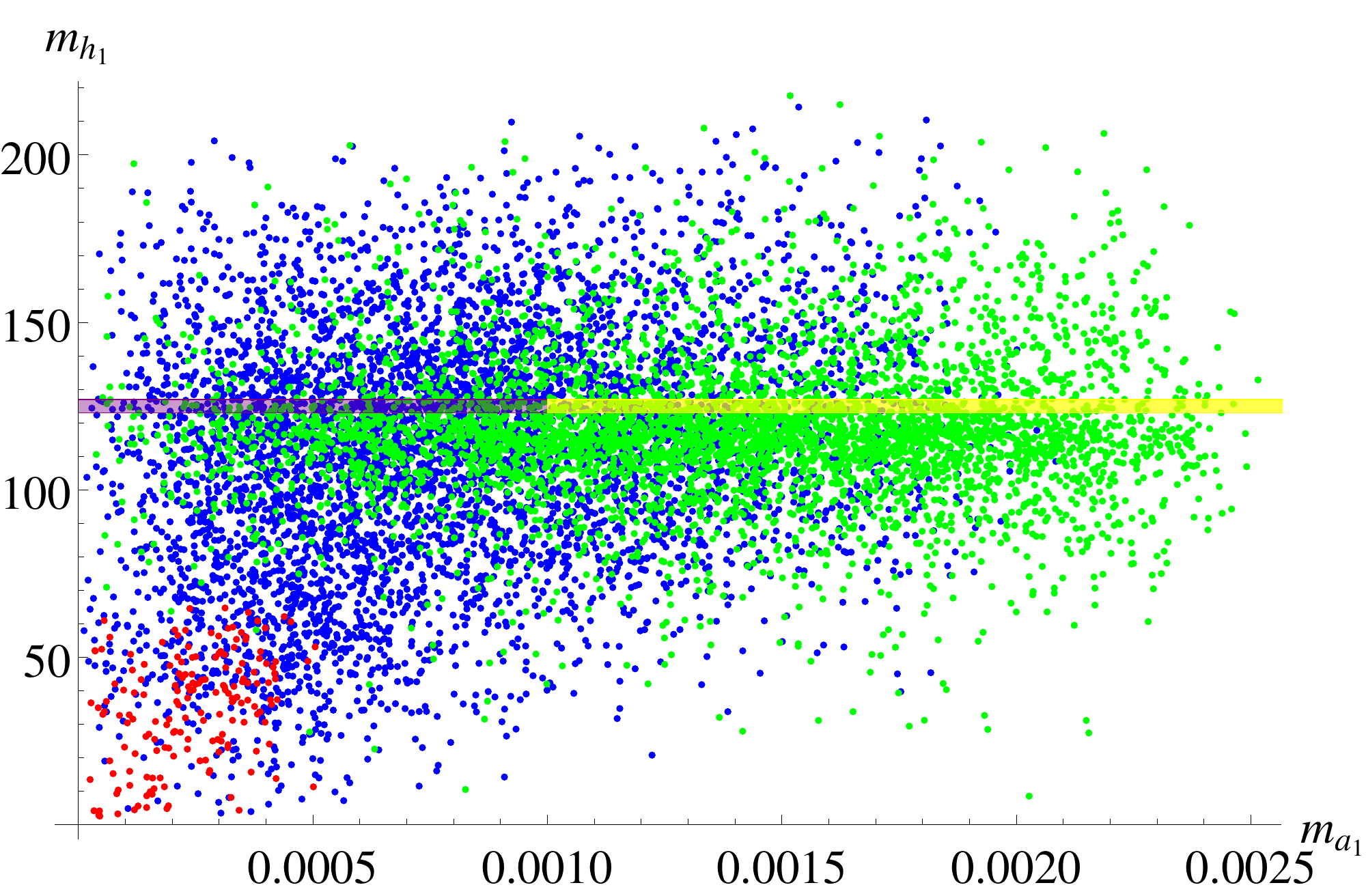}}}
\caption{The lightest CP-even Higgs boson mass $m_{h_1}$ vs the lightest pseudo-scalar mass $m_{a_1}$ at one-loop (top-stop and bottom-bottom corrections). The violet-yellow band presents the candidate Higgs mass $123\leq m_{h_1} \leq 127$ GeV.  The violet band specify the points with $m_{a_1}\leq 1$ MeV, where the $a_1 \to e^+e^-$ decay is kinematically forbidden. Red, blue and green points are defined as in Figure 2.}\label{mAA1}
\end{center}
\end{figure}
This symmetry is spontaneously broken by the vevs of the doublets, triplet and the singlet fields and should contain a physical massless pseudoscalar, $a_1$,  the Nambu-Goldstone boson of the symmetry. The soft breaking parameters will then lift the mass of $a_1$, turning it into a pseudo-Goldstone mode whose mass will depend on the $A_i$. The symmetry takes the form 
\bea\label{csmy}
(\hat{H}_u,\hat{H}_d, \hat{T},\hat{ S}) \to e^{i\phi}(\hat{H}_u,\hat{H}_d, \hat{T},\hat{ S}) .
\eea 
If this symmetry is softly broken by very small parameters $A_i$ of $\mathcal{O} (1)$ GeV, we get a very light pseudoscalar
\cite{nmssm, Agashe:2011ia} which could be investigated at cosmological level. Notice that the vector-like nature of the symmetry decouples this pseudoscalar from any anomalous behaviour.
We are going to briefly investigate some features of the $a_1$ state in the context of the recent Higgs discovery and we will consider two different realizations.
In the first case we consider a scenario where such continuous symmetry is broken very softly. In this case we choose the $A_i$ parameters to be $\mathcal{O}(1)$ GeV. We expect the pseudo-Goldstone boson to be very light, with a mass $\mathcal{O}(1)$ GeV. In Figure~\ref{mAA1} we show the mass correlation between 
the lightest CP-even neutral Higgs boson $h_1$ and the lightest massive CP-odd neutral Higgs boson $a_1$. The red points are of doublet type, the green points represent massive states of triplet/singlet type and the blue points represent the mixed contributions to the $a_1$ pseudoscalar. The violet-yellow band presents the region of parameter space where $h_1$ is the  candidate Higgs, with a mass $123\leq m_{h_1} \leq 127$ GeV.  It is rather clear from Figure~\ref{mAA1}(a) that there plenty of points in the parameter space where there could be a hidden pseudoscalar Higgs boson along with or without a CP-even hidden scalar. Such a light pseudoscalar boson gets strong experimental bounds from LEP searches \cite{LEPb} and from the bottomonium decay rates \cite{bottomium}.  Such light pseudoscalar in the mass range of 5.5 -14 GeV, when it couples to fermions, gets strong bound  from the recent CMS data at the LHC \cite{cmsab}. For triplet/singlet green points these bounds can be evaded quite easily since these states do not couple to gauge bosons (the $Z$ boson in the case of a triplet) and to fermions. Of course, for real mass eigenstates the mixing between the doublet-triplet/singlet would be very crucial in the characterization of their allowed parameter space. 

For a mass of the $a_1$ of $\mathcal{O}(100)$ MeV, the decay to $\pi \gamma \gamma$, 
$\pi \pi \pi$ could be an interesting channel to investigate in order to search for this state \cite{infnr}. The simpler 2-particle decay channel $a_1\to \pi\pi$
is not allowed due to the CP conservation of the model. 
Due to the singlet/triplet mixing nature of this state, it decays into fermion pairs $e^+e^-, \mu \bar{\mu}, \tau\bar{\tau}$, if kinematically allowed. Notice that there is no discrete symmetry to protect this state from decaying, preventing it from being a dark matter candidate \cite{Mambrini:2011ri}. 
Now, if we choose the $A_i$ parameters to be of $\mathcal{O}(1)$ MeV then we get a very light pseudoscalar
boson with mass of the same order, as shown in Figure~\ref{mAA1}(b). Such a bosons cannot decay to $\mu \bar{\mu}, \tau\bar{\tau}$ kinematically. Following the same reasoning, if its mass is $<1$ MeV, then even the $a_1 \to e^+e^-$ channel is not allowed and only the photon channel remains open to its decay.
In this case the $a_1$ resembles an axion-like particle, and can be a dark matter candidate only if its lifetime is larger than the age of the universe \cite{infnr, Bernabei:2005ca}. The pseudoscalar, in this case, couples to photons at one-loop, due to its doublet component which causes the state to have a direct interaction with the fermions.\\
We recall that the effective lifetime of a light pseduoscalar decaying into two photons is given by Eq.~(\ref{agg}) \cite{Bernabei:2005ca}

\bea\label{agg}
\tau_a =\frac{64 \pi}{g^2_{a\gamma\gamma}m^3_{a}}
\eea
where $g_{a\gamma\gamma}$ is the effective pseduoscalar /fermion coupling which is proportional 
to the doublet-triplet/singlet mixing. Notice that the $a_1$ shares some of the behaviour of axion-like particles, which carry a mass that is unrelated to their typical decay constant, and as such are not described by a standard Peccei-Quinn construction. They find 
a consistent description in the context of extensions of the SM with extra anomalous abelian symmetries \cite{ga0} \cite{ga1} and carry a direct anomalous (contact) interaction to photons. Such interaction is absent in the case of a $a_1$ state. \\
Along with the lightest neutralino of the TNMSSM, this particle can be a dark matter candidate. In the supersymmetic context a similar scenario, with two dark matter candidates 
has been discussed in \cite{ga2}. The role of this pseudoscalar state, in the context of the recent results by FERMI about the 1-3 GeV excess gamma-ray signal from the galactic center \cite{Arina:2014yna} is under investigation for this model \cite{pb3}.  

\section{$\sim 125$ GeV Higgs and LHC data}\label{Hdata}
In this section we consider the one-loop Higgs mass spectrum, including only the correction coming from quarks and squarks, in light of recent results from the LHC \cite{CMS,CMS2, ATLAS} and the existing data from LEP \cite{LEPb}. In particular, we consider the uncertainties in the decay modes of the Higgs to $WW^*$, $ZZ^*$ and $\gamma\gamma$ in a conservative way \cite{CMS, ATLAS}. We explore the scenario where one of the CP-even neutral scalars is the candidate $\sim 125$ GeV Higgs boson within the mass range $123\leq m_{h_i}\leq 127$ GeV and investigate the possibilities of having one or more light  scalars, CP-even and/or CP-odd, allowed by the LEP data and consistent with the recent Higgs decay branching fractions at the LHC. 
 
We just mention that in the TNMSSM the triplet and the singlet type Higgs bosons do not couple to the $Z$ boson but the triplet couples to the $W^\pm$ bosons, which result in a modified $h_i\, W^\pm\,W^\mp$ vertices given by 
\bea
h_i\,W^\pm\,W^\mp = \frac{i}{2}\,g_L^2\left(v_u\mathcal{R}^S_{i1} +v_d\mathcal{R}^S_{i2} +4\,v_T\mathcal{R}^S_{i4}\right),
\eea
where the rotation matrix $R^S_{ij}$ is defined in Eq.~(\ref{rot}). The vertices $h_i\,Z\,Z$ are given by
\bea
h_i\,Z\,Z = \frac{i}{2}\left(g_L\cos\theta_W+g_Y\sin\theta_W\right)^2\left(v_u\mathcal{R}^S_{i1} +v_d\mathcal{R}^S_{i2}\right),
\eea
where $\theta_W$ is the Weinberg angle. The Yukawa part of the superpotential is just the MSSM one. Hence the couplings of the CP-even sector to the up/down-type quarks and to the charged leptons are
\bea
h_i\,u \,\bar u = -\frac{i}{\sqrt2}y_u \mathcal{R}^S_{i1},\\
h_i\,d \,\bar d = -\frac{i}{\sqrt2}y_d \mathcal{R}^S_{i2},\\
h_i\,\ell \,\bar\ell = -\frac{i}{\sqrt2}y_\ell \mathcal{R}^S_{i2},
\eea
respectively. 

\begin{figure}[thb]
\begin{center}
\mbox{\subfigure[]{\hskip -15 pt
\includegraphics[width=0.55\linewidth]{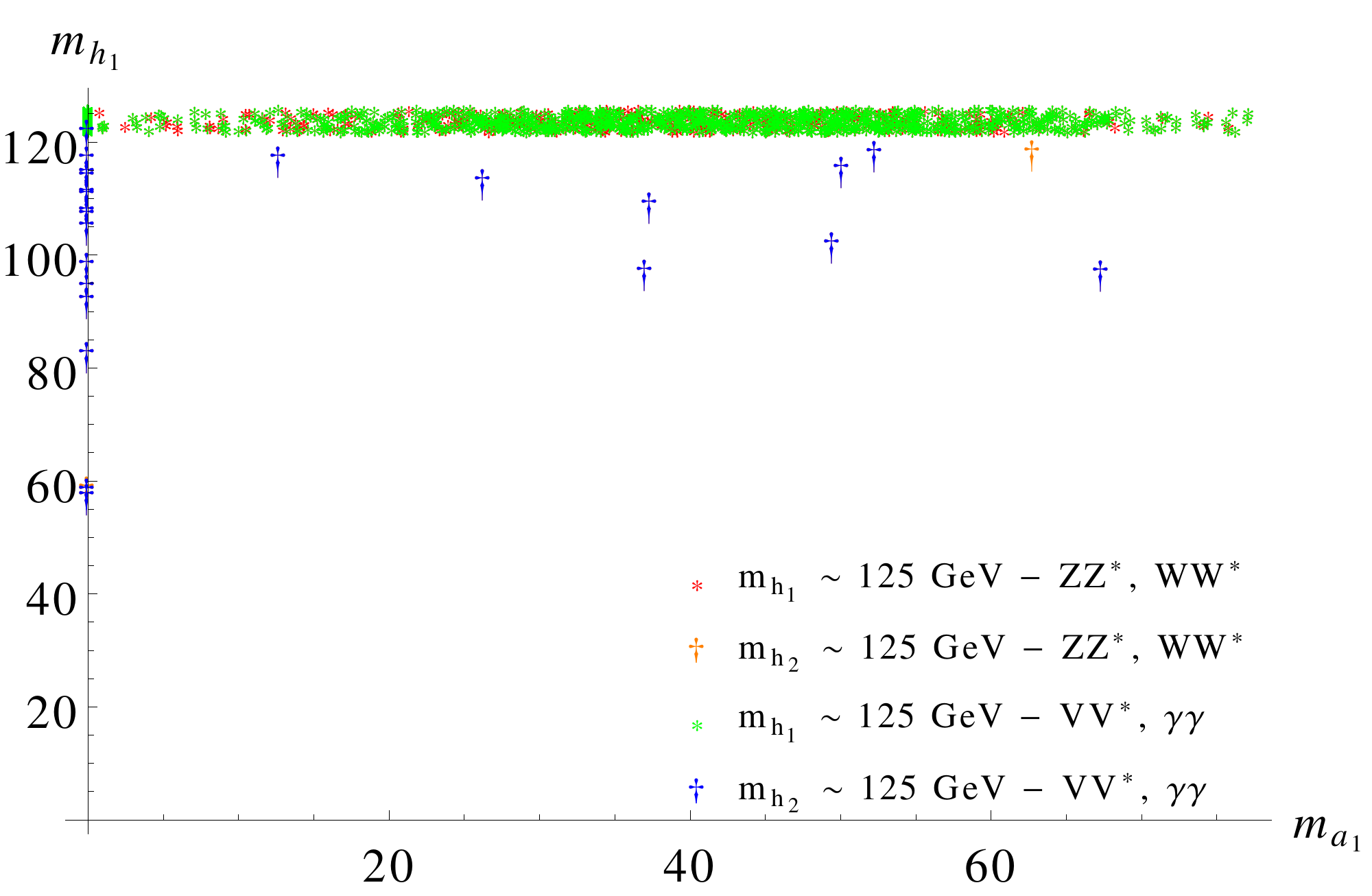}}
\subfigure[]{\includegraphics[width=0.55\linewidth]{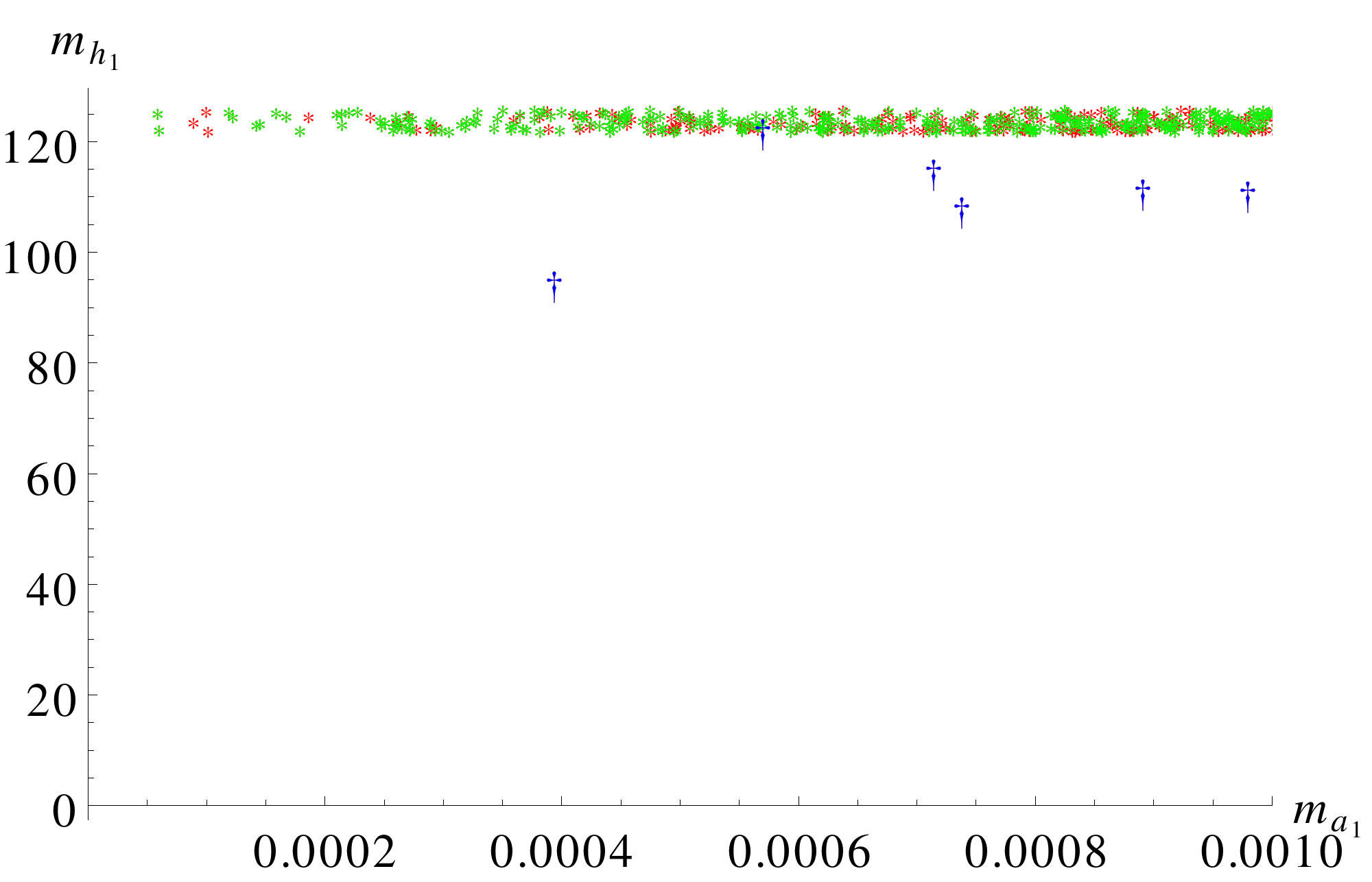}}}
\caption{The lightest CP-even Higgs boson mass $m_{h_1}$ vs the lightest pseudo-scalar mass $m_{a_1}$ at one-loop (top-stop and bottom-bottom corrections) consistent with the Higgs data from CMS, ATLAS and LEP. The red points corresponds to the case where $m_{h_1}\sim m_{125}$, the orange points correspond to mass values of $m_{h_1}$ and $m_{a_1}$  where  $m_{h_2}\sim m_{125}$ and all of them satisfy  the $ZZ^*$, $WW^*$ bounds at $1\sigma$ and $\gamma\gamma$
bound at $2\sigma$ level from both CMS and ATLAS. The red (orange) points which satisfy the  $\gamma\gamma$ result at $1\sigma$ are marked green (blue). Very light pseudoscalar masses $m_{a_1}\leq 1$ MeV are shown in panel (b), which is a zoom of the small mass region of (a).}\label{higgsdata}
\end{center}
\end{figure}

On the other hand, in the Higgs bosons decay into di-photons, 
there are more virtual particles which contribute in the loop compared to the SM. This is due to the
enlarged Higgs and strong sectors which have a non-zero coupling with 
the photon. In particular there are three charginos ($\chi_{1,2,3}^\pm$), three charged Higgs bosons ($h_{1,2,3}^\pm$), the stops ($\tilde{t}_{1,2}$) and the sbottoms ($\tilde{b}_{1,2}$). Compared to the MSSM and the NMSSM we have two additional charged Higgs bosons and one additional chargino which contribute to the decay. The decay rate in the di-photon channel is given by \begin{align}\label{gammagamma}
\Gamma(h\rightarrow\gamma\gamma)&=\frac{\alpha\,m_h^3}{1024\,\pi^3} \Big|\frac{g_{hWW}}{m_W^2}\,A_1(\tau_W)+\sum_{ \chi_i^\pm,\,t,\, b}2\frac{g_{hf\bar{f}}}{m_f}N^c_f\, Q_f^2\, A_{1/2}(\tau_f)\\
&+\sum_{h_i^\pm,\,\tilde{t}_i,\,\tilde{b}_i}\frac{g_{hSS}}{m_S^2}N_S^c\,Q_S^2\,A_0(\tau_S)\Big|^2,\nn
\end{align}
where $N_{f, S}^c$ are the color number of fermion and scalars, $Q_{f, S}$ are the electric charges, in unit of $|e|$, of the fermions and scalars, and $\tau_i=\frac{m_h^2}{4\,m_i^2}$. $A_0, \,A_{1/2}$ and $A_1$ are the spin-0, spin-1/2 and spin-1 loop functions
\bea
&&A_0(x)=-\frac{1}{x^2}\left(x-f(x)\right),\\
&&A_{1/2}(x)=\frac{2}{x^2}\left(x+(x-1)f(x)\right),\\
&&A_1(x)=-\frac{1}{x^2}\left(2\,x^2+3\,x+3(2\,x-1)f(x)\right),
\eea
with the analytic continuations 
\bea
f(x)=\left\{
\begin{array}{lr}
\arcsin^2(\sqrt{x})& x\leq1\\
-\frac{1}{4}\left(\ln\frac{1+\sqrt{1-1/x}}{1-\sqrt{1-1/x}}-i\pi\right)^2& x>1
\end{array}\right.
\eea
In the limit of heavy particles in the loop, we have $A_0\rightarrow 1/3$, $A_{1/2}\rightarrow 4/3$ and $A_1\rightarrow-7$.\\
Using the expression above, we study the discovered Higgs boson ($h_{125}$)  decay rate to di-photon in this model. We also check the consistency of light scalar(s) and/or light pseudoscalar(s)
with the current data at the LHC and the older LEP data.
Such analysis is presented in Figure~\ref{higgsdata}. Figure~\ref{higgsdata}(a) shows such hidden Higgs scenarios with one $a_1$ and/or one $h_1$ below 123 GeV, which find significant realizations. 
 We first consider the results  coming from both CMS and ATLAS in the decay of the Higgs to $WW^*$ and $ZZ^*$ modes at $1\sigma$ \cite{CMS, CMS2, ATLAS} and also consider the cross-section bounds from LEP \cite{LEPb}. The allowed mass values are shown as red points for which the lightest CP-even Higgs boson ($h_1$) is the detected Higgs at $\sim 125$ GeV. Cleary we see that there are many light pseudo-scalars ($\leq 100$ GeV) which are allowed. The orange points present the scenario where $m_{h_2}\sim m_{125}$ and which leaves both  $h_1$ and $a_1$ hidden ($< 125$ GeV).

We have performed additional tests of such points and compared them with the results from the decay of the Higgs boson to di-photons at the LHC, both from CMS \cite{cmsgamma} and ATLAS \cite{atlasgamma}. The red points (with one hidden Higgs boson)  which satisfy $h_{125}\to \gamma \gamma$ at $1\sigma$ level, are marked as green points. The orange points (with two hidden Higgs bosons) when allowed at  $1\sigma$ level, have been marked as blue points. Notice that all the points in Figure~\ref{higgsdata} are allowed at  $1\sigma$ by the $WW^*$, $ZZ^*$ channels and at $2\sigma$ by the $\gamma \gamma$ channel. These requirements automatically brings the fermionic decay modes closer to the SM expectation. Of course the uncertainties of these decay widths give us a room for $h_{125}\to a_1a_1/h_1h_1$. 

\begin{figure}[thb]
\begin{center}
\mbox{\hskip -15 pt\subfigure[]{\includegraphics[width=0.55\linewidth]{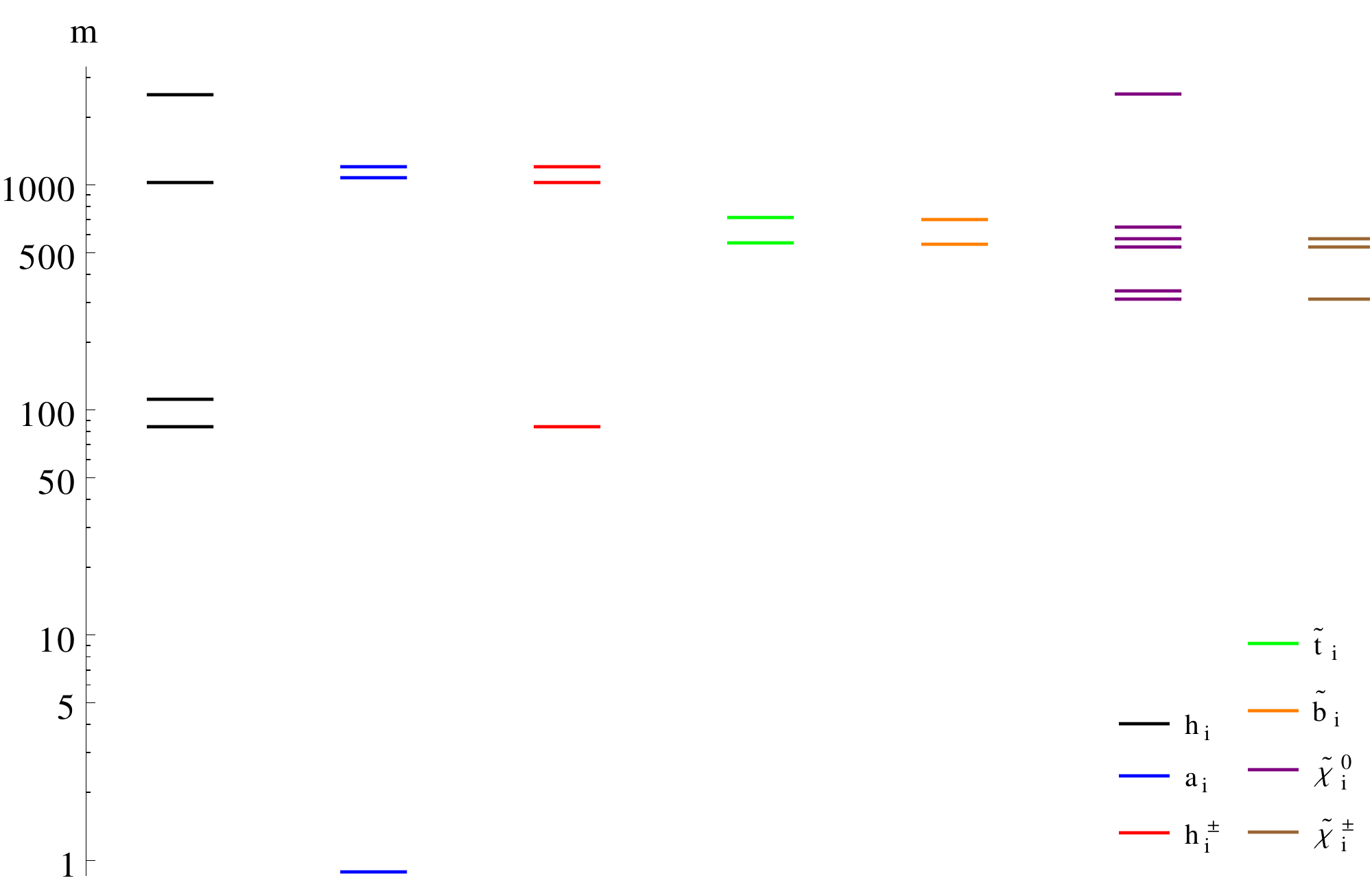}}
\subfigure[]{\includegraphics[width=0.55\linewidth]{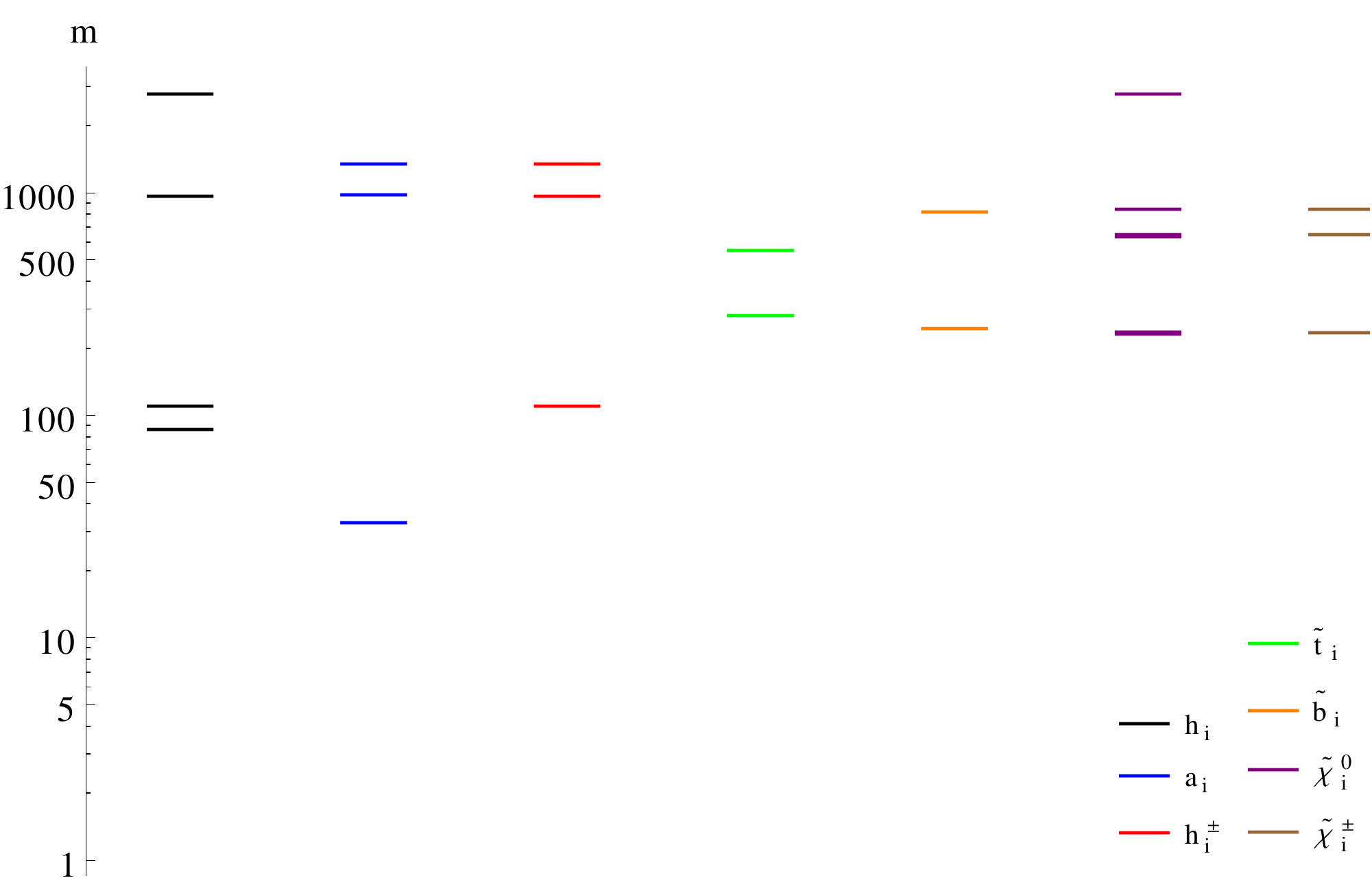}}}
\mbox{\subfigure[]{\includegraphics[width=0.55\linewidth]{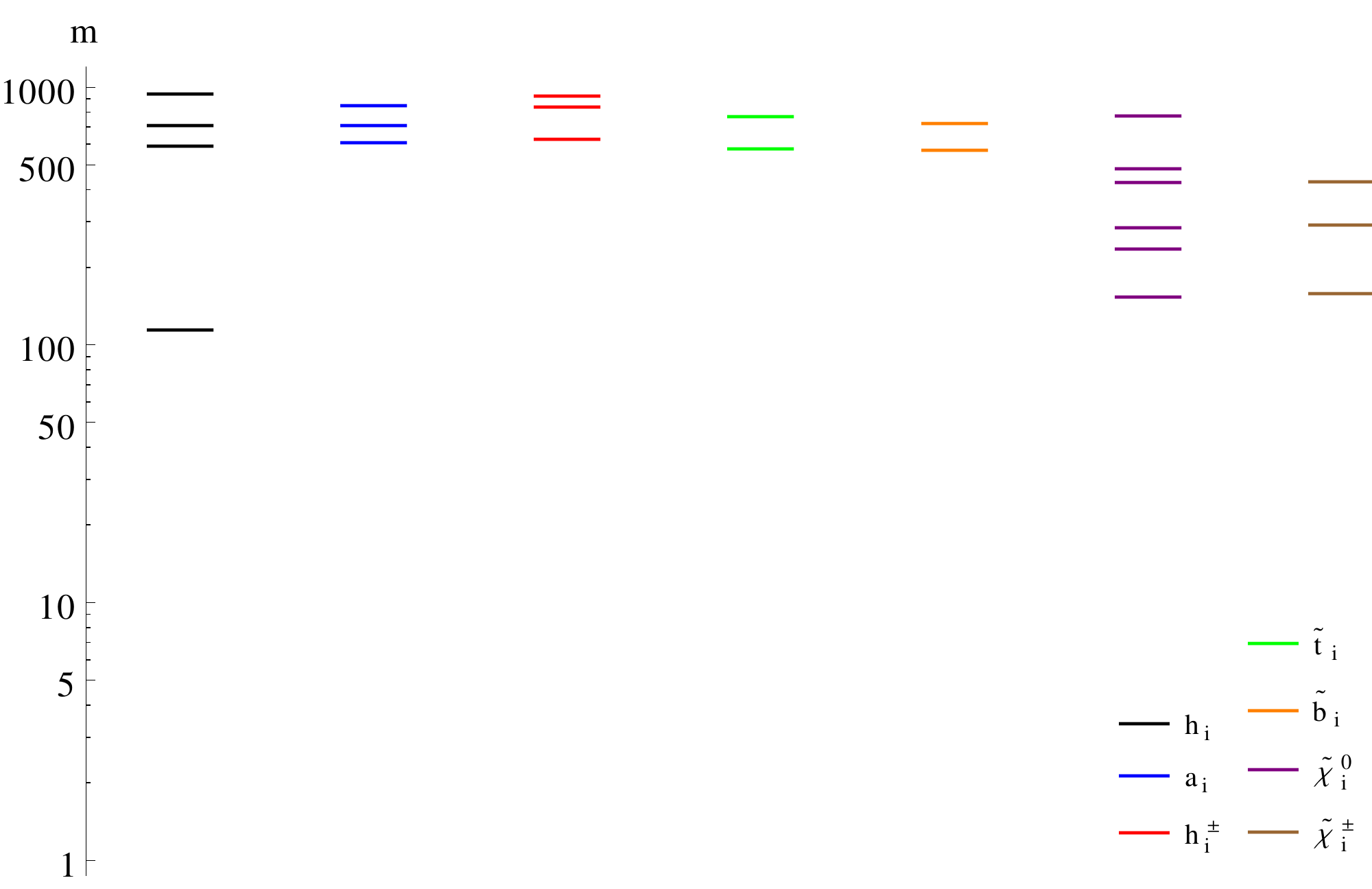}}
}
\caption{We show the benchmark points of the model which are allowed both by LHC \cite{CMS, ATLAS} and LEP\cite{LEPb} data. The neutral Higgs spectrum has been calculated at one-loop and the rest of the spectrum at tree-level.}\label{bps}
\end{center}
\end{figure}

Notice also the presence of a very light pseudoscalar mass values near $a_1 \sim 0$. Figure~\ref{higgsdata}(b) is a zoom of this region, where such solutions are shown for $m_{a_1} \leq 1$ MeV. The points in this case correspond to possible $a_1$ states which do not decay into any charged fermion pair ($m_{a_1} \leq 2m_{e}$) and have an interesting phenomenology, as briefly pointed out in section~\ref{axion}. The fact that such mass values only allow a decay of this particle to two photons via doublet mixing mediated by a fermion loop, makes the $a_1$ a possible dark matter candidate, being long lived. Two hidden Higgs bosons render the phenomenology very interesting, allowing both the $h_{125} \to a_1 a_1$ and the $h_{125} \to h_1 h_1$ decay channels \cite{hdlh, ehdc}. In Figure~\ref{bps} we show some of the points in this model as benchmark points (BMP's), which are allowed 
both by LHC \cite{CMS, ATLAS} and LEP \cite{LEPb} data.
The neutral Higgs spectrum has been calculated at one-loop order and the remaining states at tree-level. Figure~\ref{bps}(a) shows a point (BMP1) where we have a hidden pseudoscalar ($a_1$) with mass $\mathcal{O}(10^{-1})$ MeV and another triplet/singlet-like hidden CP-even scalar ($h_1$) with a mass around $\sim 93$ GeV. In this case the candidate Higgs boson is $h_2$, taken around $125$ GeV. This point also have a triplet type very light charged Higgs boson at a mass around 90 GeV, which is not excluded by the recent charged Higgs bounds from the LHC \cite{chHb}. Figure~\ref{bps}(b) shows a benchmark point (BMP2) where we have  a pseudoscalar  around 37 GeV, and the lightest scalar and charged Higgs bosons around 100 GeV. Figure~\ref{bps}(c) shows a trivial (SM-like) solution where we have a doublet-type CP-even Higgs around $\sim 125$ GeV, with the other states decoupled. In the next study we are going to analyse such points through a detailed collider simulation \cite{pb3}.

\section{Phenomenology of the TNMSSM}\label{pheno}
\begin{figure}[thb]
\begin{center}
\mbox{\subfigure[]{
\includegraphics[width=0.3\linewidth]{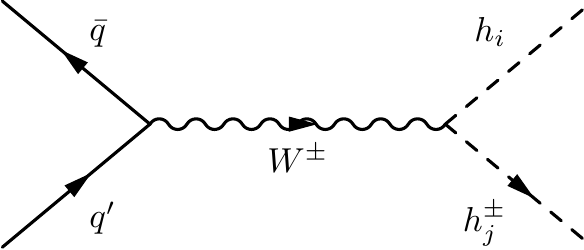}}
\hskip 25 pt
\subfigure[]{\includegraphics[width=0.3\linewidth]{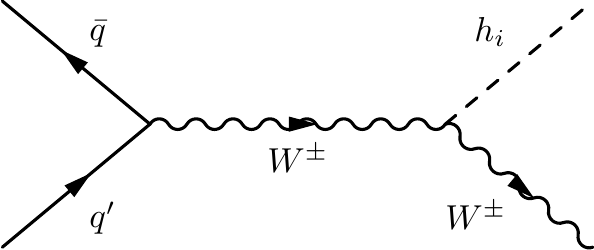}}}
\mbox{\subfigure[]{
\includegraphics[width=0.3\linewidth]{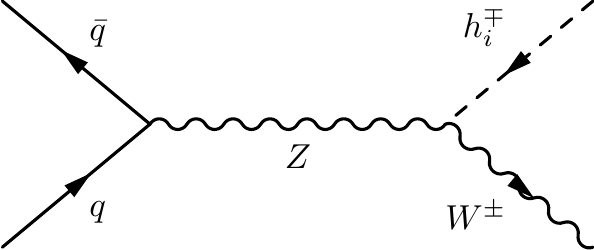}}
\hskip 25 pt
\subfigure[]{\includegraphics[width=0.25\linewidth]{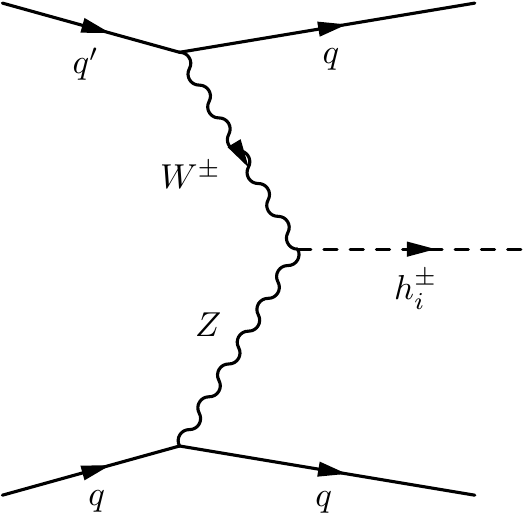}}}
\caption{The new and modified production channels for the Higgs bosons at the LHC.}\label{higgsprd}
\end{center}
\end{figure}
The TNMSSM extends the Higgs sector as well as the electroweak chargino-neutralino sectors by additional
higgsino contributions. Both the triplet and singlet fields do not couple to the fermions but affect the
phenomenology to a large extent. In the context of the recent Higgs discovery, searches for additional Higgs bosons, both neutral and charged, are timely. In particular, if an extended Higgs sector will be discovered at the LHC, it will be crucial to determine the gauge representation which such states belong to, by investigating its allowed decays modes. 

We have seen from Eq.~(\ref{spt}), that a $Y=0$ hypercharged triplet couples to the $W^\pm$ bosons and contributes to their mass. On the other hand, the singlet does not directly couple to any of the gauge bosons. In the case of Higgs mass eigenstates which carry a doublet-triplet-singlet mixing, we need to look either for their direct production processes at the LHC or take into consideration the possibility of their cascade production from other Higgses or supersymmetric particles.

\subsection{Productions}
We have detailed a model with a rich Higgs sector with additional Higgs bosons of triplet and singlet type. We recall that the relevant  production processes of a Higgs boson which is a SU(2) doublet at the LHC \cite{anatomy1, anatomy2} are the gluon-gluon fusion (GGF) and vector boson fusion channels, followed by the channels of associated production of gauge bosons and fermions. In our case, the production channels for the new Higgs bosons are different, due to their different couplings to the gauge bosons and fermions. We list below the possible additional production channels for the neutral and charged Higgs bosons at the LHC.

\begin{itemize}
\item \underline{ Neutral Higgs boson production in association with charged Higgs boson:} The triplet only couples to $W^\pm$ boson.
Thus a neutral Higgs (doublet or triplet) can be produced in association with a charged Higgs boson (doublet or triplet) via a $W^\pm$ exchange. As shown in Figure~\ref{higgsprd}(a) a light in mass and charged Higgs boson in the TNMSSM can be easily explored 
by this production channel $\bar{q} q' \to h_i h^\pm_j$.

\item  \underline{ Neutral Higgs boson production in association with $W^\pm$: } A triplet or a doublet type neutral Higgs boson
can be produced via $\bar{q}q' \to W^\pm h_i$ as shown in Figure~\ref{higgsprd}(b). A triplet admixture modifies the $h_i-h^\pm_j-W^\mp$ couplings by an additional term proportional to the vev of the triplet.

\item  \underline{Charged Higgs boson production in association with $W^\pm$: } Triplet of $Y=0, \pm2$ hypercharge 
has a non-zero tree-level coupling to $Z-W^\pm-h^\mp_i$. This leads to additional contributions to 
$q\bar{q} \to W^\pm h^\mp_i$ as shown in Figure~\ref{higgsprd}(c).

\item  \underline{Production of charged Higgs boson in vector boson fusion:} The non-zero $Z-W^\pm-h^\mp_i$ coupling leads to vector boson fusion ($Z, W$ fusion) which produces a charged Higgs boson as shown in Figure~\ref{higgsprd}(d). This mode is absent in 2-Higgs doublet models (2HDM), in the MSSM and in the NMSSM. This is a unique feature of the $Y=0, \pm2$ hypercharge, triplet-extended scenarios. 

\item \underline{Singlet Higgs production:} The singlet in this model is not charged under any of the gauge groups, and hence the direct  production of such a singlet at the LHC is impossible.
Gauging this additional singlet with the inclusion of an extra additional $U(1)'$ gauge group would open new production channels via the additional gauge boson ($Z'$). Most of the extra $Z'$ models get a bound on the $Z'$ mass, $m_{Z'} \gsim 2.79$ TeV \cite{LHCZ'} which makes such channels  less promising at the LHC. In our case such a singlet type Higgs boson would only be produced via mixing with the Higgs bosons of doublet and triplet type.

\end{itemize}
\subsection{Decays}
The smoking gun signatures for the model would be the decays of the doublet, triplet and singlet states that are produced. Different F-term contributions can generate these types of mixing and corresponding decay vertices.  We list  all the vertices in the gauge basis in appendix \ref{HVs}. The vertices in the mass eigenstate basis can be found by the rotations given in Eq.~(\ref{rot}).   

\begin{figure}[thb]
\begin{center}
\mbox{\subfigure[]{
\includegraphics[width=0.23\linewidth]{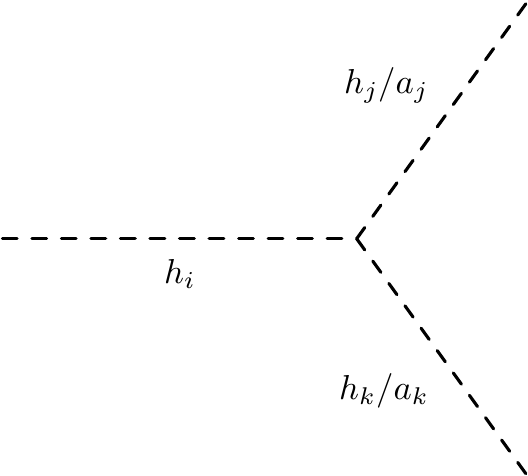}}
\hskip 25 pt
\subfigure[]{\includegraphics[width=0.23\linewidth]{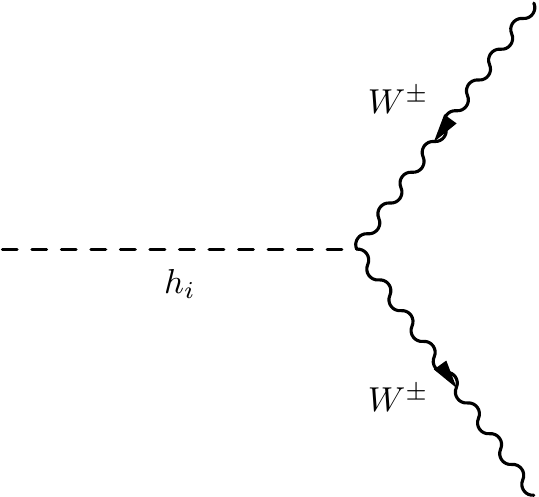}}}
\mbox{\subfigure[]{
\includegraphics[width=0.23\linewidth]{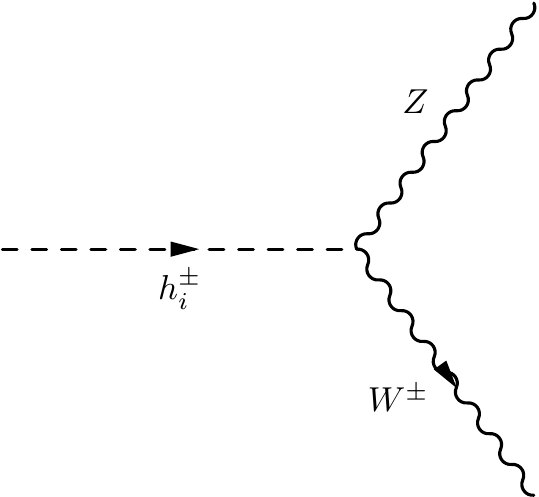}}
\hskip 25 pt
\subfigure[]{\includegraphics[width=0.23\linewidth]{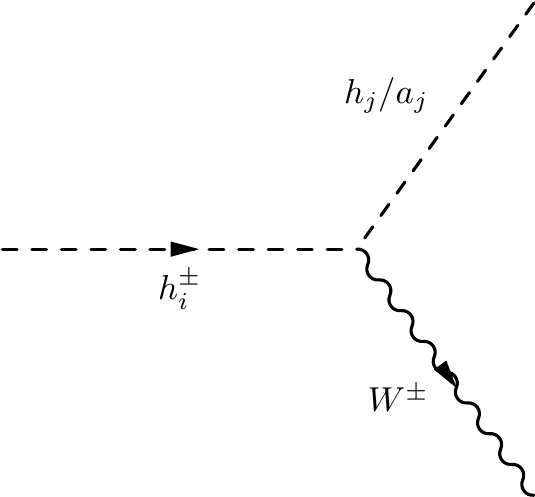}}
\hskip 25 pt
\subfigure[]{\includegraphics[width=0.23\linewidth]{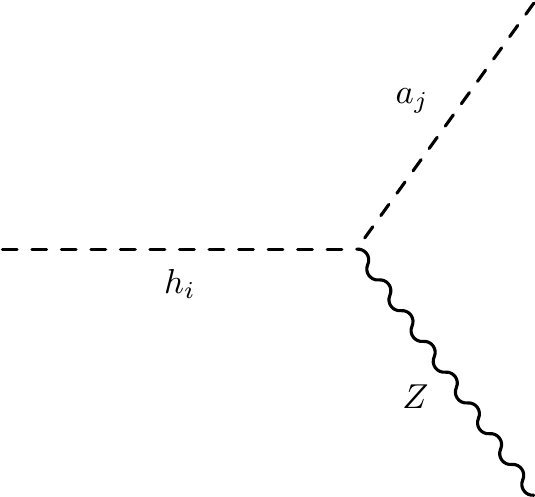}}
}
\caption{The new and modified decay channels of the Higgs bosons at the LHC.}\label{higgdcy}
\end{center}
\end{figure}
\begin{itemize}

\item \underline{Higgs decays to Higgs pairs:}  The candidate Higgs around the 125 GeV mass in this case can decay into two hidden Higgs bosons if this channel is kinematically allowed as can be seen in Figure~\ref{higgdcy}(a). Such hidden Higgs boson(s) could be both scalar and pseudoscalar in nature. The discovered Higgs is however $99\%$ CP-even \cite{CMS2}, which forbids any CP-violating decay of the nature $h_{125} \to a_i h_j$.  However, the CP-conserving decays  like  $h_{125} \to a_i a_j$ and/or $h_{125} \to h_i h_j$ are allowed. Such decays should be carefully investigated on the basis of the current Higgs data at the LHC. If the two light Higgs bosons are mostly singlet or triplet, then it is easy to evade the bounds from LEP \cite{LEPb}. As we have already pointed out, a singlet Higgs does not couple to any of SM gauge bosons and even the triplet type does not couple to the $Z$ boson. Such a light Higgs boson could decay into $\tau$ pairs only through the mixing with the doublet type Higgs bosons, since neither the singlet nor the triplet couple to fermions (see Eq.~(\ref{spt})). The mixing angle is also constrained by data on bottomonium decay, for a very light neutral Higgs boson ($\lsim 8$ GeV) \cite{bottomium}. \\
The decay of a Higgs boson into other Higgs bosons depends on the cubic coupling, which is proportional to the vevs of  Higgs fields, and thus it is very sensitive to the values of $v_i$. It is therefore requires an analysis of the allowed decay widths of the Higgs boson into a light Higgs pair using LHC data \cite{hdlh}.

\item \underline{Higgs decays to $W^\pm W^\mp$:} The triplet couples to $W^\pm$ via its non-zero SU(2) charge which is at variance respect to the analogous coupling of the doublet, as can be seen in Eq.~\ref{vrtx}. This will modify the decay width of $h_i \to W W$ (Figure~\ref{higgdcy}(b)).  The recent data show that  there is some disagreement and uncertainties  between the CMS \cite{CMS, CMS2} and  ATLAS \cite{ATLAS} results in the  $h_{125} \to WW^*$ 
channels. The measurement of this decay channel thus becomes even more crucial under the assumption of a triplet mixture. 
\item \underline{Charged Higgs decays to $Z\,W^\pm$:}   We know that the triplet type charged Higgs has a non-zero tree-level coupling to $Z\,W$, for a non-zero triplet vev, as shown in Figure~\ref{higgdcy}(c). This opens up the possible decay modes $h^\pm_i \to Z W^\pm$, which are absent in the 2HDM and in the MSSM at tree-level.

\item \underline{Charged Higgs decays to $h_j(a_j)W^\pm$:} A doublet or triplet type Higgs boson can decay to a lighter neutral Higgs and a $W^\pm$ (Figure~\ref{higgdcy}(d)). A possibility of a very light triplet-singlet-like neutral Higgs makes this decay mode more interesting compared to the case of the CP-violating MSSM \cite{cpv}. 

\item \underline{Higgs decays to $a_j Z$:} In the MSSM the odd and heavy Higgs bosons are almost degenerate, so $h_i \to a_j Z$ is not kinematically allowed. The introduction of a triplet and of a singlet 
adds two more massive CP-odd Higgs bosons, and the degeneracy is lifted. In this case we have a relatively 
lighter CP-odd Higgs state $a_i$ which makes $h_i \to a_j Z$ possible, as shown in Figure~\ref{higgdcy}(e). This scenarios is also possible in the context of the CP-violating MSSM, where we have a very light pseudoscalar Higgs boson due to the large mixing between the Higgs CP eigenstates \cite{pb2} and in the NMSSM, for  having an additional scalar \cite{nmssm}.

\item \underline{Higgs decays to fermion pairs:}  In a scenario where a  triplet or/and singlet type Higgs boson decays to gauge bosons and other Higgses are kinematically forbidden, the only permitted decays are into light fermion pairs, viz, $bb, \tau\tau$ and $\mu\mu$. Even such decays are only 
possible by a mixing with doublet type Higgs bosons. When such mixing angles are very small
this can results into some displaced charged leptonic signatures.

\end{itemize}
\subsection{Possible signatures}\label{sign}
The unusual production and decay channels lead to some really interesting phenomenology which could be tested in the next run of the LHC and at future colliders.  From the testability point of view, one could use the data form the discovered Higgs boson $\sim 125$ GeV in order to get bounds from the Higgs decaying to Higgs boson pair \cite{hdlh}, or the existing bounds from LEP \cite{LEPb} for two Higgs bosons productions. We have already taken into account these bounds by ensuring that the hidden Higgs boson is mostly of singlet or of triplet type. Given the uncertainty in the Higgs decay branching fractions in different modes and the absence of direct bounds on the non-standard decays of Higgs boson to Higgs boson pair ($h_{125}\to a_i a_j/h_i h_j$), this remains phenomenologically an interesting scenario. Below we list different possible signatures that could be tested in the LHC with 13/14 TeV. 

\begin{itemize}

\item 
The singlet and doublet F-terms  generate the doublet-triplet-triplet vertex which is proportional to $\lambda_S \lambda_{TS}$ and $\lambda^2_T$. This would provide a signature of a doublet type Higgs decaying into two triplet type Higgs bosons, which, in turn, do not decay into fermions. Similarly the F-terms of $H_u$ and $H_d$ generate vertices involving triplet-singlet-doublet  which are proportional to 
$\lambda_T \lambda_S$. The F-term of triplet type also contributes to this mixing, which is 
proportional to $\lambda_T \lambda_{TS}$. Thus the relative sign between the two contributions become important. The vertex is given in Eq.~\ref{vhts}.

In the case of a $\sim 125$ GeV Higgs boson, this can decay into two triplet-like hidden scalars or pseudoscalars, which in turn decay into off-shell $W^\pm$s only. This type of decays can be looked for by searching for very soft jets or leptons coming from the off-shell $W^\pm$s.  The signatures could be the $4\ell +\ptmiss$ or $4j+2l +\ptmiss$ channels, where the jets and the leptons are very soft. On the other hand, both the triplet and the singlet hidden Higgses can decay to fermion pairs ($b\bar{b}$, $c\bar{c}, e^+e^-, \mu\bar{\mu}, \tau \bar{\tau}$) via the mixing with doublets. The recent bounds on these non-standard decays has been calculated for the LHC \cite{ehdc}. Such decays give $4\ell, \, 2b+ 2\ell$ final states, where the leptons are very soft. For the triplet type hidden Higgs bosons it would be interesting to analyze the competition between the four-body and the two-body decays (which depend on the triplet-doublet mixing). Demanding for the presence of softer  leptons and jets in the final states, allows to reduce the SM backgrounds at the LHC. If the mixing is very small, this could lead to displaced charged leptonic final states, similar to those of a Higgs boson decay in a $R$-parity violating supersymmetric scenario \cite{pbrp}.  Due to the coupling both with the up and the down type doublets, this coupling could be tested both at a low and a high $\tan{\beta}$.

\item The singlet does not contribute to charged Higgs mass eigenstates, so the charged Higgs bosons  could be either of doublet or triplet nature. In the case of a heavy doublet type, the heavier charged Higgs can decay to a triplet type a neutral Higgs (CP even or odd) and a triplet type charged Higgs ($H^\pm_{u,d} \to T^0 T^\pm_{1,2}$) (see Appendix~\ref{HVs}). The coupling is proportional to ($g^2_L -\lambda^2_T$).  The lighter triplet type charged Higgs then mostly decays into on-shell or off-shell $Z\, W^\pm$. This is a generic signature for $Y=0, \pm2$ hypercharge triplets with non-zero triplet vev, which breaks the custodial symmetry of the Higgs potential.  The relatively lighter triplet (either CP-odd or even) neutral Higgs can decay via an on/off-shell $W^\pm$
 boson pair, which leads to leptonic final states. The final states  with multi-lepton($>3\ell$), multi-jet($>4$) and missing energy, could be the signature for this model. Depending on the off-shell decays, few leptons or a jet could be softer in energy.  

\item  In other cases a triplet type heavier charged Higgs can decay into a doublet type neutral Higgs and a triplet type charged Higgs.  These couplings are proportional  to ($\frac{g^2_L}{2} -\lambda^2_T$ ) and can give rise to $3\ell +2b +\ptmiss$ and  $3\ell +2\tau +\ptmiss$ final states. Here the $b$ and $\tau$ pairs expected from the neutral doublet type Higgs boson decay. 

\item Unlike to the neutral Higgs bosons, the up and down type charged Higgs bosons doublet only mix with the triplets. The couplings are again proportional to a combination of $\lambda_S \lambda_{T}$
and $\lambda_T \lambda_{TS}$.  In this case the doublet (triplet)  charged Higgs state will decay into a triplet (doublet) charged Higgs and a singlet neutral Higgs boson. As the singlet is not coupled to any SM particles, it can only decay through mixing with doublets and triplets.  Decays of such singlets to leptons (in the case of mixing with doublets) and off-shell or on-shell $W^\pm$-pair will be determined by the mixing only. In a fine-tuned region where such mixing is very low this decay channel can lead to a displaced vertex of charged leptons, whose measurement can give information about such a mixing. 
\end{itemize}

\section{Discussions and Conclusions}\label{dis}
In this paper we have considered a scenario with an extended Higgs sector characterized by a $Y=0$ hypercharge $SU(2)$ triplet and a gauge singlet superfields, along with the remaining MSSM superfields. The triplet vev is restricted by the $\rho$ parameter, hence the $\mu_{\text{eff}}$  is generated 
spontaneously mostly by the singlet vevs. In models with gauged $U(1)'$ symmetry the singlet could be invoked 
in the mass generation of the extra gauge boson $Z'$ by spontaneous symmetry breaking. This would require 
a large singlet vev $v_S$, due to the recent bounds on extra $Z'$ coming from the analysis at the LHC \cite{LHCZ'}.

We have first investigated the masses of the Higgs sector of the model at  tree-level. The lightest tree-level Higgs state, in this case, is not bounded to lay below $M_Z$, due to the additional contributions from the triplet and the singlet, which are proportional to their respective couplings and are enhanced at low $\tan{\beta}$. This allows to reduce the size of the quantum correction needed in order to reach the $\sim 125$ GeV  at one-loop, compared to the MSSM or to others constrained MSSM scenarios. Then we have extended our analysis at one-loop level. The one-loop Higgs with mass around $\sim 125$ GeV puts some indirect bounds on the masses of the particles contributing in the radiative corrections. For this purpose we have included the one-loop contributions using the Coleman-Weinberg potential. We have also presented results for the neutralino, and charginos spectra, together with the stop and sbottom mass matrices. We have calculated full one-loop Higgs masses considering all the weak sectors and the strong sectors. We also showed that the gauge boson-gaugino-higgsino sectors mostly contribute negatively to the mass eigenstates, while the stop-top, sbottom-bottom and Higgs sectors contribute positively. Due to the large number of scalars, seven neutral and three charged Higgs bosons, the Higgs self corrections can be larger than the strong corrections in the large $\lambda_{T,S}$ limit.  This substantially reduces the indirect lower bounds on the stop and sbottom masses. Thus in TNMSSM the discovery of a $\sim 125$ GeV Higgs boson does not put a stringent  lower bound on the stop and sbottom masses, and one has to rely on direct search results  for the lower bounds on the SUSY mass scale.

 We have implemented the model in SARAH3.5 \cite{sarah} in order to generate the vertices and other model files for CalcHEP \cite{calchep}. The beta-functions have also been generated at one-loop. We have addressed the issues of perturbativity of the couplings at the higher scale, as we have run the corresponding renormalization group equations from the electroweak scale up.  This has shown that the couplings of the model at the electroweak scale need to be restricted to certain values. For example, even with a value of $\lambda_{T,S}\sim 0.8$ at the electroweak scale, the theory remains perturbative up to $10^{8-10}$ GeV.  Setting all the couplings at a value ($\lambda_{TS} \sim 0.8$, $\kappa\sim 2.4$) the upper scale in the perturbative evolution gets lowered to $10^{4-6}$ GeV. The issue of fine-tuning at the electroweak scale has been discussed in this context.  We have seen that although the tree-level mass spectrum is highly fine-tuned for larger $\lambda_{T,S}$, the amount of fine tuning is reduced after the inclusion of the radiative corrections.
 
 The prospects for hidden Higgs(es), which are scalars and/or pseudoscalars of mass lower than the current Higgs mass, has been discussed quite thoroughly. We have seen that in the rich Higgs spectrum of the model there are several possibilities for having one or more hidden neutral Higgs bosons ($\lesssim125$ GeV) both CP-even and CP-odd. A special scenario emerges when we break the continuous $U(1)$ symmetry softly by the parameters $A_i$. This leads to the appearance of a very light pseudoscalar state of $\mathcal{O}(1)$ GeV to $\mathcal{O}(1)$ MeV in mass, which has its own interesting  phenomenology.
 
  Finally, we have discussed the doublet-triplet-singlet mixing which influences the productions and decays of neutral and charged Higgs bosons at the LHC. The existence of a $h^\pm_i-W^\mp-Z$ tree-level vertex, due to the triplet, impacts both the production as well as the decay channels of the charged Higgs bosons \cite{tssmch1}. In the presence of a light pseudoscalar, the $h_i\to Z a_j$ channel is a possibility due to the very light mass of the pseudoscalar(s). Both the triplet and the singlet states do not couple to the fermions, which leads to some very interesting phenomenology. This property also has an impact on rare decays like $b\to \mu \mu$ and $b\to s \gamma$ \cite{tssmyzero, infnr}.
Given the rich phenomenology and the specific predictions of this model, the  current analysis at the LHC and future colliders could be able to test and shed a light on this scenario by looking at its interesting signatures. 
 
\section*{Acknowledgements} 
P.B. wants to thank Eung Jin Chun and Ajit Mohan Shrivastava for the discussion regarding light pseudo-scalar as hot or cold dark matter candidate.  The work of C.C. is supported by a {\em The Leverhulme Trust Visiting Professorship} at the
University of Southampton at the STAG Research Centre and Mathematical Sciences. He thanks Kostas Skenderis and all the member of the Centre for the kind hospitality.

\appendix
\section{RG equations}
\label{RGs}
We list the RG equations at one-loop for the dimensionless coupling $\lambda_{T, S, TS}$, $\kappa, g_Y, g_L, g_c, y_{t, b}$. Here $t=\ln{\frac{\mu}{\mu_0}}$, where $\mu$ is the running scale and $\mu_0$ is the initial scale. 
\begin{align}
g_Y'(t)&=\frac{33}{80 \pi ^2}g^3_y(t),\\
g_L'(t)&=\frac{3}{16 \pi ^2}g^3_w(t),\\
g_c'(t)&=-\frac{3}{16 \pi ^2}g^3_c(t)
\end{align}
\begin{align}
y_t'(t)&=\frac{1}{16 \pi ^2}\Bigg(3
   y^3_ t(t)+y^2_b(t) y_t(t)\\
   &+\Bigg(-\frac{13}{15} g^2_Y(t)-3
   g^2_L(t)-\frac{16 g^2_c(t)}{3}+\frac{3 \lambda^2_T(t)}{2}+\lambda^2_S (t)+3y_t^2(t)\Bigg)
   y_t(t)\Bigg),\nn\\
 y_b'(t)&=\frac{1}{16 \pi ^2}\Bigg(3
   y^3_ t(t)+y^2_t(t) y_b(t)\\
   &+\Bigg(-\frac{13}{15} g^2_Y(t)-3
   g^2_L(t)-\frac{16 g^2_c(t)}{3}+\frac{3 \lambda^2_T(t)}{2}+\lambda^2_S (t)+3y_b^2(t)\Bigg)
   y_b(t)\Bigg)\nn,
\end{align}
\begin{align}
\lambda_S '(t)&=\frac{1}{16 \pi^2} \Bigg(4 \lambda^3_S (t)-\frac{3}{5}
   g^2_Y(t) \lambda_S (t)-3 g^2_L(t) \lambda_S (t)\\
   &+3 \lambda^2_T(t) \lambda_S (t)+6 \lambda^2_{TS}(t)
   \lambda_S (t)+2 \kappa^2(t) \lambda_S (t)+3
   \Big(y_t^2 (t)+y^2_b (t)\Big) \lambda_S (t)\Bigg),\nn
\end{align}
\begin{align}
\lambda_T'(t)&=\frac{1}{16 \pi^2} \Bigg(4 \lambda^3_T(t)-\frac{3}{5} g^2_Y(t) \lambda_T(t)-7
   g^2_L(t) \lambda_T(t)\\
   &+4 \lambda^2_{TS}(t) \lambda_T(t)+2 \lambda^2_S (t) \lambda_T(t)+3
   \Big(y_t^2 (t)+y^2_b (t)\Big) \lambda_T (t)\Bigg),\nn\\
\kappa '(t)&=\frac{1}{8 \pi
   ^2}3  \kappa (t) \Bigg(3 \lambda^2_{TS}(t)+\kappa^2(t)+\lambda^2_S (t)\Bigg),\\
\lambda_{TS}'(t)&=\frac{1}{8 \pi ^2}\lambda_{TS}(t)  \Bigg(-4 g^2_L(t)+\lambda^2_T(t)+7
   \lambda^2_{TS}(t)+\kappa^2(t)+\lambda^2_S (t)\Bigg).\nn
\end{align}

\section{Higgs Vertices}
\label{HVs}
The Higgs boson vertices are given in the gauge basis and the vertices in the Higgs boson mass basis can be obtained by the rotation defined in in Eq.~(\ref{rot}).  $\mathcal{R}^S_{ij}$ is the rotation matrix for CP-even neutral Higgs boson and $H_i=(H^0_{u,r}, H^0_{d,r}, S_r, T^0_r)$, $h_i=(h_1, h_2, h_3, h_4)$ are the CP-even neutral Higgs bosons in the gauge and mass basis respectively. For the pseudo-scalar we used the rotation matrix $\mathcal{R}^P$ which rotates the CP-odd neutral Higgs boson from their gauge basis, $A_i=(H^0_{u,i}, H^0_{d,i}, S_i, T^0_i)$ to the mass basis $a_i=(a_0, a_1, a_2, a_3)$. $\mathcal{R}^C$ is the corresponding rotation matrix that rotates the gauge basis $H^\pm_i=(H_u^+, T_2^+, H_d^-, T_1^-)$ to mass eigenstates $h^\pm_i=(h_0^\pm, h_1^\pm, h_2^\pm, h_3^\pm)$. Here $a_0$ and $h^\pm_0$ are the neutral and charged Goldstone bosons which give masses to the $Z$ and $W^\pm$ bosons respectively. 

\bea\label{rot}
h_i= \mathcal{R}^S_{ij} H_j\nn\\
a_i= \mathcal{R}^P_{ij} A_j\\
h^\pm_i= \mathcal{R}^C_{ij} H^\pm_j\nn
\eea
\bea\label{vrtx}
T^0_{r}\, H^0_{u,\, r/i}\, H^0_{d,\, r/i}  &: &\frac{A_T}{2} + \frac{\lambda_{TS}\lambda_T v_S}{\sqrt 2}-\lambda_S\,\lambda_{TS}\,v_T;\\
T^0_{i} \, H^0_{u,\, r/i}\, H^0_{d,\, i/r}  &: &\frac{\lambda_{TS}\lambda_T v_S}{\sqrt 2}-\frac{A_T}{2}  - \lambda_{TS}\lambda_S v_T;\\
T^0_{r}\, H^0_{u/d,\, r/i} \,H^0_{u/d,\, r/i}  &: &\lambda_T\,(\frac{v_T\lambda_T}{4}-\frac{v_S\,\lambda_S}{2 \sqrt 2});\\
T^0_{i}\, H^0_{u/d,\, r} \,H^0_{u/d,\, i}  &:&0;\\
T^0_{r} \,H^0_{u/d, \,r}\, S_{r} &:&\frac{1}{\sqrt 2}(v_{d/u}\lambda_{TS}-v_{u/d}\lambda_S)\lambda_T;\label{vhts}
\eea
\bea\label{vrtx1}
T^0_{r} \,H^0_{u/d, \,i}\, S_{i} &:&\frac{1}{\sqrt 2}v_{u/d}\,\lambda_T\lambda_{TS};\\
T^0_{i}  \,H^0_{u/d, \,r/i} \,S_{ i/r}  &:&-\frac{1}{\sqrt 2}
(v_{u/d}\lambda_S+v_{d/u}\lambda_{TS})\lambda_T;\\
T^0_{r} \,S_{r} \,S_{r} &:&v_T\,\lambda_{TS}(\kappa+2\lambda_{TS});\\
T^0_{i} \,S_{r} \,S_i &:& 2\kappa \lambda_{TS} v_T;\\
T^0_{r} \,S_{i} \,S_{i} &:&v_T\,\lambda_{TS}(2\lambda_{TS}-\kappa);
\eea
\bea
T^0_{r} \,T^0_{r} \,S_{r} &:&\lambda_{TS}\,v_S\,(\kappa+2\lambda_{TS})+ \frac{A_{TS}}{\sqrt 2};\\
T^0_{r} \,T^0_{i} \,S_{i} &:& 2\kappa\,v_S\,\lambda_{TS}-\sqrt 2 A_{TS};\\
T^0_{i} \,T^0_{i} \,S_{r} &:&\lambda_{TS}\,v_S\,(2\lambda_{TS}-\kappa)- \frac{A_{TS}}{\sqrt 2};\\
T^0_{r} \,T^0_{r/i} \,T^0_{r/i} &:&v_T\,\lambda_{TS}^2; \\
S_{r} \,S_{r} \,S_{r} &:&\frac{A_{\kappa}}{3\sqrt 2}+\kappa^2v_S;
\eea
\bea
S_{r} \,S_{i} \,S_{i} &:&\kappa^2v_S-\frac{A_{\kappa}}{\sqrt 2}; \\
S_{r} \,H^0_{u/d,\, r/i} \,H^0_{u/d,\, r/i} &:&\frac{1}{2}v_S\lambda_S^2-\frac{v_T \lambda_T \lambda_S}{2\sqrt 2}; \\
S_{r} \,H^0_{u/d,\, r/i} \,H^0_{d/u,\, r/i} &:&\frac{v_T \lambda_T \lambda_S}{\sqrt 2}-\kappa \lambda_S v_S  -  \frac{A_S}{\sqrt 2}; \\
S_{i} \,H^0_{u/d,\, r/i} \,H^0_{d/u,\, i/r} &:&\frac{A_S}{\sqrt 2} -\kappa \lambda_S v_S + \frac{v_T \lambda_T \lambda_{TS}}{\sqrt 2}; \\
S_{i} \,H^0_{u/d,\, r} \,H^0_{u/d,\, i} &:&0;
\eea
\bea
H^0_{u/d,\, r} \,H^0_{u/d,\, r/i} \,H^0_{u/d,\, r/i} &:&\frac{1}{8}(g^2_y+g^2_w)\,v_{u/d};\\
H^0_{u/d,\, r} \,H^0_{d/u,\, r/i} \,H^0_{d/u,\, r/i} &:&\frac{1}{2}(\lambda_S^2+\frac{1}{2}\lambda_T^2-\frac{1}{4}(g_y^2+g_w^2))v_{u/d};\\
H^0_{u/d,\, i} \,H^0_{u/d,\, r} \,H^0_{d/u,\, i} &:&0;\\
H^0_{u/d,\, r} \,T^0_{r} \,T^0_{r} &:&\frac{1}{4}v_{u/d}\lambda^2_T -\frac{1}{2}v_{d/u}\lambda_S\lambda_{TS};\\
H^0_{u/d,\, r} \,T^0_{i} \,T^0_{i} &:&\frac{1}{4}v_{u/d}\lambda^2_T +\frac{1}{2}v_{d/u}\lambda_S\lambda_{TS};
\eea
\bea
H^0_{u/d,\, i} \,T^0_{r} \,T^0_{i} &:&-v_{d/u}\lambda_S\lambda_{TS};\\
H^0_{u/d,\, r} \,S_{r} \,S_{r} &:&\frac{1}{2}\lambda_S(v_{u/d}\lambda_S- v_{d/u}\kappa);\\
H^0_{u/d,\, r} \,S_{i} \,S_{i} &:&\frac{1}{2}\lambda_S(v_{u/d}\lambda_S+ v_{d/u}\kappa);\\
H^0_{u/d,\, i} \,S_{r} \,S_{i} &:&- v_{d/u}\kappa\lambda_S;
\eea

\bea
T^0\, H^{+}_{u}\, H^{-}_{d}  &: &\frac{A_T}{\sqrt 2};\\
T^0\, (H^{+}_{u})^\dagger\, (H^{-}_{d})^\dagger  &: &\lambda_{TS}(v_S\lambda_T+\sqrt 2 v_T\lambda_S);\\
T^0\, H^{+/-}_{u/d}\, (H^{+/-}_{u/d})^\dagger  &: &\frac{\lambda_T}{2}(\lambda_S v_S+\frac{\lambda_T v_T}{\sqrt 2});\\
T^0\, H^{+/-}_{u/d}\, (H^{-/+}_{d/u})^\dagger  &: & 0;\\
T^0\, T^{-/+}_{1/2}\, (T^{-/+}_{1/2})^\dagger  &: & \frac{g_L^2}{\sqrt 2}v_T;\\
T^0\, T^{-}_{1}\, T^{+}_{2}  &: & 0;\\
T^0\, (T^{-}_{1})^\dagger\, (T^{+}_{2})^\dagger  &: & \sqrt 2 v_T(2 \lambda_{TS}-g^2_L);\\
T^0\, T^{-/+}_{1/2}\, (T^{+/-}_{2/1})^\dagger  &: & 0;\\
T^0\, T^{-/+}_{1/2}\, T^{-/+}_{1/2}  &: & 0;\\
T^0\, T^{-/+}_{1/2}\, H^{+/-}_{u/d}  &: & 0;\\
T^0\, (T^{-/+}_{1/2})^\dagger\, (H^{+/-}_{u/d})^\dagger  &: &\mp \frac{v_{u/d}}{2}(\lambda_T^2-g_L^2);\\
T^0\, T^{-/+}_{1/2}\, (H^{+/-}_{u/d})^\dagger  &: & 0;\\
S \,  H^{+/-}_{u/d}\, (H^{+/-}_{u/d})^\dagger  &: & \lambda_S(\lambda_T\frac{v_T}{2}+\lambda_S\frac{v_S}{\sqrt 2});\\
S \,  T^{+/-}_{2/1}\, (H^{+/-}_{u/d})^\dagger  &: & \frac{1}{\sqrt 2}\lambda_S\lambda_T v_{u/d};
\eea


\end{document}